\documentclass[preprint]{aastex}

\usepackage{graphicx}
\usepackage{color}
\usepackage{multirow}
\usepackage{epsfig}
\usepackage{natbib}
\usepackage{wasysym}
\usepackage{verbatim}

\newcommand{\LEC}{low-energy cutoff}
\newcommand{\LECs}{low-energy cutoffs}
\newcommand{\FIES}{flare-injected electron flux spectrum}
\newcommand{\MEFS}{mean electron flux distribution}
\newcommand{\chimap}{$\chi^{2}$-mapping}
\newcommand{\MCMC}{Markov chain Monte Carlo}
\newcommand{\bmcmc}{Bayesian/MCMC}
\newcommand{\MC}{Monte Carlo}
\newcommand{\covmat}{covariance matrix}
\newcommand{\sobs}{\mbox{$\mathbf{D}$}}
\newcommand{\sfit}{\mathbf{C}}
\newcommand{\varv}{\theta}
\newcommand{\vars}{\mathbf{\varv}}
\newcommand{\varm}{N_{\vars}}
\newcommand{\varhat}{\hat{\vars}}
\newcommand{\varihat}{\hat{\varv_{i}}}
\newcommand{\varvhat}{\hat{\varv}}
\newcommand{\brem}{bremsstrahlung}
\newcommand{\pdf}{probability density function}
\newcommand{\pdfs}{probability density functions}

\newcommand{\surfs}{hypersurfaces}
\newcommand{\surf}{hypersurface}
\newcommand{\cmat}{\alpha}
\newcommand{\unitsEM}{10^{49}\mbox{cm}^{-3}}
\newcommand{\keV}{keV}
\newcommand{\ssize}{scale-size}
\newcommand{\ssizes}{scale-sizes}
\newcommand{\janfive}{19 January 2005}
\newcommand{\jultwo}{23 July 2002}
\newcommand{\chit}{\chi^{2}}
\newcommand{\responsemat}{M_{ij}}
\newcommand{\ratename}{R}
\newcommand{\ratedef}[1]{\ratename^{#1}}
\newcommand{\ratedata}{\ratedef{D}}
\newcommand{\ratemodel}{\ratedef{C}}
\newcommand{\information}{\mathcal{I}}
\newcommand{\notcalc}{{\it not calculated}}
\newcommand{\notdet}{{\it not determined}}
\newcommand{\freqd}{F}
\newcommand{\fdy}{\varv}
\newcommand{\varlim}[1]{1\le #1 \le \varm}
\newcommand{\nelossbins}{n}
\newcommand{\nefluxbins}{m}
\newcommand{\NormGauss}{Normal}
\newcommand{\NormGaussly}{Normally}
\newcommand{\Xray}{X-ray}
\newcommand{\flRuxes}{fluxes}
\newcommand{\flRux}{flux}

\newcommand{\lapprox} {\, \lower3pt\hbox{$\sim$}\llap{\raise2pt\hbox{$<$}}\,}
\newcommand{\gapprox} {\, \lower3pt\hbox{$\sim$}\llap{\raise2pt\hbox{$>$}}\,}

\shorttitle{Estimating the properties of hard X-ray solar flare}
\shortauthors{Ireland et al.}

\begin{document}


\title{Estimating the properties of hard X-ray solar flares by
  constraining model parameters}


\author{J. Ireland}
\affil{ADNET Systems, Inc. at NASA Goddard Spaceflight Center, Greenbelt, MD 20771}

\author{A. K. Tolbert and R. A. Schwartz}
\affil{Catholic University of America at NASA Goddard Spaceflight Center, Greenbelt, MD 20771.}

\author{G. D. Holman and B. R. Dennis}
\affil{NASA Goddard Spaceflight Center, Code 671, Greenbelt, MD 20771}




\begin{abstract}
  We wish to better constrain the properties of solar flares by
  exploring how parameterized models of solar flares interact with
  uncertainty estimation methods.  We compare four different methods
  of calculating uncertainty estimates in fitting parameterized models
  to {\it Ramaty High Energy Solar Spectroscopic Imager} (RHESSI)
  \Xray\ spectra, considering only statistical sources of error.
  Three of the four methods are based on estimating the scale-size of
  the minimum in a \surf\ formed by the weighted sum of the squares of
  the differences between the model fit and the data as a function of
  the fit parameters, and are implemented as commonly practiced.  The
  fourth method is also based on the difference between the data and
  the model, but instead uses Bayesian data analysis and Markov chain
  Monte Carlo (MCMC) techniques to calculate an uncertainty
  estimate. Two flare spectra are modeled: one from the GOES
  (Geostationary Operational Environmental Satellite) X1.3 class flare
  of \janfive, and the other from the X4.8 flare of \jultwo.  We find
  that the four methods give approximately the same uncertainty
  estimates for the \janfive\ spectral fit parameters, but lead to
  very different uncertainty estimates for the \jultwo\ spectral fit.
  This is because each method implements different analyses of the
  \surf, yielding method-dependent results that can differ greatly
  depending on the shape of the \surf.  The \surf\ arising from the
  \janfive\ analysis is consistent with a \NormGauss\ distribution;
  therefore, the assumptions behind the three non-Bayesian uncertainty
  estimation methods are satisfied and similar estimates are found.
  The \jultwo\ analysis shows that the \surf\ is not consistent with a
  \NormGauss\ distribution, indicating that the assumptions behind the
  three non-Bayesian uncertainty estimation methods are not satisfied,
  leading to differing estimates of the uncertainty.  We find that the
  shape of the \surf\ is crucial in understanding the output from each
  uncertainty estimation technique, and that a crucial factor
  determining the shape of \surf\ is the location of the
  \LEC\ relative to energies where the thermal emission dominates.
  The \bmcmc\ approach also allows us to provide detailed information
  on probable values of the \LEC, $E_{c}$, a crucial parameter in
  defining the energy content of the flare-accelerated electrons.  We
  show that for the \jultwo\ flare data, there is a 95\% probability
  that $E_{c}$ lies below approximately 40~keV, and a 68\% probability
  that it lies in the range 7--36 keV.  Further, the \LEC\ is more
  likely to be in the range 25-35 keV than in any other 10 keV wide
  energy range.  The \LEC\ for the \janfive\ flare is more tightly
  constrained to $107 \pm 4$~keV with 68\% probability.  Using the
  \bmcmc\ approach, we also estimate for the first time \pdfs\ for the
  total number of flare accelerated electrons and the energy they
  carry for each flare studied.  For the \jultwo\ event, these
  \pdfs\ are asymmetric with long tails orders of magnitude higher
  than the most probable value, caused by the poorly constrained value
  of the \LEC.  The most probable electron power is estimated at
  $10^{28.1} \mbox{erg sec}^{-1}$, with a 68\% credible interval
  estimated at $10^{28.1-29.0}\mbox{erg sec}^{-1}$, and a 95\%
  credible interval estimated at $10^{28.0-30.2}\mbox{erg sec}^{-1}$.
  For the 19 January 2005 flare spectrum, the \pdfs\ for the total
  number of flare accelerated electrons and their energy are much more
  symmetric and narrow: the most probable electron power is estimated
  at $10^{27.66 \pm 0.01}\mbox{erg sec}^{-1}$ (68\% credible
  intervals).  However in this case the uncertainty due to systematic
  sources of error is estimated to dominate the uncertainty due to
  statistical sources of error.
\end{abstract}





\section{Introduction}\label{sec:intro}

The detailed understanding of solar flares requires an understanding
of the physics of accelerated electrons, since electrons carry a large
fraction of the total energy released in a flare
\citep{1971SoPh...17..412L, 1976SoPh...50..153L, 2004JGRA..10910104E,
  2005JGRA..11011103E}.  Since we cannot measure the electron flux
      {\it in situ}, the behavior of the flare-accelerated electrons
      is inferred from the photons emitted by their interaction with
      the ambient plasma.  For a general inhomogeneous optically thin
      source of plasma density $n(\mathbf{r})$ and electron flux
      density\footnote{In this paper, ``flux density'' refers to an
        amount per unit area per unit time.} energy spectrum $F(E,
      \mathbf{r})$ (electrons $\mbox{cm}^{-2}\mbox{s}^{-1}
      \mbox{keV}^{-1}$) in volume $V$ for electron energy $E$, the
      \brem\ photon flux density energy spectrum $I(\epsilon)$ (photons
      $\mbox{cm}^{-2}\mbox{s}^{-1} \mbox{keV}^{-1}$ at Earth distance
      $R$) can be written \citep{1971SoPh...18..489B,
        2003ApJ...595L.115B} as
\begin{equation}\label{eqn:basic}
I(\epsilon) =
\frac{ \overline{n}V }{4\pi R^{2}}
\int_{\epsilon}^{\infty}
\overline{F}(E)Q(\epsilon, E) dE,
\end{equation}
where $\overline{n} = \int_{V} n dV/V$, $\overline{F}(E)$ is the
\MEFS, $\overline{F}(E) = \int_{V}
n(\mathbf{r})F(E,\mathbf{r})dV/(\overline{n}V)$, and $Q(\epsilon, E)$
is the \brem\ cross-section differential in photon energy
$\epsilon$. In this paper we model the photon flux density energy
spectrum as the sum of emission due to a \FIES\ interacting with a
target, and emission from hot plasma with a Maxwellian distribution of
speeds corresponding to some temperature $T$.

The {\it Ramaty High Energy Solar Spectroscopic Imager} (RHESSI, Lin et
al. 2002) flags all photons detected in any one of the nine germanium
detectors by the time of occurrence (to 1 microsecond), the amount of
energy lost by the photon in the detector (in 0.3-keV-wide pulse
height analyzer (PHA) bins), and the detector number. For spatially
integrated spectral analysis, the counts can be combined arbitrarily
over different detectors and PHA bins.

We define \sobs = $(D_{1},...,D_{i},...D_{n})$ as the number of counts
observed in a given set of energy-loss bins labeled in the range $1\le
i\le \nelossbins$ in a given time interval.  These counts are noisy,
and are assumed to be drawn from a Poisson distribution with a mean of
$C_{i}$,
\begin{equation}\label{eqn:poisson}
p(D_{i}) = \frac{C_{i}^{D_{i}}}{D_{i}!}e^{-C_{i}}.
\end{equation}

The measured count rate $\ratedata_{i}$ in energy-loss bin `$i$' is
determined from the measured counts $D_{i}$ divided by the live
time\footnote{The live time is the observation time minus the dead
  time.  The dead time is the amount of time that the detector cannot
  respond to an incoming photon.} $t_{LT}$.  The predicted count rate
$\ratemodel_{i}$ arises from the incident photon \flRux\ rate via
\begin{equation}\label{eqn:crm}
\ratemodel_{i} = \responsemat I_{j};
\end{equation}
that is, the predicted count rate in an energy-loss bin `$i$' is
modeled via a detector response matrix $\responsemat$ for an incident
photon \flRux\ spectrum $I_{j}$, where the index $j$, $1\le j \le
\nefluxbins$, labels energies at which the incident photon spectrum is
calculated.  The response matrix $\responsemat$ is calculated by
RHESSI Solarsoft routines once the count energy-loss bins (indexed by
`$i$') and incident photon energies (indexed by `$j$') are defined.
The incident photon \flRux\ energy spectrum is deduced by comparing the
observed with the predicted count rates in all energy bins assuming a
model for the photon \flRux\ energy spectrum until some criterion for
agreement is met.



One goal of RHESSI data analysis is to recover the electron \flRux\
energy spectrum $F(E, \mathbf{r})$ from the detected counts $D_{i}$ in
a given time interval.  In general, this requires detailed knowledge
of the energy losses suffered by the \brem-producing electrons in the
emitting volume.  It is often only practical to recover
$\overline{F}(E)$; to do this, two approaches are commonly taken.

Since the rates are measured, and everything other than
$\overline{F}(E)$ is known (either calculated, measured or assumed),
$\overline{F}(E)$ can be obtained through Equations \ref{eqn:basic}
and \ref{eqn:crm}.  This approach is known as inversion.  The
advantage of inversion is that one does not make an assumption as to
the nature of the \MEFS.  The disadvantage of this approach is that
noise in the observed data and errors in instrument calibration can
lead to the creation of spurious features in the solution.  This
effect can be mitigated by adding extra constraints to the inversion
process which forces the solution to be smooth across energy bins
(note that this is required by the \brem\ process and RHESSI's energy
resolution).  Consider discretizing Equation \ref{eqn:basic} by energy
bins to yield a matrix expression,
\begin{equation}\label{eqn:tik}
\mathbf{J} = \mathbf{A}\mathbf{F}
\end{equation}
where $\mathbf{J}$ is a $m$-element vector representing the observed
number of photons $I(\epsilon)$, $\mathbf{A}$ is a $m \times n$=matrix
representing $Q(\epsilon, E)$ and $\mathbf{F}$ is a $n$-element vector
representing the mean electron spectrum $\overline{F}(E)$.  The
standard approach is to minimize the residual
\[
||\mathbf{A}\mathbf{F}-\mathbf{J}||
\]
for $\mathbf{F}$ where $||\cdot||$ is the Euclidean norm.  This matrix
problem can be ill-posed due to the noise sources discussed above, or
by $\mathbf{A}$ being ill-conditioned or singular.  Regularization
mitigates these issues by imposing extra constraints on the solution
for $\mathbf{F}$. Tikhonov regularization does this by adding an extra
term $||\mathbf{\Gamma}\mathbf{F}||$ for some choice of Tikhonov matrix
$\mathbf{\Gamma}$, to the above minimization problem, yielding
\begin{equation}\label{eqn:tikh}
\min_{\mathbf{F}}\left(
\mathbf||\mathbf{A}\mathbf{F}-\mathbf{J}|| + ||\mathbf{\Gamma}\mathbf{F}||
\right).
\end{equation}
\citet{2003ApJ...595L.127P} demonstrate a Tikhonov-regularized
inversion algorithm that takes the observed counts and finds
$\overline{F}(E)$ and the uncertainty on $\overline{F}(E)$.
\citet{2003ApJ...595L.127P} show that this method led to an unexpected
`dip' in the mean electron spectrum which is thought (in most cases)
to arise from the presence of a significant photospheric albedo \flRux\
contributing to the observed \Xray\ \flRux\
\citep{2006A&A...446.1157K,2008SoPh..252..139K}.

In the second approach, known as forward fitting, a parameterized
model for the \MEFS\ $\overline{F}(E)$ is used to describe the photon
\flRux\ $I_{j}$ incident at RHESSI. The photon emission, parameterized by
$\vars$ ($\varm$ variables) is
\begin{equation}\label{eqn:param}
I_{j} = I_{j}(\vars).
\end{equation}
A fitting process is then used to find values of the parameters that
best reproduce the counts $D_{i}$ observed by
RHESSI.  The disadvantage of this method is that the spectral model is
prescribed rather than derived, and so features that are not in the
model cannot be described by it, although their presence in the data
may be indicated by the residuals \citep{2006ApJ...643..523B}.  The
advantages of this method are that by judicious choice of
parameterization the major features of the spectrum can be modeled,
and values to the parameters with uncertainty estimates can be
obtained.

In this paper, we use the forward fitting approach and consider four
different methods of estimating a range of ``acceptable'' model
parameter values that describe our understanding of the flare within
the confines of the model.  By comparing different methods, we seek to
understand the differences in the final answer that may be brought
about by the way the estimates were obtained.  Further, by comparing
two different spectra we can better understand how, for a given model,
the estimated parameter values and errors are influenced by the data.
It is assumed that the only source of noise is the Poisson
distribution that follows naturally from independent photon events
(Eq.  \ref{eqn:poisson}).

Systematic error sources are undoubtedly important in determining the
uncertainties in the model parameters \citep{2011ApJ...731..126L}, but
they are not explicitly included in the uncertainty determination
methods described below. Two types of systematic uncertainties are
common in this type of spectroscopy, integral and
differential. Integral uncertainties are basically the uncertainties
in the overall sensitivity of a given detector. Based on comparisons
of flare spectra measured with different detectors, they are known to
be smaller than approximately 10\%. They affect primarily the absolute
value of the emission measure in the thermal model and the total
electron \flRux\ in the nonthermal electron spectrum. The differential
uncertainties are basically the uncertainties in the sensitivity in
each energy bin with respect to its neighbors. They affect primarily
model parameters that depend on the slope of the measured
spectrum. They are therefore important for the temperature in the
thermal model and the low energy cutoff and power-law index of the
nonthermal electron spectrum. For RHESSI, the differential
uncertainties are less than 1\% and are generally negligible compared
to the statistical uncertainties. \citet{2009ApJ...699..968M} (using
detectors 1, 3, 4, 5, 6, and 9) and \citet{2011ApJ...731..106S} (using
detectors 1, 3, 4, 6, 8 and 9) show that there is scatter in the
best-fit parameter values determined from different individual
detectors for the flare models they considered but that the range of
the scatter indicates that the systematic errors are not significantly
greater than the statistical errors. The systematic uncertainties are
not important in developing a basic understanding of how each
uncertainty determination method behaves in the presence of noisy data
and consequently they have not been included in the analysis done for
this paper.

The observations and spectral models are described in Section
\ref{sec:obs}.  Section \ref{sec:parest} describes the parameter and
uncertainty estimation methods used.  Section \ref{sec:res}
describes the results and Section \ref{sec:discussion} discusses the
implications of these results for fitting spectral models to RHESSI
data.

\section{Spectral model and observations}\label{sec:obs}

In the \Xray\ energy range covered by RHESSI
\citep{2002SoPh..210....3L} -- generally from $\sim$3 keV up to a few
hundred keV -- the emitted photon spectrum is modeled as the sum of a
thermal component that generally dominates at the lower \Xray\
energies, typically below $\sim$10--20 keV, and a non-thermal
component that dominates at higher energies. The thermal component is
the line and continuum emission from the flare-heated plasma. The line
emission is mainly from transitions in highly ionized iron --
primarily FeXXV -- that appears in the RHESSI spectrum as an
unresolved peak at 6.7 keV with a much weaker feature at $\sim$8
keV. The continuum emission is a combination of free-free emission
(\brem) and free-bound emission (recombination radiation).

For our spectral analysis, we have used the thermal
line-plus-continuum spectra provided by CHIANTI
\citep{1997A&AS..125..149D, 2009A&A...498..915D} assuming an
isothermal plasma with the ionization balance given by
\citet{1998A&AS..133..403M} and the ``sun coronal'' abundances given
by \citet{1992ApJS...81..387F}. The only free parameters are the
temperature ($kT$ in keV) and the volume emission measure ($EM$ in
$cm^-3$).

The thermal continuum emission is made up of the sum of \brem\ (or
free-free) emission and free-bound emission. The form of the
bremsstrahlung contribution as a function of photon energy $\epsilon$
is approximately
\begin{equation}\label{eqn:thermal}
  I_{thermal}(\epsilon) \propto \frac{[EM]}{\epsilon T^{1/2}}
  \exp(-\epsilon/kT),
\end{equation}
where $k$ is Boltzmann's constant  and $I_{thermal}$ is in units of
photons $s^{-1} \mbox{erg}^{-1}$ \citep{1988psf..book.....T}.  The
free-bound continuum spectrum has a similar dependency on EM and T.

The non-thermal component of the measured \Xray\ spectrum is
\brem\ from flare-accelerated electrons interacting with the ambient
medium. Following \citet{1971SoPh...18..489B}, we assume a cold, thick
target, meaning that the electrons collisionally lose their energy in
cold, fully ionized plasma as they radiate.  The energy loss rate per
unit distance $x$ as an electron with speed $v$ streams through the
ambient plasma is $dE/dx = -2Kn_e(x)/(mv^2)$, where $m$ is the
electron mass, $n_e(x)$ is the number density of plasma electrons, and
$K$ is approximately constant \citep[see][]{2011SSRv..159..107H}.
Using this result, the mean electron \flRux\ becomes 
\begin{equation} \label{eqn:coll_mef}
\overline{F}(E) = {1 \over \overline{n}V} \frac{mv^2}{2K}\int_E^{\infty} F_0(E_0) dE_0,
\end{equation}
where $F_0(E_0)$ is now the injected electron \flRux\ energy spectrum
(electrons s$^{-1}$ keV$^{-1}$).  We use the following broken
power-law functional form for the spectrum of injected electrons:

\begin{equation}
     \label{eqn:powerlaw}
     F_{0}(E_0) = A\left\{
         \begin{array}{ll}
             0 & E_0 < E_{c} \\
             (E_0/E_{p})^{-\delta_{1}} & E_{c} \le E_0 < E_{b} \\
             (E_0/E_{p})^{-\delta_{2}} (E_{b}/E_{p})^{\delta_{2}-\delta_{1}} & E_{b} \le E_0 < E_{h}\\
             0 & E_0 \ge E_{h} \\
         \end{array}
     \right.
     .
\end{equation}

\noindent The seven parameters of this nonthermal component are the
normalization parameter $A$, the low- and high-energy cutoffs, $E_{c}$
and $E_{h}$, the pivot energy $E_{p}$, the break energy $E_{b}$, and
the power-law indices below and above the break energy, $\delta_{1}$
and $\delta_{2}$, respectively.  The radiated \Xray\ spectrum is
modeled as the sum of the isothermal component and
Equation~\ref{eqn:basic}, where $\overline{F}(E)$ is given by
Equations \ref{eqn:coll_mef} and \ref{eqn:powerlaw}.  The \Xray\
emission is assumed to be isotropic and, with this assumption, the
contribution \flRux\ from photospheric albedo to the total incident \Xray\
at the instrument can be estimated \citep[see][]{2011SSRv..159..301K}.

We model RHESSI spectral data from two flares -- the GOES class X1.3
flare on \janfive\ starting at 08:03 UT, and the X4.8 flare starting
at 00:18 UT on \jultwo.  We choose these flares because previous
studies have shown that the \LEC\ - $E_{c}$ - is estimated to lie in
very different portions of the spectrum.  In the \jultwo\ event, the
\LEC\ of the flare-accelerated electrons is estimated to have an
energy in the region where the observed hard \Xray\ emission is
thermally dominated.  This makes it difficult to place limits on the
\LEC\ since it is difficult to determine the signal of the
flare-accelerated electrons against the dominant thermal
\brem\ emission.  Most flares are thought to have \LECs\ close to or
in the region where the emission is dominated by thermal \brem.  In
contrast, \citet{2009ApJ...699..917W} studied the \janfive\ event, and
found that late in the impulsive phase, the \LEC\ energy much higher
than energies at which the thermal \brem\ dominates.  Therefore,
thermal \brem\ cannot be a significant factor in determining the
uncertainty in the \LEC\ for this flare.  The \LEC\ is one of the most
important properties of a flare as its value strongly influences the
estimated flare-accelerated electron energy content.  Therefore,
knowledge of the uncertainty in the \LEC\ directly influences
knowledge of the energy content of the flare.  Hence, these two flares
and the models used to study them are good test-beds for understanding
how different uncertainty estimation methods operate when generating
uncertainties for parameters that are crucial for understanding the
properties of solar flares.

Table \ref{tblbrd} has details of the two flares and the two spectral
accumulation times chosen, the models used, and the best-fit parameter
values obtained that fit the spectral models to the data (see Section
\ref{sec:nlls}). These two spectra were chosen because they were both
well observed with RHESSI and they highlight the excellent spectral
capabilities of the cooled germanium detectors of this instrument
\citep{2002SoPh..210...33S}. Both flares have been extensively
analyzed previously -- see for example \citet{2009ApJ...699..917W} for
the \janfive\ flare and \citet{2003ApJ...595L..97H} for the
\jultwo\ flare. The most notable difference between the two spectra is
that the first has a \LEC\ in the electron spectrum of over 100 keV,
well above the thermal component. This is in contrast to the second
flare where the \LEC\ is estimated to be below $\sim$ 40 keV
\citep{2003ApJ...595L..97H} and consequently difficult to determine
because of the dominance of the thermal component at lower energies.
This difference between these two flares motivates their selection for
this study.  These two flare events are good candidates that allow us
to explore how well we can determine the value of the crucial
\LEC\ parameter (and flare properties that depend on it) given the
data, the model, and the uncertainty estimation methods used.

Traditionally, RHESSI spectral analysis involves summing data from
multiple RHESSI detectors to improve counting statistics -- see for
example \citet{2009ApJ...705.1584S}. Instead of this usual approach,
we chose to use data from just one detector with good energy
resolution and sensitivity -- detector \#4. This allowed us to apply
the most accurate corrections for energy resolution and calibration,
pulse pile-up, and background subtraction. In the time periods
selected, the count rates were sufficiently high that selecting a
single detector did not seriously degrade the spectroscopy capability
up to the highest energies considered of $\sim$500 keV. The energy bin
widths were chosen to be as narrow as possible to preserve spectral
details resolvable with the detector's $\sim$1~keV FWHM spectral
resolution while maintaining $>$30 counts in each bin as required for
the $\chi^2$ analysis procedure to be approximately valid
\cite{citeulike:418876}.  The only part of the spectral data that is
affected by small numbers is at the high energy part of the spectrum,
well away from the low energy part of the spectrum. At these energies,
the simple \NormGauss\ approximation to the Poisson distribution --
($Poisson(\lambda)\approx N(\lambda,\sqrt{\lambda})$ for $\lambda$
'large') -- is no longer appropriate.  However, the gross properties
we are most interested in - flare energy content, the number of
flare-accelerated electrons and the \pdf\ of the \LEC, are largely
unaffected by a biassed fit of a spectral model to the data at high
energies, since these properties are largely determined by the flare
spectrum at energies where the \NormGauss\ distribution can be used.
We can assert this for the flares studied in this analysis because
these are relatively large flares with large numbers of counts.  The
vast majority of flares are smaller than the ones studied here, and
therefore fits or parameterized models to the data are more likely to
suffer from biassed fits over more extensive energy
ranges\footnote{It should also be noted that even although a
  substantial part of the spectrum have large enough counts, biassed
  values to the fit are still possible when minimizing a
  $\chi^{2}$-like expression - see \citet{1979ApJ...228..939C} and
  also \citet{2009ApJ...693..822H} and references therein.}.

\begin{figure}[pht]
\begin{center}
     \includegraphics[width=0.4\textwidth]{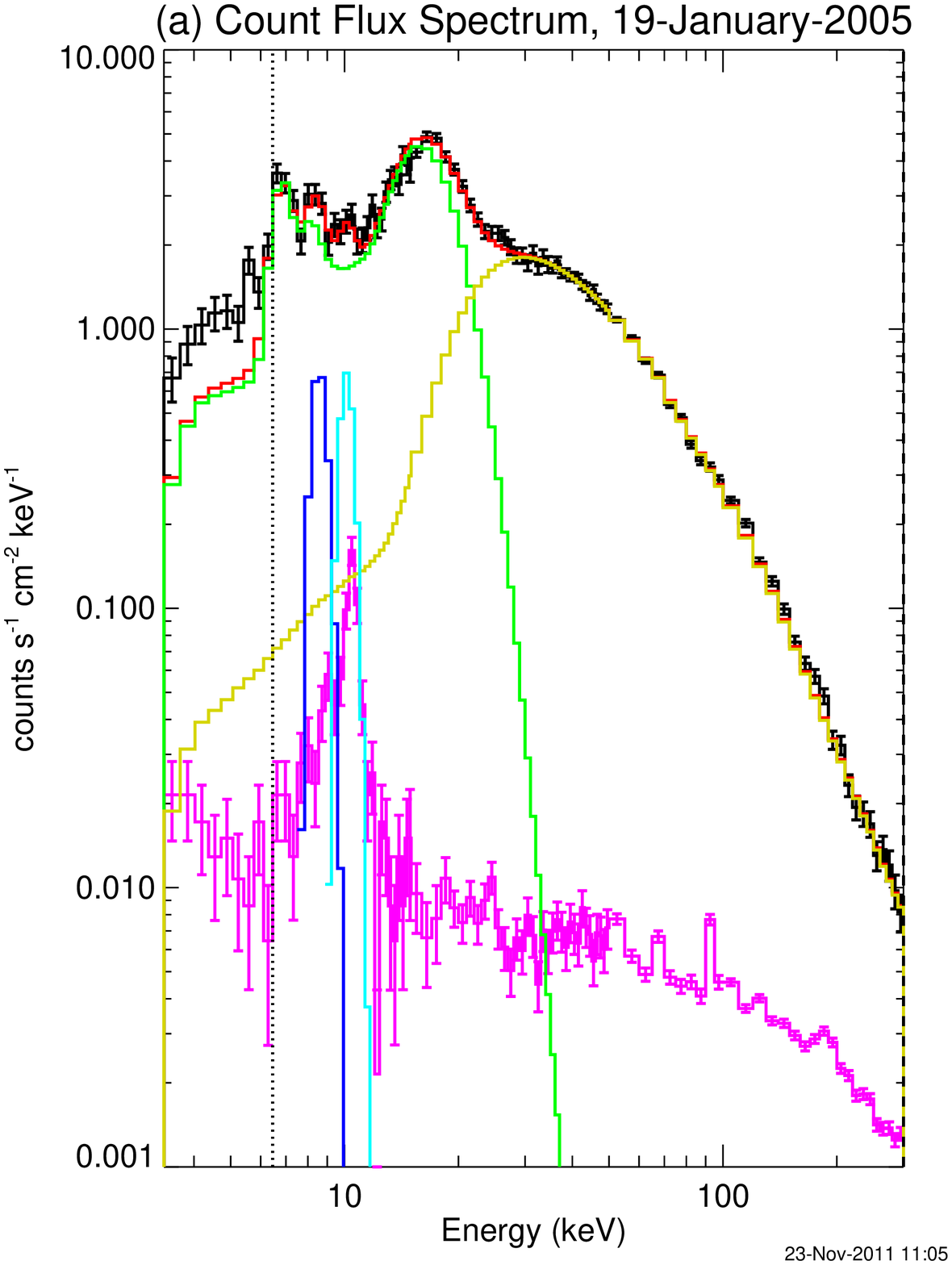}
     \includegraphics[width=0.4\textwidth]{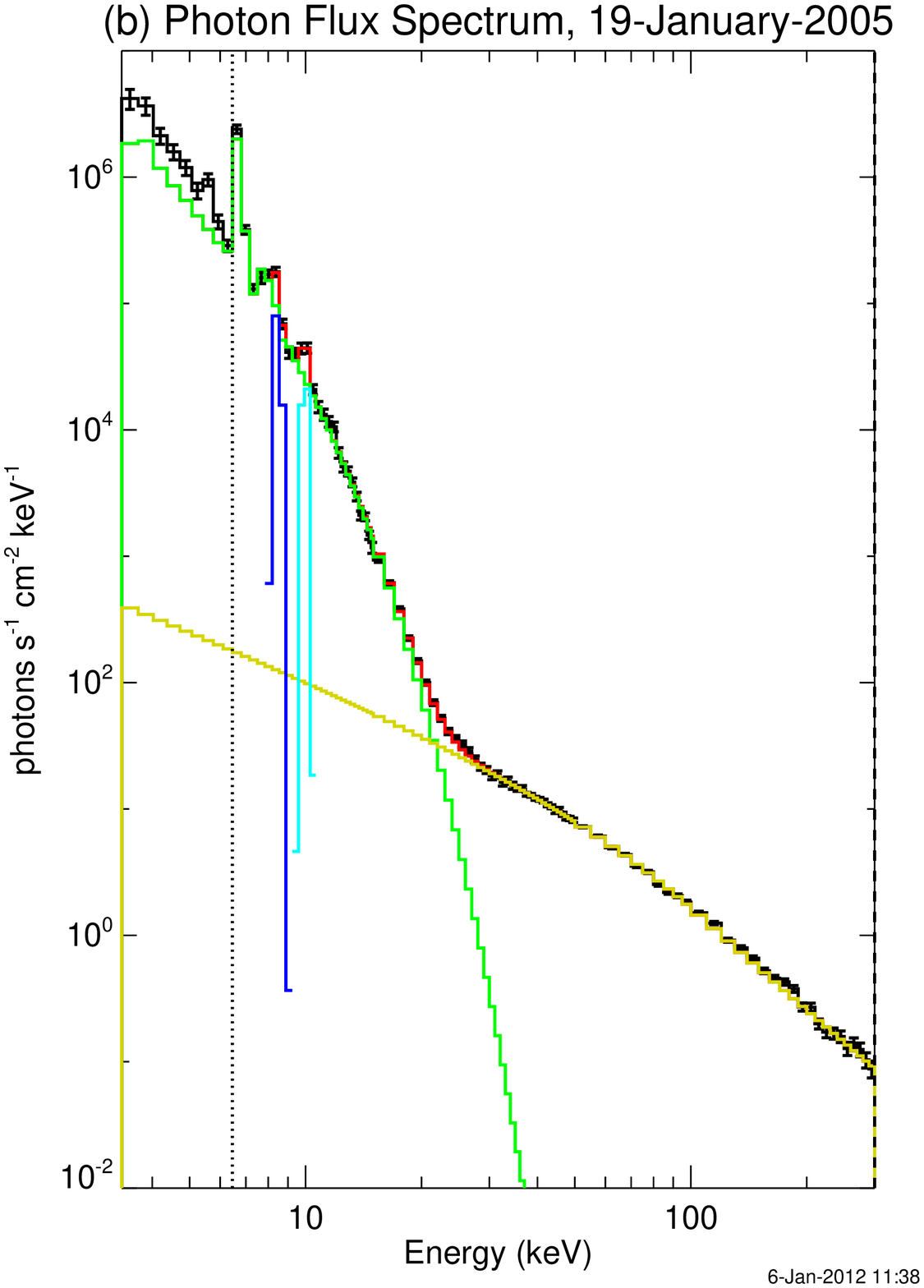}
\end{center}
     \caption{Count and photon spectra for the \janfive\ flare in the
       analysis period 08:26:00 to 08:26:20 UT. (a) The histograms
       with $\pm1 \sigma$ statistical error bars represent the
       background-subtracted count \flRuxes\ (black) and the background
       \flRuxes\ (pink) vs. energy loss in the detector. The smooth
       curves represent the different components of the model used to
       fit the data as follows: isothermal (green), thick-target
       \brem\ (yellow), Gaussians (blue and cyan). The sum of all the
       components is shown in red. (b) Incident photon flux (in units
       of photons $s^{-1} cm^{-2} keV^{-1}$) vs. photon energy with
       the different components of the model shown in the same colors
       as in (a).  The energy range used for the spectral fits lies
       between the vertical line at 6.45 keV and the edge of the plot
       at 300 keV.}
     \label{jan_spectra}
\end{figure}

Both flares have been extensively analyzed previously –- see for
example \citet{2009ApJ...699..917W} for the 19 January 2005 flare and
\citet{2003ApJ...595L..97H} for the 23 July 2002 flare.  For ease in
comparing results in each case, we have generally followed their lead
in choosing background spectra, energy ranges, model components,
fitting procedures, etc. in the spectral analysis.  Table \ref{tblbrd}
has details of the two flares and the two spectral accumulation times
chosen, the models used, and the best-fit parameter values obtained
that fit the spectral models to the data (see Section
\ref{sec:nlls}). Corresponding count flux\footnote{By count flux we
  mean the measured count rate per keV divided by a nominal detector
  area corrected for grid transmission, equal to $38 \mbox{cm}^{2}$
  for the single detector used in our analysis.} and photon flux
spectra are shown in Figures \ref{jan_spectra} and \ref{july_spectra}.
The model count flux spectrum is computed by taking the best fit
photon spectrum and convolving it with the instrument response matrix.
Figures \ref{jan_spectra}b and \ref{july_spectra}b shows the best fit
photon spectrum and the photon spectrum derived from the measured
count flux spectrum using the ratio of the best fit photon spectrum to
the measured counts in each energy bin.  The units in Figures
\ref{jan_spectra} and \ref{july_spectra}b are photons $s^{-1} cm^{-2}
keV^{-1}$.)

\subsection{19 January 2005}
\label{19Jan2005}

The first flare considered was the GOES X1.3 flare that peaked at
08:22 UT on \janfive~on the solar disc at N15W51. We used the RHESSI
observations of this flare from 08:26:00 - 08:26:20 UT, the same time
interval when \cite{2009ApJ...699..917W} found an unusually hard
spectrum during the final peak of the impulsive phase, possibly
resulting from a \LEC\ in the electron spectrum as high as 120 keV
(see their Figure 1 for RHESSI light-curves of this event).

We used the standard procedures that form OSPEX, the standard spectral
analysis package used in RHESSI data analysis, to determine the
best-fit parameters of the thermal and nonthermal components of the
incident photon spectrum. As is common in RHESSI data analysis, the
background spectrum that was subtracted from the measured count rate
spectrum was calculated by linear interpolation in time between
spectra measured before and after the flare. The estimated background
spectrum is about an order of magnitude less than the flare spectrum
at all energies considered.  The background can therefore be
considered as having very little influence on the final \pdfs\ of the
model parameters and the gross properties of the flare such as its
energy content and the number of flare accelerated
electrons. Following \cite{2009ApJ...699..917W}, we included two
narrow Gaussian-shaped emission lines in the model photon spectrum to
accommodate features in the count-rate spectra that are believed to be
instrumental in origin.

We included the standard corrections for energy calibration
adjustments and pulse pile-up, but these did not play a significant
role for the selected time interval since the attenuators were in the
A3 state (both thick and thin attenuators in place) resulting in
relatively low counting rates. The albedo component was not included
here, although it was included by \cite{2009ApJ...699..917W}. We found
that adding the albedo component did not significantly alter the
fitted parameters or the estimates of the uncertainties. We used the
following energy bins for this flare: $1/3$ keV from 3 to 15 keV, 1
keV from 15 to 50 keV, 5 keV from 50 to 100 keV, and 10 keV from 100
to 300 keV. The photon spectrum was extended above the fitted energy
range up to 600 keV to allow for non-photopeak response of the
detector.

Again, following \cite{2009ApJ...699..917W}, we modeled the thermal
component with a single-temperature function from CHIANTI using
coronal abundances and a \cite{1998A&AS..133..403M} ionization
balance. The nonthermal component was modeled assuming thick-target
interactions of electron with a single-power-law spectrum at energies
above $E_c$.  This is accommodated in Eq. \ref{eqn:powerlaw} by fixing
both $\delta_{2}$ at the default value of 6.0 and $E_{b}$ at 32 MeV so
that they have no significant effect on the \brem\ \Xray\ spectrum in
the fitted photon energy range below 300 keV.  $E_{h}$ was fixed at 32
MeV so that, like $E_{b}$, it has negligible effect on the
\brem\ \Xray\ spectrum in the fitted energy range, and so is equivalent
to having no cutoff at all.  For this flare, the normalization was
taken to be $F_0$, the total integrated electron flux over the
electron energy range from $E_{c}$ to $E_{h}$ with $E_{p}$ fixed at 1
keV, instead of $A$ in Eq. \ref{eqn:powerlaw}.  The advantage in
normalizing to $F_0$ is that this is a physically interesting
quantity.  The disadvantage is that it is strongly dependent on the
value of both the \LEC\ and the spectral index. For the conditions
described here, $F_0 = AE_c^{1-\delta_1}/(\delta_1 - 1)$.  The package
OSPEX was configured to use this implementation of Equation
\ref{eqn:powerlaw} for this flare.  An alternate implementation was
required for the \jultwo\ event (see Sections \ref{23July2002} and
\ref{sec:res:2002jul23}).

In our detailed spectral analysis and assessment of uncertainties, we
had a total of seven free parameters -- EM, kT, $F_0$, $E_c$,
$\delta_1$, $G_1$ and $G_2$ -- (see Table \ref{tblbrd}). Other
parameters covering the instrumental effects - energy calibration,
pulse pile-up, and Gaussian features below 10 keV - were determined
from the analysis of the count-flux spectra for other time intervals
and other flares, and then fixed for the subsequent determination of
uncertainties in this time interval. The amplitudes of the two
Gaussians ($G_{1}$ and $G_{2}$) were free during the spectral fits.


\subsection{23 July 2002}
\label{23July2002}

The second flare considered was the GOES X4.3 flare\footnote{Many more
  details concerning this flare can be found in the special issue of
  the Astrophysical Journal Letters (vol. 595) dedicated to its
  study.} that peaked at 00:35 UT on \jultwo\ from a location closer
to the limb at S13E72 than the first event. Following
\cite{2003ApJ...595L..97H}, we chose to analyze the time interval from
00:30:00 to 00:30:20.250 UT during the first peak of the impulsive
phase (see their Figure 1 for RHESSI \Xray\ light-curves of this event;
see also \citet{2003ApJ...595L..69L}, their Figure 1 for a lightcurve
of the GOES \Xray\ flux).  The measured \Xray\ spectrum was again
assumed to be the sum of an isothermal spectrum and the thick-target
\brem\ spectrum from non-thermal electrons with the broken power-law
of Eq.\ \ref{eqn:powerlaw}.  In this case, the full double power-law
was assumed with the break energy, $E_b$, and the second power-law
index, $\delta_2$, both free parameters. The normalization constant
for this flare, A in Eq.\ \ref{eqn:powerlaw}, was defined as the
electron \flRux\ at the pivot energy $E_p$ that was fixed at 50 keV. As
with the first flare, the high energy cutoff to the electron spectrum
$E_h$ was set at 32 MeV to ensure that it had no significant effect in
the fitted photon energy range.

\begin{figure}[pht]
\begin{center}
     \includegraphics[width=0.4\textwidth]{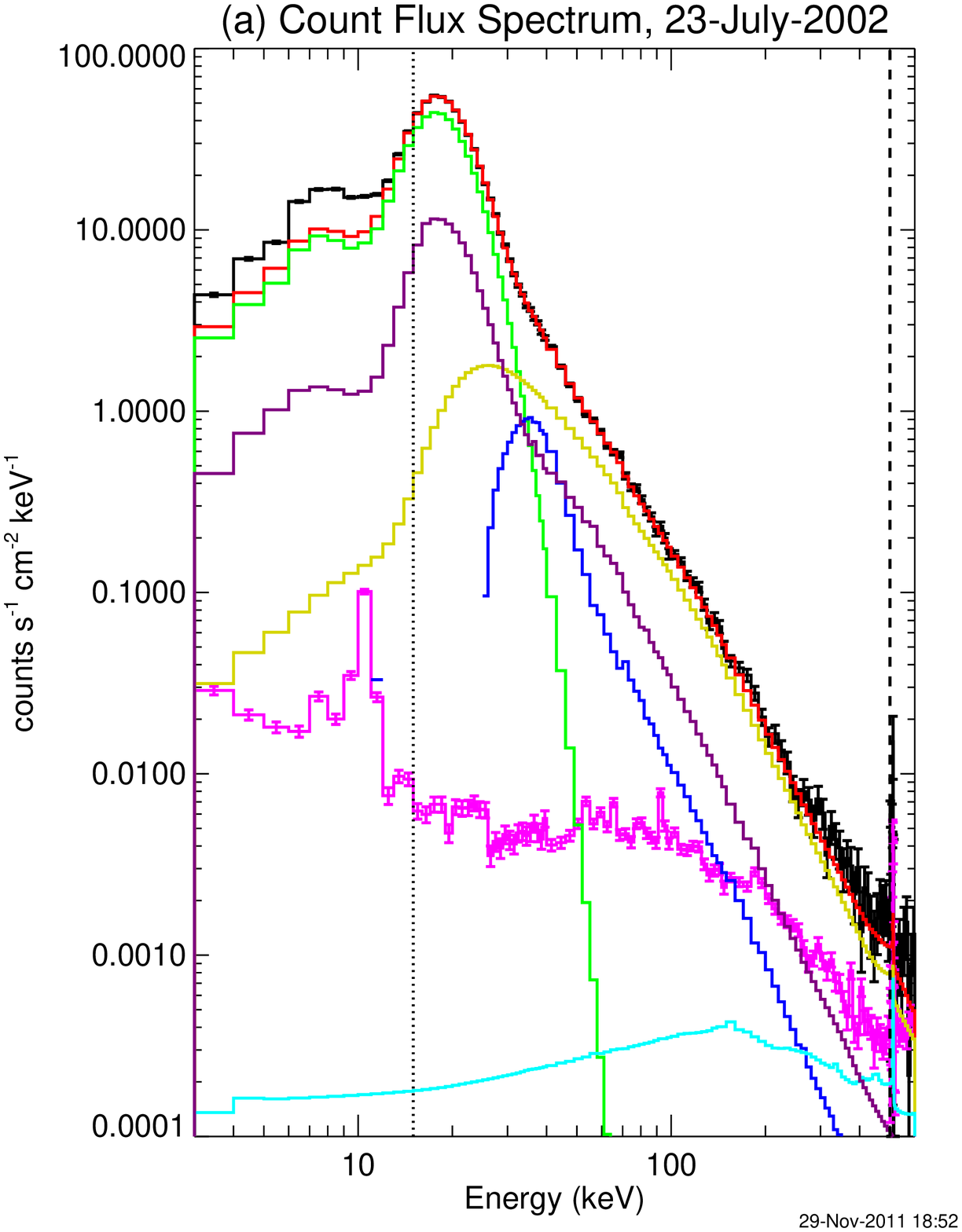}
     \includegraphics[width=0.4\textwidth]{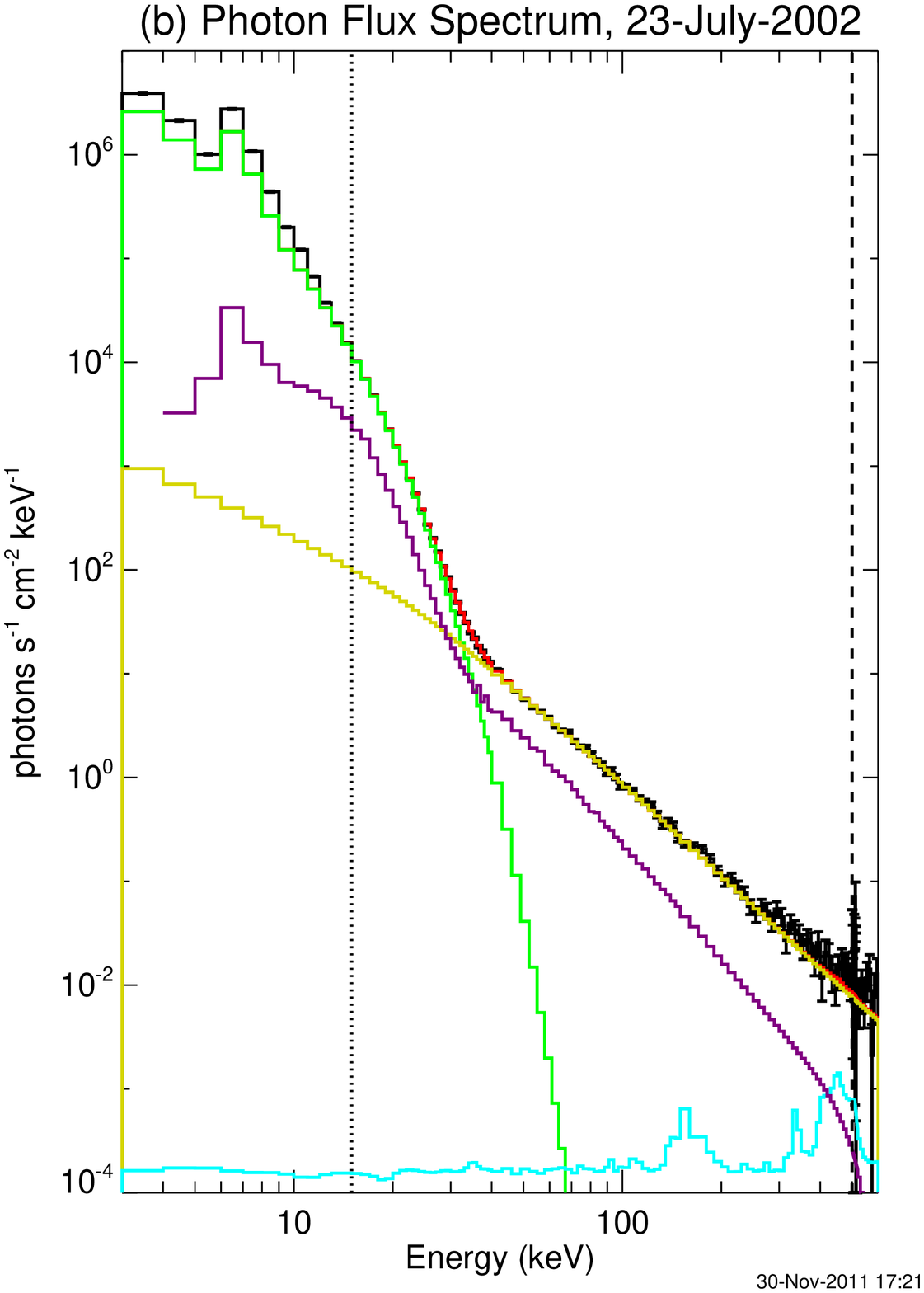}
\end{center}
     \caption{Similar to Fig.\ \ref{jan_spectra} for the \jultwo\
       flare. The following three additional components are included in
       this plot: albedo (purple), pulse pile-up (blue), and the
       nuclear template (cyan). The two Gaussians shown in
       Fig. \ref{jan_spectra} were not used for this fit. The energy
       range used for the spectral fits lies between the two vertical
       lines at 15 and 500 keV.}
     \label{july_spectra}
\end{figure}

The following 130 energy bins were used for this event: 1-keV wide
bins from 3.0 to 40 keV, 3-keV from 40 to 100 keV, 5-keV bins from 100
to 150 keV, 10-keV bins from 150 to 500 keV, 1-keV bins from 501 to
520 keV, and 10-keV bins from 520 to 600 keV. We extended the energy
range of the assumed photon spectrum up to 20 MeV to allow for the
off-diagonal elements of the instrument response matrix due to the
non-photopeak response of the detector. The fitted photon energy range
was restricted to be above 15 keV to avoid the need for the two
Gaussian emission line sources to accommodate the supposed
instrumental features below 10 keV used for the first flare.  The
upper energy of the fit range was extended up to 500 keV to provide
more information on the power-law spectrum above the break energy.
This increase in the upper energy limit also necessitated adding in a
nuclear component in the form of a template appropriate for a
power-law ion spectrum \citep{1991ApJ...371..793M} with the
normalization parameter fixed at the value obtained to give a best fit
to the data. This nuclear component (shown in Fig. \ref{july_spectra})
contributes $<$10\% to the photon \flRux\ at all energies below $\sim$400
keV and hence has only marginal significance in the subsequent
analysis.

Other parameters were determined from least-squares fits to the
count-flux spectrum and then fixed for the subsequent determination of
uncertainties. These included parameters to characterize the
instrumental effects of pulse pile-up that is a more important
component for this flare since the count rates were a factor of
$\sim$10 higher than in the first flare. Also, although it is not
significant for flares at the solar limb, the albedo spectrum was
included for this flare assuming isotropic \Xray\ emission using the
the procedure described in \cite{2006A&A...446.1157K} and implemented
in OSPEX. Both the pile-up and albedo components are shown in
Fig. \ref{july_spectra}.

For our detailed spectral analysis and assessment of uncertainties for
this flare, there was a total of seven free parameters -- EM, kT, A,
$E_c$, $E_b$, $\delta_1$, $\delta_2$ (see Table \ref{tblbrd}). The
background-subtracted count \flRux\ and photon spectra are shown in
Fig.\ \ref{july_spectra} along with the best-fit model components.
Note that the implementation of the normalization used for this
analysis is different from that used for the \janfive\ flare.  In this
analysis using the normalization $A$ at the pivot energy $E_{p}$ is
preferred.  The reason for this choice is given in Section
\ref{sec:res:2002jul23}.

\begin{deluxetable}{ccccccccc}

     \tabletypesize{\footnotesize}
     \tablecaption{Flare characteristics and model parameters.\label{tblbrd}}

     \tablehead{
& & & \multicolumn{2}{c}{Flare 1} & \hspace*{0.1in} & \multicolumn{2}{c}{Flare 2}\\
         \multicolumn{3}{l}{Date} &   \multicolumn{2}{c}{\janfive} & &
\multicolumn{2}{c}{\jultwo}\\
          \multicolumn{3}{l}{GOES Start/Peak/End Times} &
\multicolumn{2}{c}{08:03/08:22/08:40 UT} & &
\multicolumn{2}{c}{00:18/00:35/00:47 UT}\\

          \multicolumn{3}{l}{GOES Class} & \multicolumn{2}{c}{X1.3} & & \multicolumn{2}{c}{X4.8} \\
          \multicolumn{3}{l}{Location on the Sun} & \multicolumn{2}{c}{N15W51} & & \multicolumn{2}{c}{S13E72} \\
          \multicolumn{3}{l}{Radial distance from Sun center\tablenotemark{1}} & \multicolumn{2}{c}{763''} & & \multicolumn{2}{c}{904''} \\
          \multicolumn{3}{l}{Time Interval Analyzed} &
        \multicolumn{2}{c}{08:26:00 -- 08:26:20 UT} & &
        \multicolumn{2}{c}{00:30:00 -- 00:30:20.250 UT}\\
        \multicolumn{3}{l}{Fitted Photon Energy Range} & \multicolumn{2}{c}{6.45 to 300 keV} & & \multicolumn{2}{c}{15 to 500 keV}\\
        \multicolumn{3}{l}{Fitted Photon Energy Bins} & \multicolumn{2}{c}{90} & & \multicolumn{2}{c}{90}\\
        &&&&\\
         \colhead{} & \colhead{Parameter} & \colhead{Units} & \colhead{Value\tablenotemark{2} $\varhat$} & \colhead{Free/Fixed\tablenotemark{3}}
        & & \colhead{Value\tablenotemark{2}$\varhat$}& \colhead{Free/Fixed\tablenotemark{3}}
         }

     \startdata

     \multicolumn{2}{l}{\textbf{Thermal Plasma}} \\
     & EM & $10^{49}$ cm$^{-3}$ & 2.31 & free & & 2.16 & free  \\
     & Temp. (kT) & keV & 2.03 & free & & 3.18 & free \\
     & Abundance & coronal & 1 & fixed & & 1 & fixed \\

     \multicolumn{2}{l}{\textbf{Non-thermal Electrons}} \\
     & $F_0$, integrated \flRux\tablenotemark{4} & $10^{35}$~s$^{-1}$ & 0.17 & free & & \multicolumn{2}{c}{not used} \\
      & $A$, \flRux\tablenotemark{5}~~at $E_p$ & $10^{35}$~s$^{-1}$~keV$^{-1}$ & \multicolumn{2}{c}{not used} & & 0.028 & free \\
      & $E_c$ & keV & 105 & free & & 32.0 & free \\
       & $E_p$\tablenotemark{6} & keV & 1      & fixed & & 50  & fixed \\
       & $E_b$ & keV & 32,000 & fixed & & 256 & free \\
       & $E_h$ & keV & 32,000 & fixed & & 32,000 & fixed \\
       & $\delta_1$ && 3.57 & free & & 3.40 & free \\
       & $\delta_2$ && 6.0 & fixed & & 3.92 & free \\

     \multicolumn{2}{l}{\textbf{Nuclear Template}} \\
     & Normalization & photons cm$^{-2}$ & \multicolumn{2}{c}{not used} & & 2.11 & fixed \\

     \multicolumn{2}{l}{\textbf{Gaussians}}\\
    & $G_1$ peak E & keV & 8.44 & fixed & & \multicolumn{2}{c}{not used} \\
    & $G_2$ peak E & keV & 9.95 & fixed & & \multicolumn{2}{c}{not used} \\
    & $G_1$ amplitude & photons cm$^{-2}$~s$^{-1}$ & 33,300 & free & & \multicolumn{2}{c}{not used} \\
    & $G_2$ amplitude & photons cm$^{-2}$~s$^{-1}$ & 12,800 & free & & \multicolumn{2}{c}{not used}\\
    & $G_{1,2}$~FWHM & keV & 0.1 & fixed & & \multicolumn{2}{c}{not used} \\

     \enddata

     \tablenotetext{1}{As measured in the heliocentric-cartesian
       (heliographic) co-ordinate system \citep{2006A&A...449..791T}.}
     \tablenotetext{2}{Best-fit value of parameter computed using
       OSPEX - see Section \ref{sec:nlls}.}
     \tablenotetext{3}{Parameter fixed or allowed to go free in OSPEX
       least-squares fitting. Parameters noted as `fixed' are frozen
       at their values in subsequent uncertainty analyses.}
     \tablenotetext{4}{Total electron \flRux\ integrated over all
       energies from $E_c$ to $E_h$.}  \tablenotetext{5}{Electron \flRux\
       at $E_p$.}  \tablenotetext{6}{The use of the pivot value in the
       implementation of Equation \ref{eqn:powerlaw} is explained in
       Sections \ref{19Jan2005} and \ref{23July2002}.}

\end{deluxetable}
\section{Parameter and Uncertainty Estimation Methods}\label{sec:parest}
Four different methods of uncertainty estimation are described below.
The first three methods - `\covmat', `\chimap' and
`\MC' sampling (Sections \ref{sec:lscov}, \ref{sec:chimap} and
\ref{sec:mc} respectively) are widely used to estimate errors in
parameter values.  The fourth method is based on Bayesian probability
and the Markov chain Monte Carlo (MCMC) method (Section
\ref{sec:mcmc}).  Each of these methods is applied to the spectral
model and data described in Section \ref{sec:obs}, and the results are
tabulated in Table \ref{tab:2005jan19} (\janfive) and Table
\ref{tab:2002jul23} (\jultwo).

\subsection{Methods 1-3: Parameter and uncertainty estimation via
  nonlinear least-squares fitting}\label{sec:nlls} 

The first three methods are based on finding a local minimum
$\chit_{min}$ to the quantity
\begin{equation}\label{eqn:rchi}
\chit = \sum_{i=1}^{\nelossbins}\frac{[\ratedata_{i}-\ratedef{C(\vars)}_{i}]^{2}}{w_{i}^{2}}.
\end{equation}
for some value of $\vars = \varhat$ and $w_{i}$.  The quantity $\chit$
is a \surf\ parameterized by $\vars$. The quantity $\varhat$ is found
by performing a nonlinear weighted least squares fit minimizing
$\chit$ with respect to $\vars$. There are many different ways of
implementing this minimization.  The minimization was achieved using
the OSPEX spectral analysis package which uses the IDL/Solarsoft
routine MCURVEFIT.pro.  This routine is based on the nonlinear
least-squares Levenburg-Marquardt fitting algorithm of
\citet{1992nrfa.book.....P} (pages 675-683).  This implementation of
the algorithm ignores the second derivative of the fitting function
$\ratedef{C(\vars)}_{i}$ with respect to $\vars$, and is therefore
equivalent to assuming that the fitting function is linear with
respect to $\vars$ near the best-fit value $\varhat$.

The value of $\varhat$ is derived as follows.  The process is begun
with an initial estimate of $\varhat$, $\vars^{0}$.  The corresponding
\flRux\ rate spectrum $\ratedef{C(\vars^{0})}_{i}$ is calculated and
$w_{i}$ is set to $\sqrt{C_{i}(\vars^{0})}/t_{LT}$.  This value of
$w_{i}$ is passed to MCURVEFIT.pro.  This routine refines the estimate
of the values of the spectral parameters, stopping when the
termination condition is met\footnote{MCURVEFIT.pro stops iterating
  the Levenburg-Marquardt fitting algorithm when the relative change
  of $\chi^{2}$ from its current value to its previous value is less
  than 0.001.}.  This first estimate is to $\varhat$ is labeled
$\vars^{1}$.  The fitting routine is run again this time using
$\vars^{1}$ as the initial estimate to $\varhat$ and with $w_{i}$ set
to $\sqrt{C_{i}(\vars^{1})}/t_{LT}$, yielding a second estimate
$\vars^{2}$.  The routine is run a third and final time using
$\vars^{2}$ as the initial estimate to $\varhat$ and with $w_{i}$ set
to $\sqrt{C_{i}(\vars^{2})}/t_{LT}$, yielding a final parameter
estimate, labeled $\varhat$.

Estimates of the uncertainty in the value $\varhat$ are found by
defining a \ssize\ of variation in the $\chit$-\surf\ around
$\varhat$ in different ways.  Three different methods of defining
and estimating the uncertainty in the value $\varhat$ are
described below.

\subsubsection{Method 1: Uncertainty Estimation by Estimating the
  Covariance Matrix}\label{sec:lscov}

This method uses the curvature matrix of the $\chit$-\surf\ evaluated
at $\varhat$ to estimate the uncertainty in each parameter, via
the assumptions that the measurement errors in the data $\mathbf{D}$
are \NormGaussly\ distributed and that either the model
$\ratedef{C(\vars)}_{i}$ is linear in its parameters, or the region over
which the uncertainty estimate spans can be replaced by a linear
approximation to the original model.

The curvature matrix $\cmat$ of the the $\chit$-\surf\ arises in
linear and nonlinear least-squares fitting algorithms and is defined
as $\cmat_{ij} = \partial^{2} (\chit)/(\partial \varv_{i} \partial
\varv_{j})$ for $1\le i,j\le \varm$.  The implementation of
MCURVEFIT.pro gives an uncertainty estimate to each of the free
parameters based on the curvature matrix \citep{1992nrfa.book.....P}.
The uncertainty for $\varihat$ (for $\varlim{i}$) is
\begin{equation}\label{eqn:cmat}
\delta \varv_{i} = \pm\sqrt{{\cmat}^{-1}_{ii}},
\end{equation}
when evaluated at $\vars = \varhat$ (the value that minimizes
$\chit$, Equation \ref{eqn:rchi}).  The quantity $\alpha^{-1}$ in the
right-hand side of Equation \ref{eqn:cmat} is the matrix inversion of
the curvature matrix and is an estimate of the \covmat\ of the fit
parameters, evaluated at $\varhat$. Its diagonal elements are the
covariance scale-sizes that defines the uncertainty estimates in this
method.  Full details of the derivation of Equation \ref{eqn:cmat} are
given in \citet{1992nrfa.book.....P}, pages 690--692.  The assumptions
in this derivation also imply that the probability distribution for
$\delta\vars_{obs}$ (the expected error in the value of $\varhat$)
is a multivariate \NormGauss\ distribution around $\varhat$.  The
uncertainty estimate given by Equation \ref{eqn:cmat} is quoted as the
68\% value in Tables \ref{tab:2005jan19} and \ref{tab:2002jul23}.

\subsubsection{Method 2: Uncertainty Estimation using $\chit$-mapping
}\label{sec:chimap}

In this method, parts of the shape of the $\chit$-\surf\ around
$\chit_{min}$ are explicitly calculated.  It is assumed that the value
of the $\chit$-\surf\ as defined by Equation \ref{eqn:rchi}, at a
particular point $\vars$, follows a $\chi^{2}$-distribution.  By
fixing a probability and finding where that probability occurs as a
function of the parameters, one can measure \ssizes\ in the
$\chit$-\surf\ that define an estimate of the uncertainty in the value
of $\varhat$ with that probability.  The procedure is described below.

One of the parameters $\varv$ in the set $\vars$ is stepped through a
range of values while the others are allowed to vary so as to minimize
$\chit$, yielding a value $\chit_{1}$.  The quantity
$\delta\chit=\chit_{1} - \chit_{min}$ is assumed to have a
$\chit$-distribution with one degree of freedom
\citep{1992nrfa.book.....P}.  For such a distribution one can
therefore expect that $\delta\chit<1$ occurs approximately 68\% of the
time and $\delta\chit<4$ occurs approximately 95\% of the time.
Values for the 68\% and 95\% confidence intervals are found where
\begin{equation}\label{eqn:chicond}
\delta\chit(\varv^{68\%}) =1, \delta\chit(\varv^{95\%}) =4,
\end{equation}
respectively. The uncertainty estimates defined by this method are
quoted as differences from $\varhat$ in Tables \ref{tab:2005jan19}
and \ref{tab:2002jul23}, that is,
\begin{equation}\label{eqn:uncert:chimap}
\varv^{100q\%}_{i}-\varihat
\end{equation}
for $\varlim{i}$ where $q=0.68$ and $q=0.95$ and
$\varv^{100q\%}_{i}$ is defined by Equation \ref{eqn:chicond}.
Typically there are two values of $\varv^{100q\%}_{i}$ that satisfy
Equation \ref{eqn:chicond} corresponding to the upper and lower
confidence limits of the parameter value $\varihat$.  When no
value of $\varv$ can be found that satisfies the conditions of
Equation \ref{eqn:chicond}, this is reported as `\notdet' in Tables
\ref{tab:2005jan19} and \ref{tab:2002jul23}.  Finally, this method
uses the same underlying assumptions as those in Section
\ref{sec:lscov} \citep{1992nrfa.book.....P}.

\subsubsection{Method 3: Uncertainty Estimation using the \MC\ method}
\label{sec:mc} 
This method of obtaining uncertainty estimates on $\varhat$ is
commonly called the ``Monte Carlo'' method.  This method begins by
assuming that the value $\varhat$ found in method 1 best describes
the observation via the parameterized model.  By Equation
\ref{eqn:crm}, this defines an estimated count \flRux\ rate spectrum of
$\ratedef{C(\varhat)}_{i}$ that is assumed to be a good estimate
of the true count \flRux\ spectrum.  Estimates of the errors in
$\varhat$ are found by generating a new spectrum such that counts
in energy-loss bin {\it i} are drawn from a Poisson distribution with
mean value $\ratedef{C(\varhat)}_{i}$ for all $1\le i\le
\nelossbins$.  This new spectrum is now fit using the same physical
model and fit process as the original fit generating $\varhat$.
The sampling and fitting process is repeated; the distribution of
values found is centered at $\varhat$, and the width of
distribution estimates the uncertainty in $\varhat$.  The sample
and fit process is repeated 10,000 times, from which normalized
frequency distributions $\freqd(\fdy_{i})$ $(\varlim{i})$ are
calculated.  The uncertainty estimate used excludes the tail values in
a frequency distribution $\freqd(\fdy_{i})$.  The $100q$\% uncertainty
estimate for $0\le q \le 1$ is defined as
$[\fdy^{L}|_{q},\fdy^{H}|_{q}]$ where
\begin{equation}\label{eqn:credint}
\int_{-\infty}^{\fdy^{L}|_{q}}\freqd(\fdy_{i})d\fdy_{i} =
\int^{\infty}_{\fdy^{H}|_{q}}\freqd(\fdy_{i})d\fdy_{i}
= (1-q)/2.
\end{equation}
This definition finds an interval $[\fdy^{L}|_{q},\fdy^{H}|_{q}]$ such
that $100q$\% of the measurements are within the interval and an equal
percentage of the measurements are both above and below the interval.
This definition of the interval is also guaranteed to contain the
median value (which can be found from Eq. \ref{eqn:credint} by setting
$q=0$).  The uncertainty estimates found by this method are quoted in
Tables \ref{tab:2005jan19} and \ref{tab:2002jul23} as differences
\begin{equation}\label{eqn:uncert:mc}
\fdy^{L}|_{q}-\varihat, \fdy^{H}|_{q}-\varihat
\end{equation}
for $q=0.68$ and $q=0.95$ $(\varlim{i})$.


\subsection{Method 4: Parameter and uncertainty estimation using
  Bayesian data analysis}
This method uses parameter and uncertainty estimation based on
Bayesian data analysis methods
\citep{2003prth.book.....J,2005blda.book.....G}.  In Bayesian data
analysis, the probability of a hypothesis $H$ is calculated via Bayes'
theorem.  Denoting by $p(a|b,c)$ the conditional probability that
proposition $a$ is true given that propositions $b$ and $c$ are true,
Bayes' theorem is
\begin{equation}
p(H|\sobs,\information) = \frac{p(H|\information)p(\sobs|H,\information)}{p(\sobs|\information)}
\label{eqn:bayes}
\end{equation}
where $H$ is the hypothesis to be tested, $\sobs$ is the observation,
and $\information$ is any other applicable information we have prior to
calculating the posterior.

The left hand side $p(H|\sobs,\information)$ is called the posterior
probability of the hypothesis, given the data and the prior
information, and it encapsulates the available knowledge about the
hypothesis.  The quantity $p(H|\information)$ is called the prior
distribution and represents what we know about $H$ prior to
calculating the posterior. Often a prior describes a \pdf\ of likely
parameter values.  The sampling distribution or likelihood,
$p(\sobs|H,\information)$, represents the likelihood of the data given
the hypothesis $H$ and information $\information$.  The quantity
$p(\sobs|\information)$ is the unconditional distribution of $\sobs$
and is a constant which ensures that the posterior integrates to 1.

In this paper, the hypothesis $H$ is that a model count spectrum
$\sfit$ parameterized by $\vars^{B}$ explain the observations $\sobs$.
Since the counts in each energy bin are Poisson distributed, the
likelihood of measuring a certain set of counts $C_{i}(\vars^{B})$
becomes
\begin{equation}\label{eqn:expect} 
p(\sobs|\vars^{B},\information) = \prod_{i=1}^{\nelossbins}\frac{C_{i}(\vars^{B})^{D_{i}}}{D_{i}!}e^{-C_{i}(\vars^{B})}.
\end{equation}
Each parameter in the fit has its own prior
$p(\varv_{k}|\information), \varlim{k}$ so that
$p(H|\information)=\prod_{j=1}^{m}p(\varv_{k}|\information)$.  Each
parameter is given a {\it flat} or {\it uniform} prior in a fixed
range, that is, there is an equal probability that the parameter can
take any value in the fixed range.  Table \ref{tab:priors:props}
tabulates the permitted range of values for each parameter for each
model.


The Bayesian posterior probability that a set of values $\vars^{B}$
explains the observations $\sobs$ is proportional to the product of the
likelihood and the prior.  The posterior summarizes the complete state
of knowledge of $\vars$.  Values that give rise to higher posterior
probability are better explanations of the data, and vice versa.  The
best explanation of the data is the {\it maximum a posteriori} (MAP)
value $\vars^{MAP}$ which maximizes the value of the posterior.  Under
the Bayesian interpretation of probability, values
$\vars^{B}\ne\vars^{MAP}$ are less probable explanations of the data.
The full posterior \pdf\ $p(\vars^{B}|\sobs,\information)$ is used to generate
summaries that estimate the uncertainty of each parameter of the model
(see Section \ref{sec:marginal}).

The observed counts above background \sobs\ in the RHESSI data for
both flares are large enough ($\gtrsim 30$ counts in all but the very
highest energy-loss bins, \citeauthor{citeulike:418876},
\citeyear{citeulike:418876}) that the Poisson distributions in
Equation \ref{eqn:expect} can be approximated by
\NormGauss\ distributions with mean and variance both equal to
$C_{i}(\vars^{B})$.  Therefore, the logarithm of the posterior is
approximately
\begin{equation}\label{eqn:bsurf}
\ln p(\vars^{B}|\sobs,\information) \propto
\sum^{\nelossbins_{h}}_{i=1}\frac{\left(D_{i}-C_{i}(\vars^{B})\right)^{2}}{C_{i}(\vars^{B})}
\end{equation}
where $\nelossbins_{h}<\nelossbins$ is the number of energy loss bins
at which the number of counts is large enough that the Gaussian
approximation is valid. Therefore the \surf\ formed by the Bayesian
posterior \pdf\ is closely related to the $\chit$-\surf\ of Equation
\ref{eqn:rchi}.  To estimate $\vars^{MAP}$ and the less probable
explanations of the data we turn to \MCMC\ methods to efficiently
explore the posterior \pdf.  Note that the full posterior assuming the
Poisson likelihood Equation \ref{eqn:expect} was used in the analysis,
and not Equation \ref{eqn:bsurf}, since Equation \ref{eqn:expect} is
more appropriate and the \MCMC\ method applied to Bayesian data
analysis does not require \NormGauss\ distributions in order to generate
uncertainty estimates.


We note that a similar application of Bayesian data analysis
techniques was implemented to generate values and uncertainty
estimates in the recovery of the differential emission measure (DEM)
from emission line spectra. \citet{1998ApJ...503..450K} recast the DEM
recovery problem using Bayes' theorem and modeled the full DEM as a
set of emissivities and elemental abundances in a fixed number of
temperature bins.  This model is convolved with the contribution
functions of the emission lines observed to generate a predicted
emission.  The parameter space describing the DEM is explored using a
Markov chain Monte Carlo technique.  The advantage of the Bayesian
data analysis approach in DEM recovery is that it provides confidence
limits on the most probable DEM at each temperature, thus allowing a
determination of the significance of apparent structures that may be
found in a typical reconstruction.

\subsubsection{Markov chain Monte Carlo methods for posterior
  sampling}\label{sec:mcmc} 
Having written down the posterior, the remaining step in the
calculation is to sample from the posterior and calculate posterior
probabilities.  A brute force calculation of posterior probabilities
can be prohibitively computationally expensive in medium or high
dimensional spaces.  For example, explicitly calculating the posterior
probability density using ten different values in each of the seven
parameters for either of the two flare models used here would require
$10^7$ evaluations of the posterior function.  We adopt a more
practical approach by using a Markov chain Monte Carlo method to find
samples from the posterior \pdf.  MCMC methods allow for the efficient
mapping of Bayesian posterior \pdfs\ in multi-dimensional parameter
space.  After some initial period (known as ``burn-in''), the Markov
chain returns samples directly proportional to their probability
density as defined by the Bayesian posterior, that is, the equilibrium
distribution of the Markov chain is the same as the posterior
probability density function \citep{2005blda.book.....G}.  In general,
it is desirable for the Markov chain to have ``rapid mixing'', that
is, it quickly reaches its equilibrium distribution.  Many different
MCMC algorithms have been designed in order to achieve rapid mixing.
In this paper, we implement a parallel tempering MCMC algorithm (see
Appendix \ref{sec:pt} for more details).  Table \ref{tab:priors:props}
show the priors used for each variable and the range of values of
$\vars^{B}$ for each flare.  Assessing when the post burn-in state has
been achieved can be found by examining the samples.  In this paper,
the Gelman $R$ diagnostic is used to assess convergence
(\citeauthor{citeulike:105949} \citeyear{citeulike:105949}, see
Appendix \ref{sec:imp}).

\subsubsection{Summaries of the posterior \protect\pdf}\label{sec:marginal}
The \bmcmc\ summary \pdfs\ for a single parameter $\varv_{i}$ in the set
$\vars, (\varlim{i})$ are found by integrating the posterior
probability distribution (Eq. \ref{eqn:bayes}) over all the other
variables, i.e.,
\begin{equation}\label{eqn:marginal}
p(\varv_{i}) = \int p(H|\sobs,\information) d\varv_{1}...d\varv_{i-1}d\varv_{i+1}...d\varv_{\varm}.
\end{equation}
This distribution is called a {\it marginal distribution}, and it is
the \pdf\ for the variable $\varv_{i}$ given all the likely values of
all the other variables.  The marginal distribution is used to
calculate uncertainty estimates to $\varv_{i}$.  Values to the 68\%
and 95\% uncertainty are calculated using the definition of the
uncertainty interval given by Equation \ref{eqn:credint}, with the
function $\freqd(\fdy_{i})$ substituted with the marginal distribution
$p(\varv_{i})$.  The uncertainties quoted for this method in
Tables \ref{tab:2005jan19} and \ref{tab:2002jul23} are given as
\begin{equation}\label{eqn:uncert:mcmc}
\fdy^{L}|_{q} - \mbox{median}\left[p(\varv_{i})\right], \fdy^{H}|_{q}- \mbox{median}\left[p(\varv_{i})\right]
\end{equation}
where $\fdy^{L}|_{q}$, $\fdy^{H}|_{q}$ are defined using Equation
\ref{eqn:credint} (substituting $p(\varv_{i})$ for $F(\varv_{i})$) for
$q=0.68$ and $q=0.95$ and $\mbox{median}\left[p(\varv_{i})\right]$ is
the median value of the marginal \pdf\ $p(\varv_{i})$.  Note that this
definition of the interval does not necessarily include the mean or
the mode. 

\section{Results} \label{sec:res}

\subsection{\janfive} \label{sec:res:2005jan19} 

\begin{figure}
  \centerline{
  \includegraphics[width=0.501\textwidth,clip=]{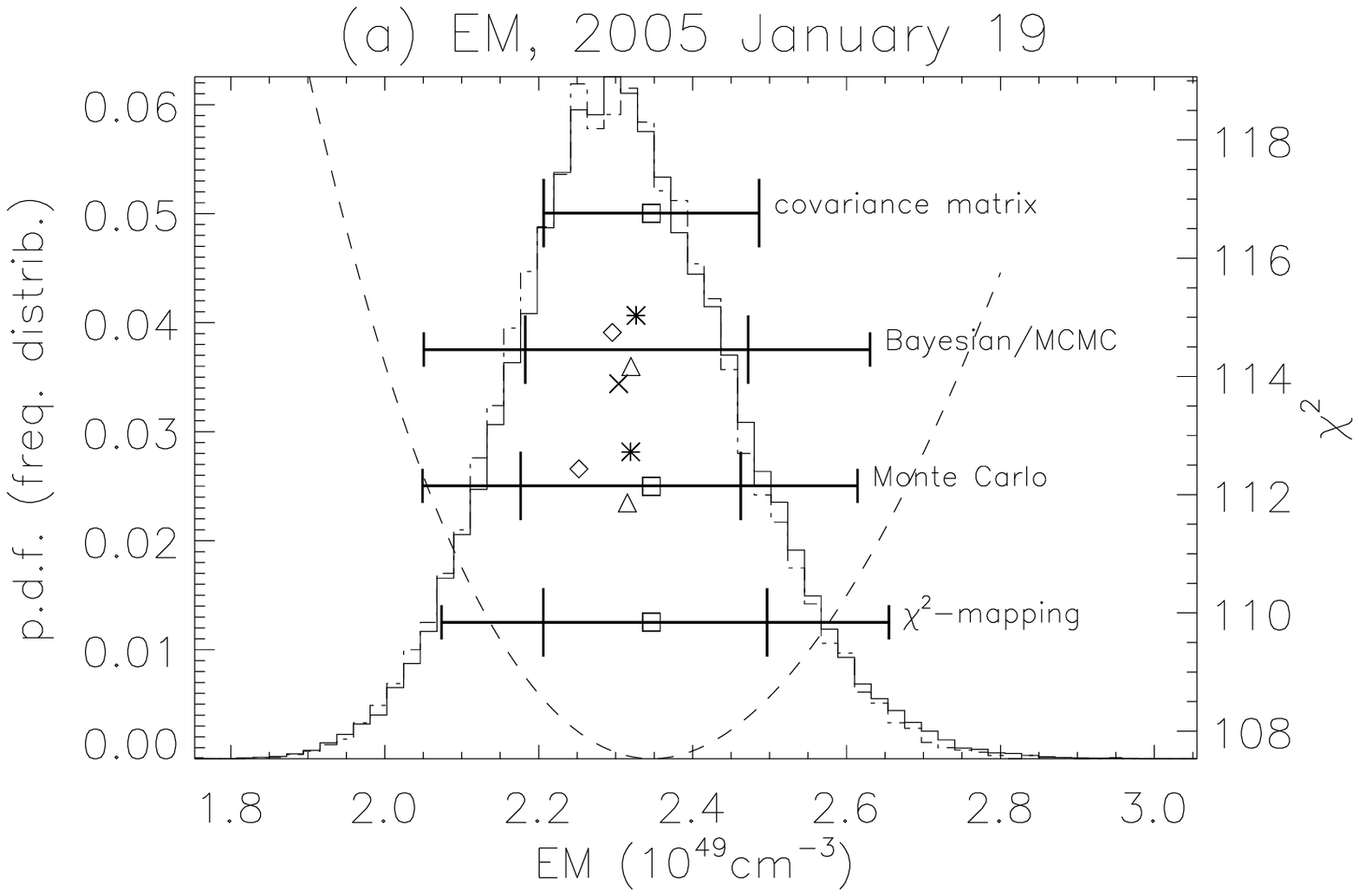}
}
  \centerline{
  \includegraphics[width=0.501\textwidth,clip=]{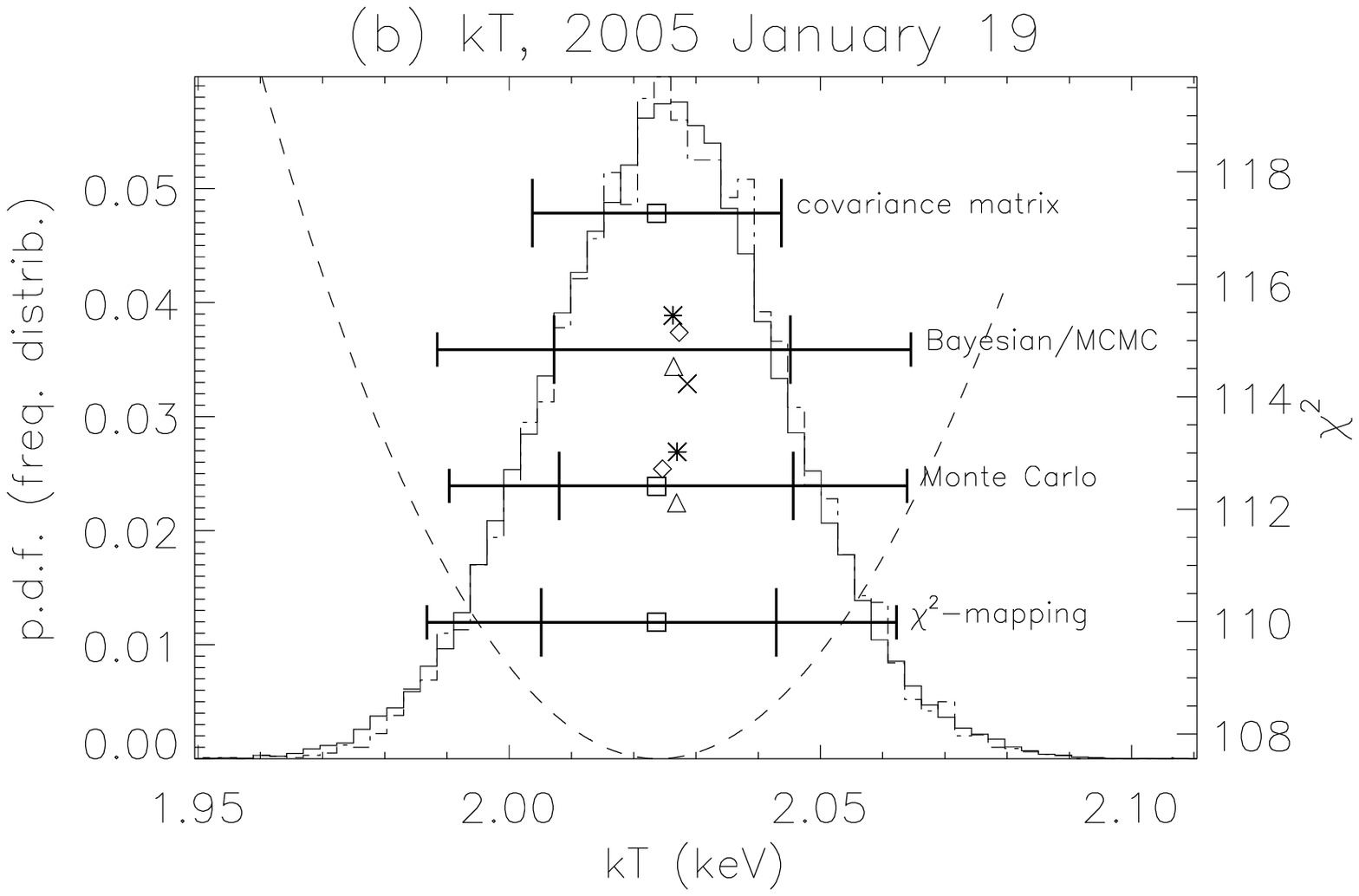}
  }
  \caption{Results from each of the four uncertainty analysis methods
    (Section \ref{sec:parest}) for the parameters of the thermal
    component of the total emission (a) $EM$ and (b) $kT$, from the
    model fit to the \janfive\ flare data.  The dashed curve is the
    value of $\chi^{2}$ found by the \chimap\ method (values are
    indicated by the right-hand plot axis).  The normalized frequency
    distribution of values found by the \MC\ method is shown as a
    histogram (dot-dashed line).  The marginal probability density
    function arising from the \bmcmc\ method is shown as a histogram
    (solid line).  Values to these histograms are indicated by the
    left-hand plot axis.  The horizontal lines show the uncertainty
    estimates calculated via the methods indicated (from top to bottom
    - \covmat, \bmcmc, \MC, and \chimap), with the 68\% and 95\%
    uncertainty estimates indicated by larger and smaller vertical
    lines that cross those lines.  The best-fit value $\varhat$ found
    via nonlinear least-squares minimization (Section \ref{sec:nlls})
    is indicated by square plot symbols.  The MAP value $\vars^{MAP}$
    is indicated by a $\times$-symbol.  The mean, mode and median
    values calculated for each of the two distributions (arising from
    the \bmcmc\ and \MC\ analyses) are indicated by asterisks,
    diamonds and triangles respectively.  These symbols are separated
    vertically scattered for clarity.  }
  \label{fig:2005jan19thermal}
\end{figure}

\begin{figure}
  \centerline{
    \epsscale{1.2}\plottwo{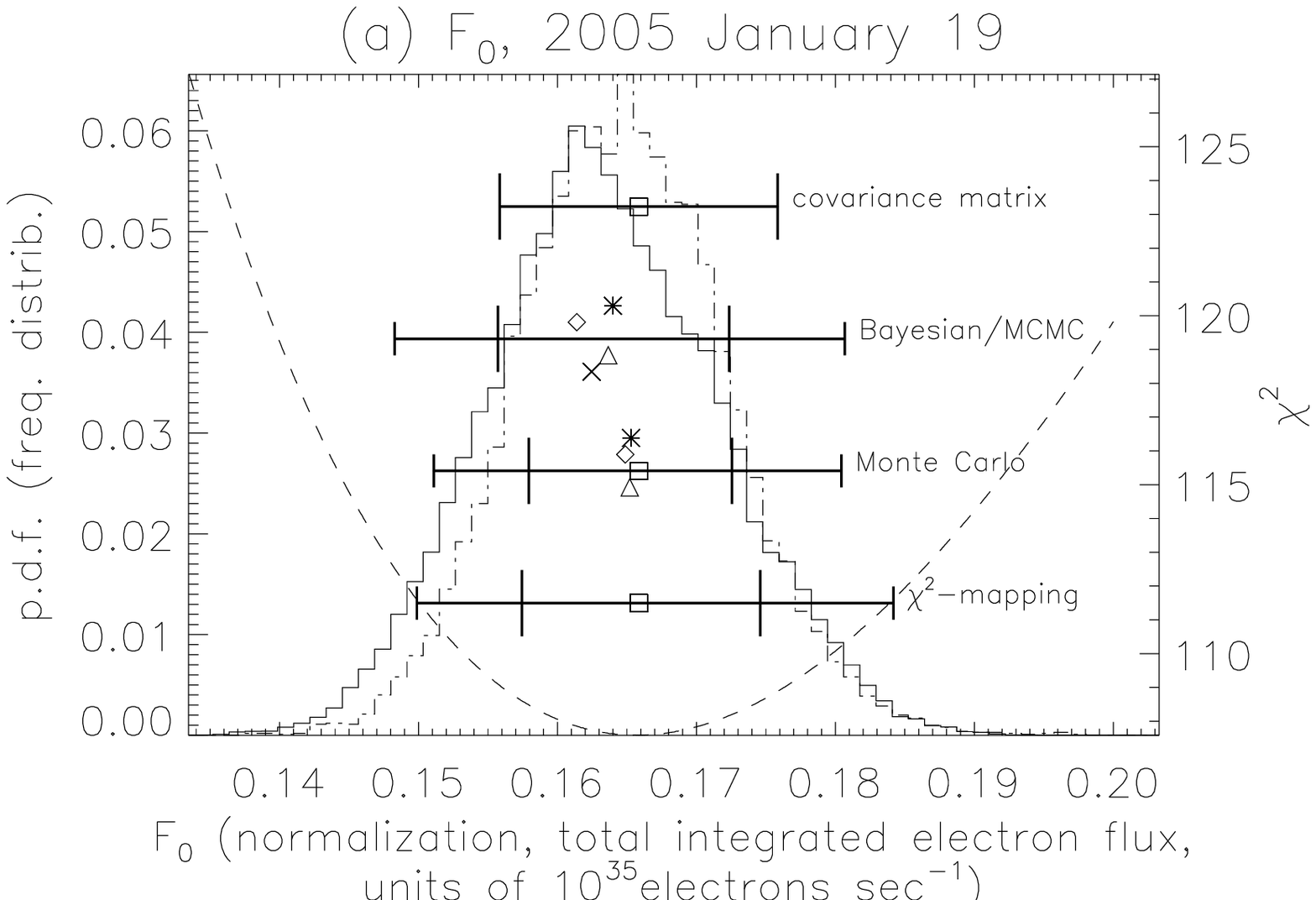}{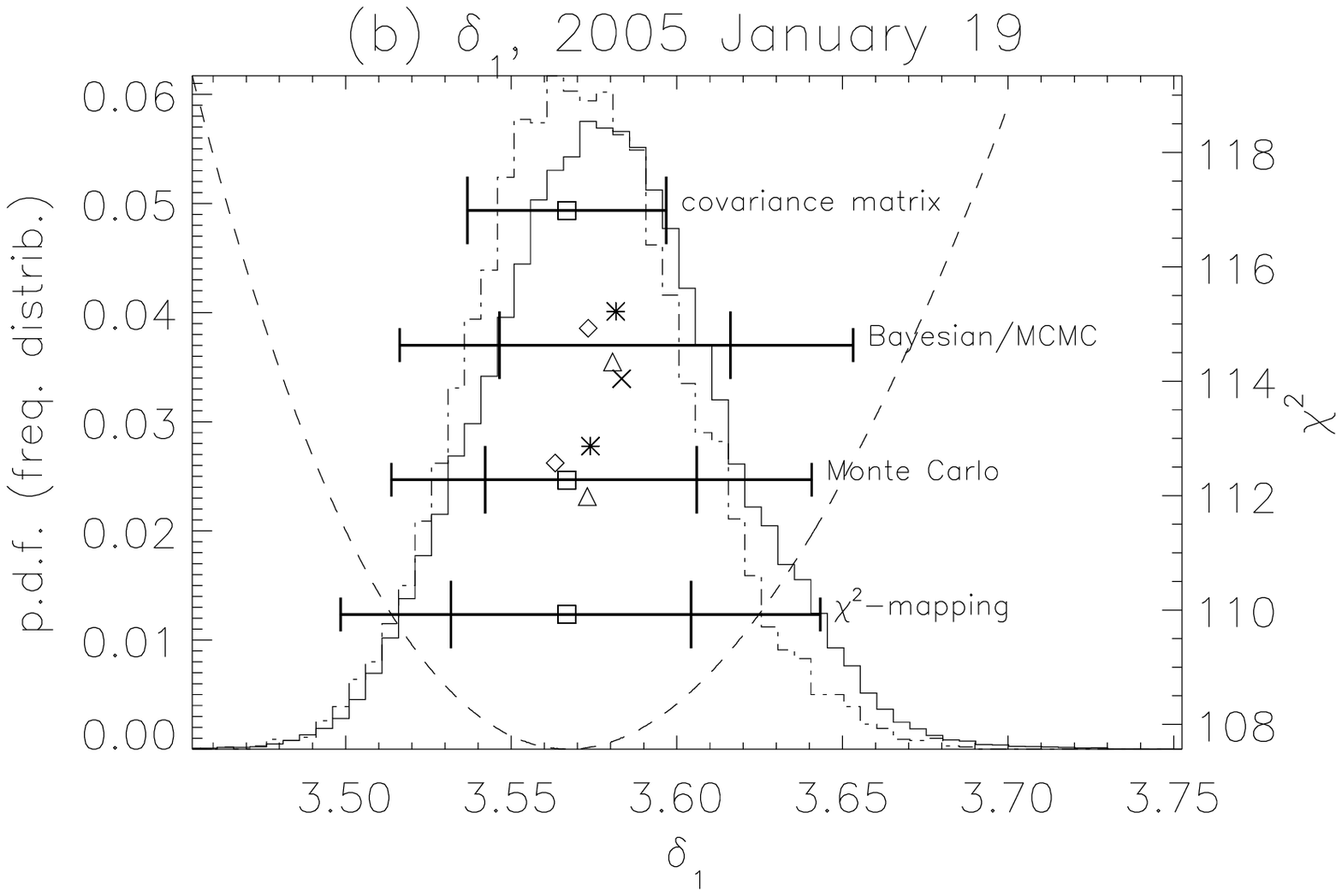}
  }
  \centerline{
  \includegraphics[width=0.501\textwidth,clip=]{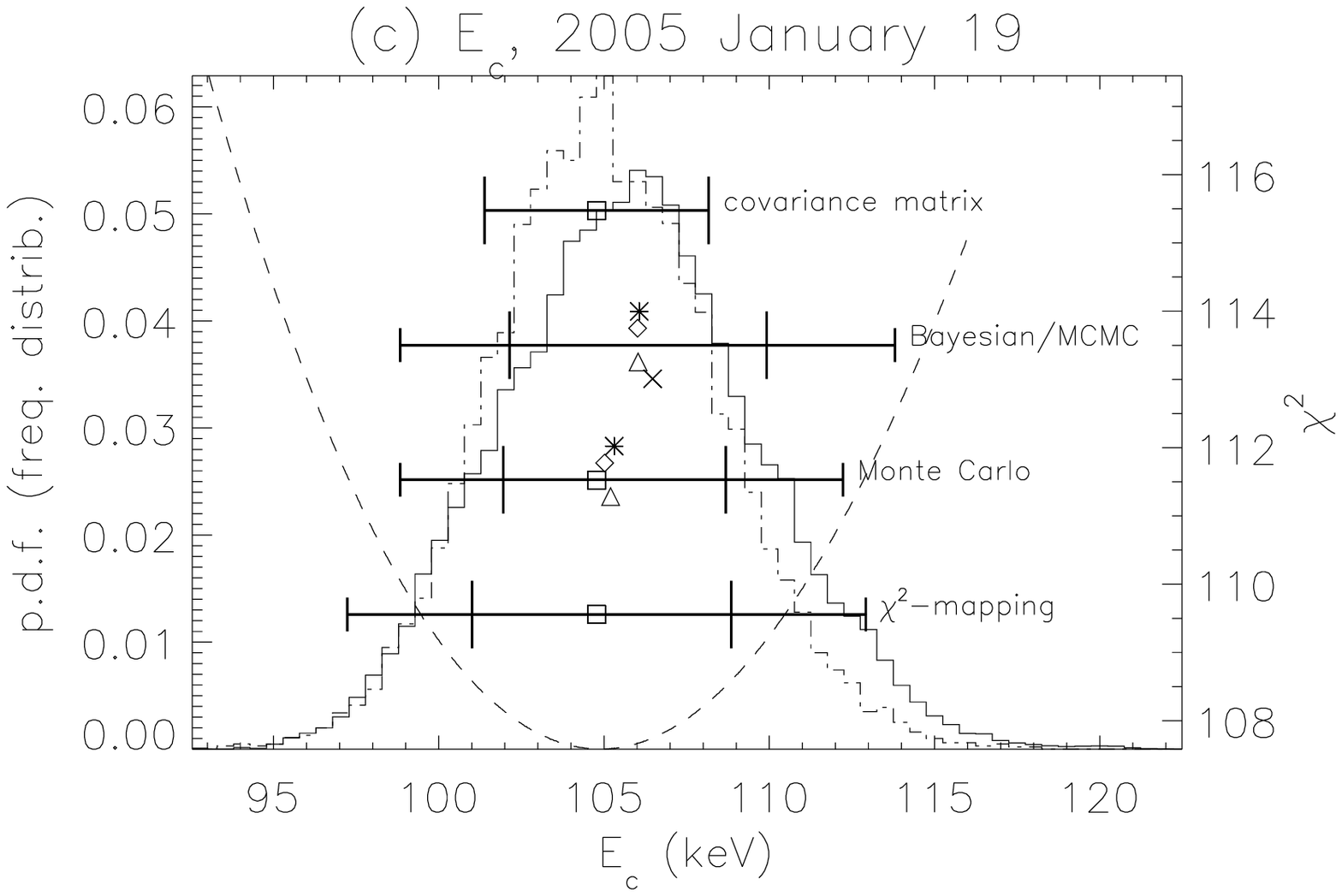}
  }

  \caption{Results from each of the four uncertainty analysis methods
    (Section \ref{sec:parest}) for the parameters of the nonthermal
    component of the total flare emission (a) $F_0$, (b) $\delta_{1}$
    and (c) $E_{c}$ (see Eq. \ref{eqn:powerlaw}) from the model fit to
    \janfive\ flare data. The type of data plotted, plot symbols and
    lines have the same meaning as in Figure
    \ref{fig:2005jan19thermal}.  }
  \label{fig:2005jan19flare}
\end{figure}

\begin{figure}
\centerline{
\includegraphics[width=0.90\textwidth,clip=]{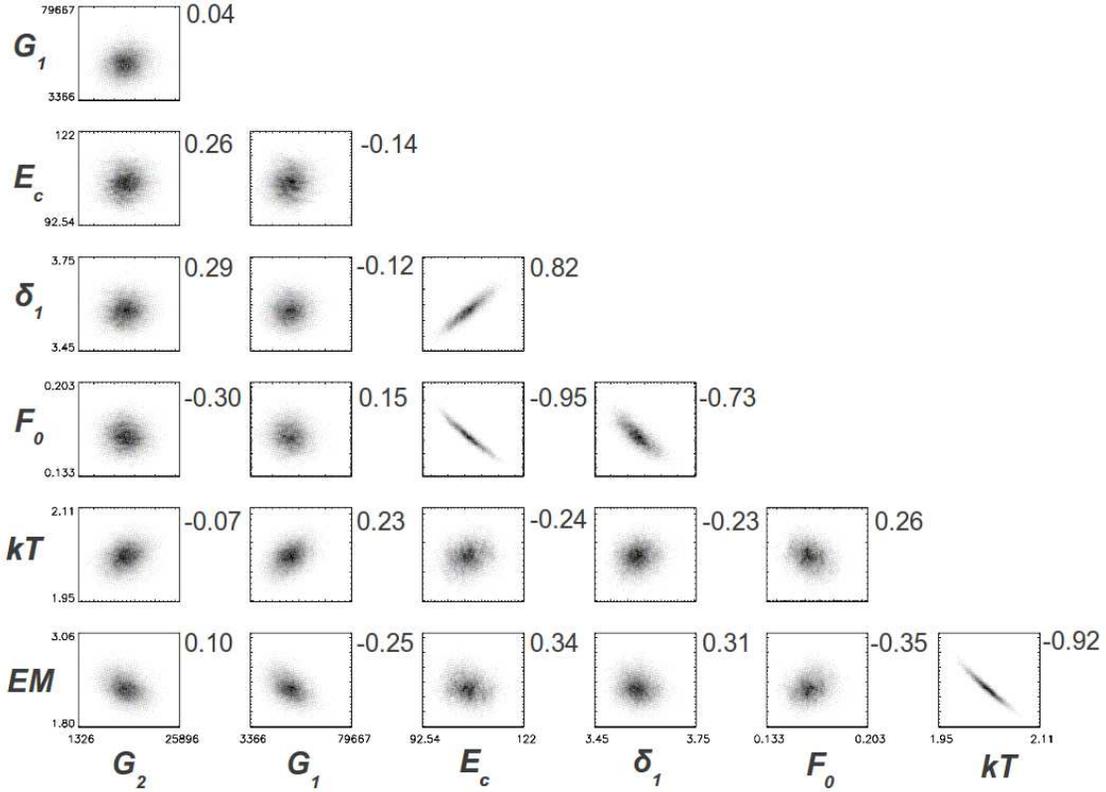}
}
\caption{Two dimensional marginal \protect\pdfs\ for the parameters of
  the model used to fit the spectrum of the \janfive\ flare.  These
  plots are found by integrating the posterior probability density
  function (found by the \bmcmc\ algorithm) over all the parameters
  excepting those indicated on the $x$- and $y$-axes.  This is the
  extension into two dimensions of the definition of the
  one-dimensional marginal distribution function given by Equation
  \ref{eqn:marginal} in Section \ref{sec:marginal}.  Each of these
  plots in this figure shows how the posterior probability density of
  the value of a given parameter depends on the value of another
  parameter, and so help visualize the shape of the full posterior
  \pdf.  Indicated parameter ranges are the lowest and highest values
  found by the Bayesian/MCMC algorithm.  Darker tones indicate a
  greater probability density.  The number on the upper right of each
  plot is the Spearman rank correlation coefficient for the two
  parameters.  For the \janfive\ flare, the distributions are all
  approximately elliptical.  The majority of the distributions are
  weakly correlated; a minority ($EM$ versus $kT$, and $F_0$,
  $\delta_{1}$ versus $E_{c}$, $F_0$ versus $\delta_{1}$) show a high
  degree of correlation.  The reasons for these strong correlations
  are discussed in Section \ref{sec:res:2005jan19}.}
  \label{fig:all2d2005jan19}
\end{figure}

Figures \ref{fig:2005jan19thermal}, \ref{fig:2005jan19flare},
\ref{fig:all2d2005jan19} and Table \ref{tab:2005jan19} show the
results for each of the four uncertainty estimation methods under
consideration using the data and electron spectral model for the
\janfive\ flare, as described in Section \ref{sec:obs}.  Figures
\ref{fig:2005jan19thermal}, \ref{fig:2005jan19flare} and Table
\ref{tab:2005jan19} show that the difference between the $\vars^{MAP}$
and $\varhat$ values are much less than the 68\% uncertainty
estimates.  For each variable, the lower and upper 68\% (and 95\%)
uncertainty estimates found by each uncertainty estimation method have
approximately the same magnitude.  Comparing across methods, it can be
seen that each also gives approximately the same uncertainly
estimates.  The ratio of the 95\% uncertainty estimate to the 68\%
uncertainty estimate are all close to 1.96, as expected from
distributions of measurements which are close to being \NormGaussly\
distributed.  In addition, Q-Q plots of all seven marginal
distributions obtained from the Bayesian analysis (see Appendix
\ref{app:normaltests}) show that each of them is approximately
\NormGaussly\ distributed.


Figure \ref{fig:all2d2005jan19} plots two-dimensional marginal
distributions arising from the \bmcmc\ analysis for every pair of
parameters in the spectral model (the priors used in the
\bmcmc\ approach can be found in Table \ref{tab:priors:props}).  It
shows the effect each parameter has on the value of the other when
finding highly probable parameter values to $\vars$.  Next to each
plot the Spearman rank correlation coefficient for the indicated
variables is shown.  It can be seen that all the two-dimensional
marginal distributions are elliptical, and the majority of them show
that the probability of getting a particular parameter value is weakly
correlated with the value of any other parameter.  The exceptions to
this for this flare are the emission measure ($EM$) and plasma
temperature ($kT$) dependency, the dependency of the spectral
normalization $F_0$ on the \LEC\ $E_{c}$ and the power law index
$\delta_{1}$, and the $E_{c}$ versus $\delta_{1}$ correlation.

The first of these dependencies is anticipated through the definition
of the thermal emission of the plasma (Equation \ref{eqn:thermal}),
and the second two arise from the definition of the normalization.
The normalization factor $F_0$ for this flare is defined as the {\it
  total integrated electron \flRux\ } over all energies, and therefore
clearly depends on the values of $E_{c}$ and $\delta_{1}$ (see Section
\ref{sec:obs}).  Figure \ref{fig:all2d2005jan19} also shows a
correlation between $E_{c}$ and $\delta_{1}$.  This is obtained
because the rate at which the \Xray\ spectrum flattens below $E_{c}$
depends on the value of $\delta_{1}$.  The spectrum flattens more
rapidly with decreasing photon energy for a steeper electron
distribution (larger $\delta_{1}$) than for a flatter electron
distribution.  Therefore, for a given \Xray\ spectrum, a larger value
of $\delta_{1}$ requires a higher value of $E_c$ to obtain the best
fit to the spectrum.  A similar correlation, for the same reason, is
found between $E_b$ and $\delta_{2}$ in the fit to the July 23 flare
spectrum (Figure~\ref{fig:all2d2002jul23}).

\begin{figure}
\centerline{
  \includegraphics[width=0.450\textwidth,clip=]{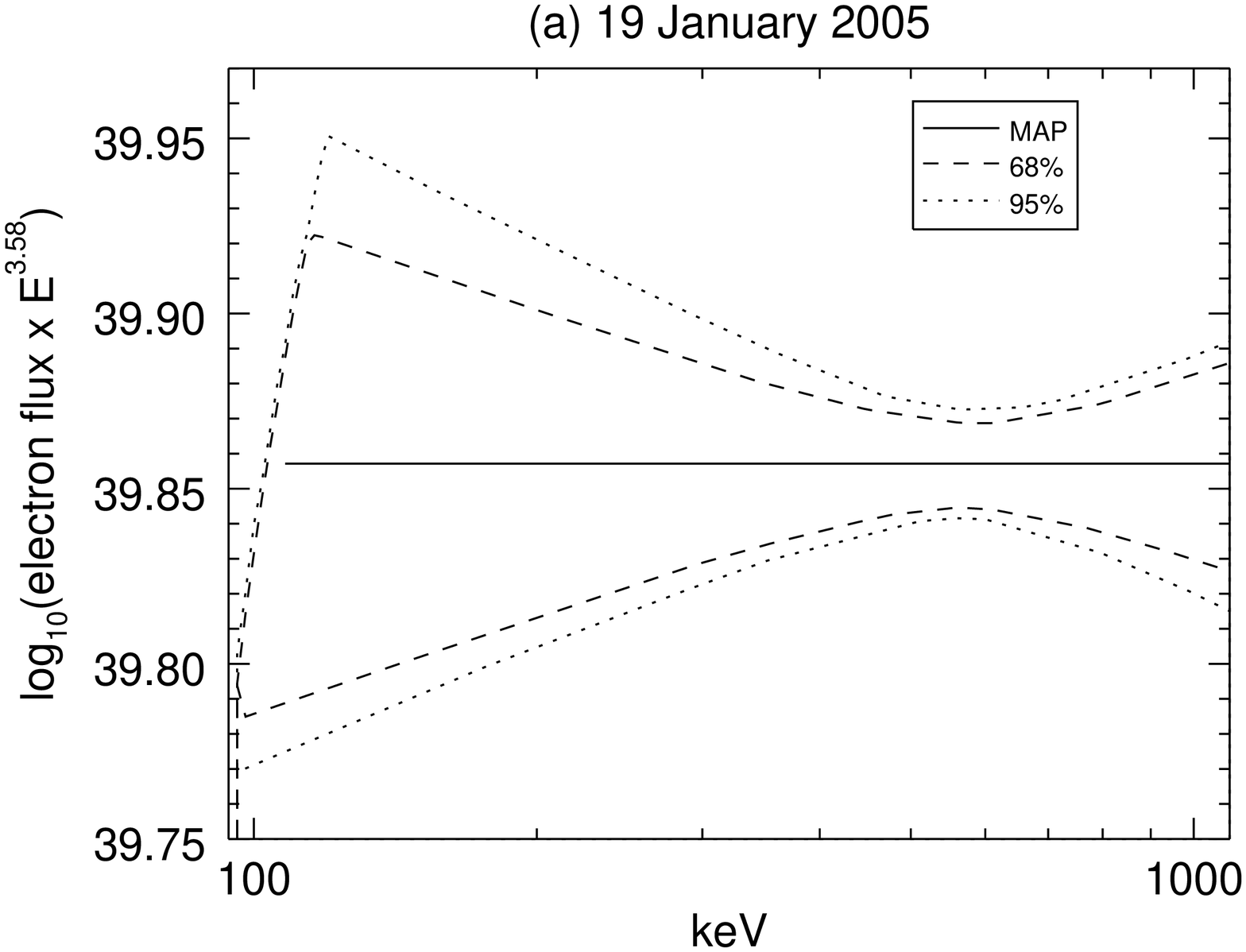}
  \includegraphics[width=0.450\textwidth,clip=]{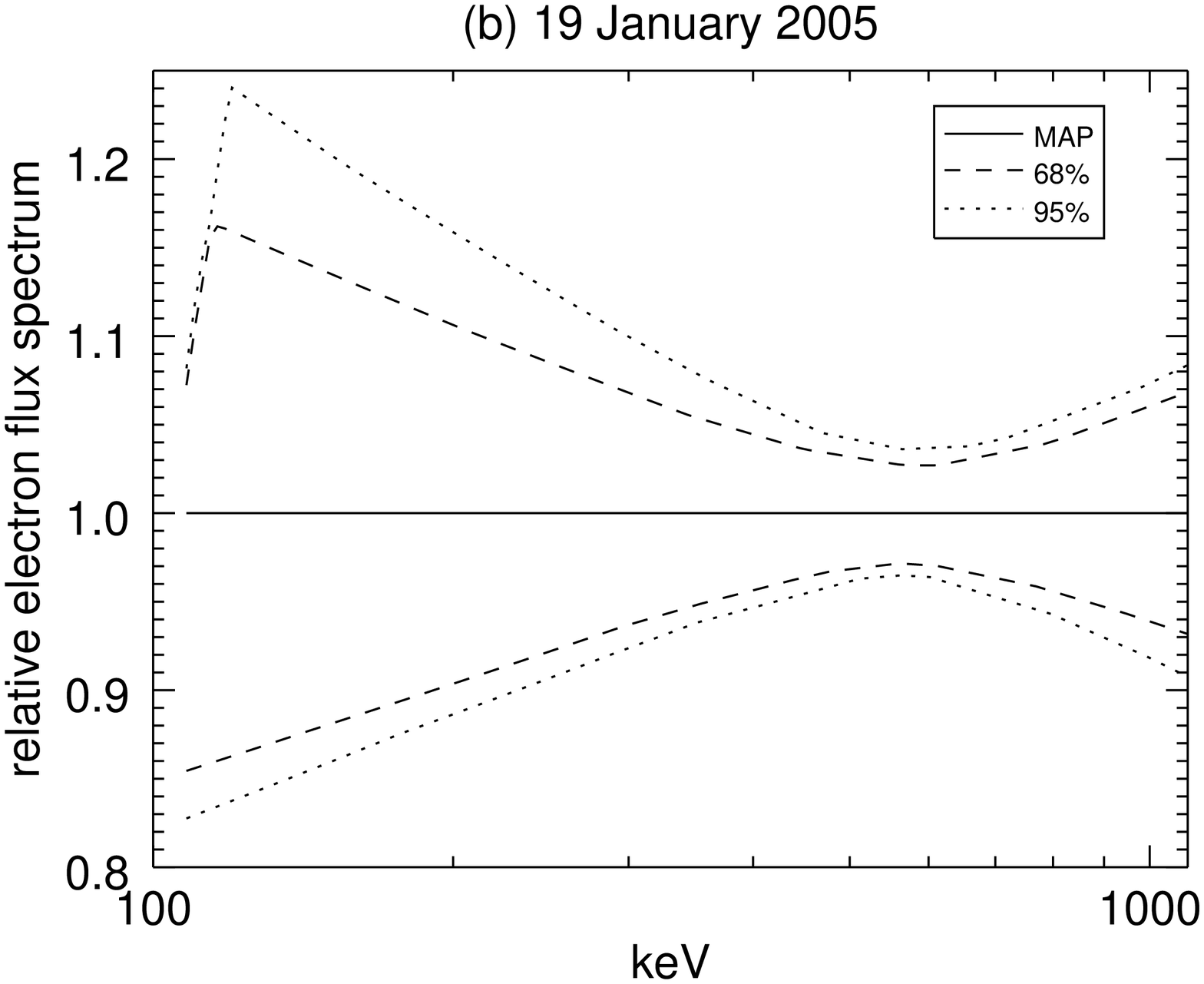}
}
  \centerline{
  \includegraphics[width=0.450\textwidth,clip=]{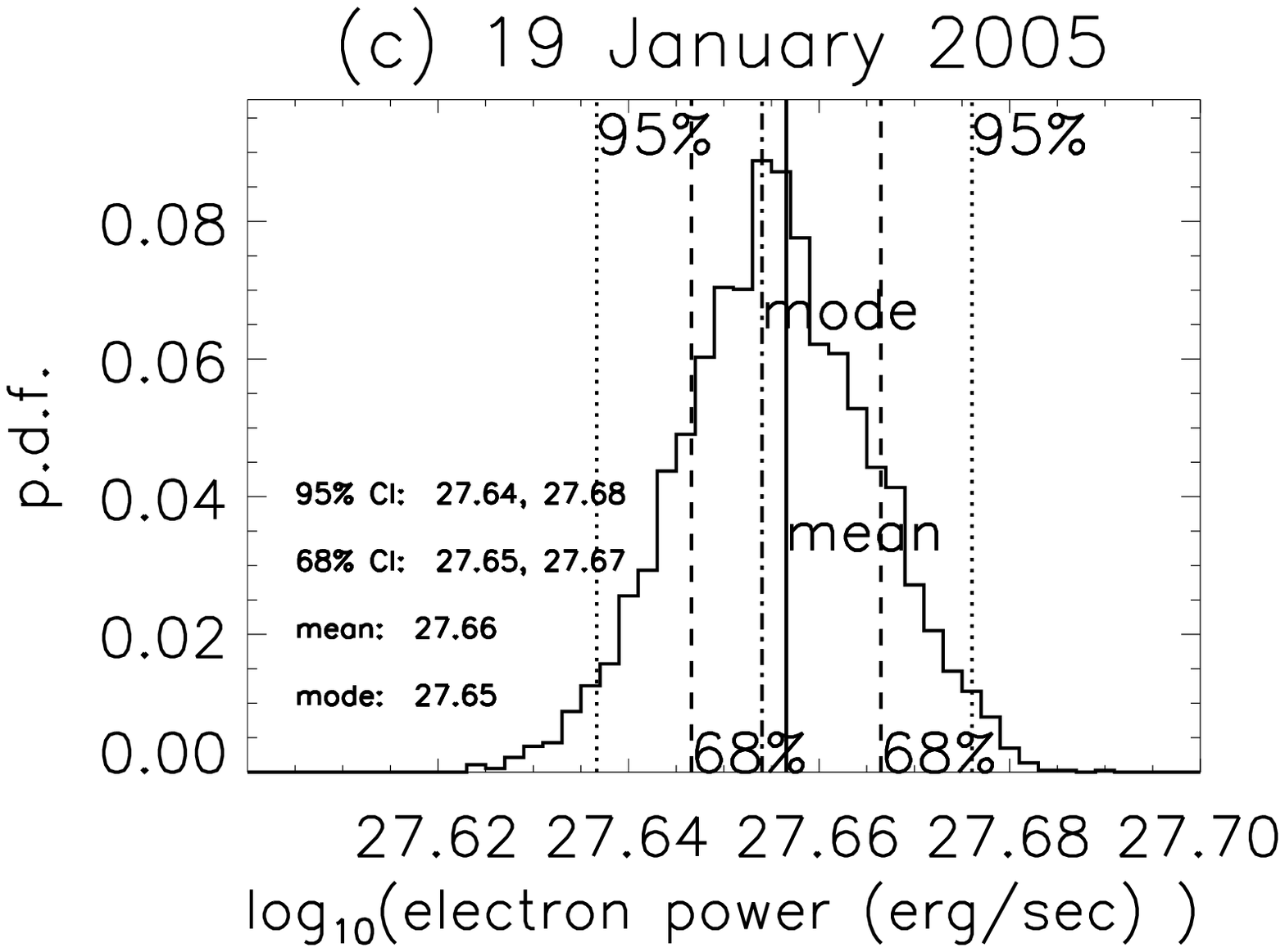}
  }
  \caption{Electron spectrum results for the flare-injected electrons
    arising from the \bmcmc\ method for the \janfive\ flare. (a)
    Electron spectrum (\flRux\ (in units of
    $\mbox{erg}$ $\mbox{keV}^{-1}$ $\mbox{s}^{-1}$) multiplied by
    $E^{3.58}$, where 3.58 is the MAP estimate $\delta_{1}$, the power
    law index of the flare-injected electron spectrum - see Table
    \ref{tab:2005jan19}) with 68\% and 95\% credible interval spectra
    indicated by the dashed and dotted lines, respectively.  The
    electron \flRux\ spectrum corresponding to $\varv^{MAP}$ is
    indicated by the solid line. (b) 68\% and 95\% credible intervals
    relative to the $\varv^{MAP}$ electron \flRux\ spectrum. In plots
    (a) and (b) curves with negative gradients indicate a behavior
    steeper than $E^{-\delta_{1}}$ and positive gradients indicate a
    behavior shallower than $E^{-\delta_{1}}$.  Note also that the MAP
    spectrum extends to its low energy cutoff value; other lower
    probability spectra extend to values of $E_{c}$, which may be
    different to the MAP value of $E_{c}$.  (c) Flare injected
    electron power \pdf, with 68\% and 95\% credible intervals
    indicated; the distribution mean/mode is indicated by the
    solid/dot-dashed vertical line.  The total integrated electron
    \flRux\ injected by the flare is given in Figure
    \ref{fig:2005jan19flare}(c).}
  \label{fig:electron:2005jan19}
\end{figure}

Figure \ref{fig:electron:2005jan19}(a) shows the (scaled) electron
\flRux\ spectrum as a function of energy for the \bmcmc\ analysis.
Figure \ref{fig:electron:2005jan19}(b) shows the ratio of the best fit
electron spectrum to the 68\% and 95\% uncertainty estimates.  Figures
\ref{fig:2005jan19flare}(a) and \ref{fig:electron:2005jan19}(c) show
  the \pdfs\ for total integrated electron number \flRux\ and electron
  power derived from the \bmcmc\ results.  Uncertainty estimates for
  the electron \flRux\ spectrum as a function of energy are found in
  the following way.  The electron \flRux\ spectrum for each
  \bmcmc-derived sample is calculated.  The spectra are then ranked
  according to their posterior probability.  The 68\% curves are found
  by finding the highest and lowest values to the electron
  \flRux\ spectrum in each energy bin for the top 68\% most probable
  samples (the 95\% curves are found similarly), yielding the
  uncertainty estimates as shown in Figure
  \ref{fig:electron:2005jan19}(a). In each energy bin, the upper and
  lower uncertainties are approximately symmetric around the best
  ($\vars^{MAP}$) value.  Further, the \pdfs\ for the electron number
  \flRux\ and power (Figures \ref{fig:2005jan19flare}(a) and
    \ref{fig:electron:2005jan19}(c)) are also approximately
    symmetrical around the mean and mode.  This is not too surprising
    since the \pdfs\ (Figures \ref{fig:2005jan19thermal},
    \ref{fig:2005jan19flare}) for each parameter in the fit are also
    approximately symmetrical.  Finally, the uncertainties in the
    values to the electron number and power are also well constrained.

\begin{deluxetable}{clcccc}
  \tabletypesize{\scriptsize} \tablecolumns{5} \tablewidth{0pt}
  \tablecaption{Parameter values and uncertainties derived for
    the four uncertainty estimation methods applied to the \janfive\
    flare spectrum, as described in Section \ref{sec:obs}.
    The final
    column ``Ratio'' is defined as the ratio of the $\pm95$\%
    uncertainties to the $\pm68$\% uncertainties; for an exact \NormGauss\
    distribution the entry in this column would be $1.96, 1.96$. Two
    ratios are quoted in order to reveal the presence of any relative
    asymmetry in the upper and lower uncertainty estimates, if
    present. See Section \ref{sec:parest} for a detailed description
    of how the uncertainty estimates are found for each method. \label{tab:2005jan19}}
\tablehead{
\colhead{Parameter} & 
\colhead{Method}    & 
\colhead{Value\tablenotemark{a}}     & 
\multicolumn{2}{c}{Uncertainties} & \colhead{Ratio} \\ 
\colhead{} & 
\colhead{} &
\colhead{} & 
\colhead{68\%} & 
\colhead{95\%} & 
\colhead{} }
\startdata 
$EM$ $(\unitsEM)$                 & \covmat\tablenotemark{b} & 2.31 & $\pm$0.14 & \notcalc\ & \notcalc\ \\ 
                      & \chimap\tablenotemark{c} & `` & -0.14, +0.15 & -0.27, +0.31 & 1.94, 2.05 \\ 
                      & \MC\tablenotemark{d} &`` &  -0.17, +0.12 & -0.30, +0.27 & 1.75, 2.31 \\ 
                      & \bmcmc\tablenotemark{e} & 2.30 & -0.14, +0.15 & -0.27, +0.31 & 1.96, 2.04 \\
                      & &   &   &   &  \\

$kT$ (\keV)  & \covmat\        & 2.03  & $\pm$0.02         & \notcalc\ &    \notcalc\       \\
                       & \chimap\             &   ``      & $\pm$0.02  & $\pm$0.04 & 1.99, 2.01 \\
                       & \MC\                   &  ``       & $\pm$0.02  & -0.03, +0.04 & 2.13, 1.84 \\
                       & \bmcmc              & 2.03 & $\pm$0.02  & $\pm$0.04 & 1.98, 2.03 \\
                      & &   &   &   &  \\

$F_0$    & \covmat\    & 0.17       & $\pm$0.01         & \notcalc\  &    \notcalc\        \\
(total integrated electron \flRux\,              & \chimap\             &``   & $\pm$0.01    &   $\pm$0.02 & 1.90, 2.10 \\

$10^{35} \mbox{ electrons} \mbox{ sec}^{-1}$)   & \MC\                   &``    &  $\pm$0.01  & $\pm$0.01 & 1.87, 2.17 \\
                       & \bmcmc               & 0.16  & $\pm$0.01     & $\pm$0.02 & 1.94, 1.96 \\
                      & &   &   &   &  \\

$\delta_{1}$           & \covmat\       &  3.57      & $\pm$0.03         & \notcalc\  &    \notcalc\        \\
                       & \chimap\               & ``   & $\pm$0.04     & -0.07, +0.08 & 1.95, 2.04 \\
                       & \MC\                   & ``   &  -0.02,  +0.04   & -0.05, +0.07 & 2.15, 1.89 \\
                       & \bmcmc                 & 3.58  & -0.03, + 0.04    & -0.06, +0.07 & 1.88, 2.04 \\
                      & &   &   &   &  \\

$E_{c}$ (\keV)    & \covmat\          & 105       &$\pm$3         & \notcalc\  &    \notcalc\        \\
                      & \chimap\               &``   &  $\pm$4 & $\pm$8 & 2.00, 2.00   \\
                      & \MC\                   & ``    & -3, 4 & -6 ,  +7 & 2.10, 1.91 \\
                      & \bmcmc                 & 107  & $\pm$4    & -7, +8 & 1.85, 2.00     \\
\enddata

\tablenotetext{a}{The \covmat, \MC\ and
  \chimap\ methods all start from the same parameter value
  $\varhat$ where $\chit$ is minimized. For the
  \bmcmc\ approach, the ``maximum a posteriori'' value $\vars^{MAP}$ is
  quoted.}
\tablenotetext{b}{See Section \ref{sec:lscov} and Equation \ref{eqn:cmat} for the
  definition of the parameter uncertainty for the \covmat\ method.}
\tablenotetext{c}{See Section \ref{sec:mc} and Equation \ref{eqn:uncert:mc} for the
  definition of the parameter uncertainty for the \chimap\ method.}
\tablenotetext{d}{See Section \ref{sec:chimap} and Equation \ref{eqn:uncert:chimap} for the
  definition of the parameter uncertainty for the \MC\ method.}
\tablenotetext{e}{See Section \ref{sec:mcmc} and Equation \ref{eqn:uncert:mcmc} for the
  definition of the parameter uncertainty for the \bmcmc\ method.}

\end{deluxetable}

\subsection{\jultwo} \label{sec:res:2002jul23}

Figures \ref{fig:2002jul23thermal}, \ref{fig:2002jul23flare},
\ref{fig:all2d2002jul23} and Table \ref{tab:2002jul23} show the
results for each of the four uncertainty estimation methods under
consideration using the data and electron spectral model for the
\jultwo\ flare, as described in Section \ref{sec:obs}.  It is clear
from Figures \ref{fig:2002jul23thermal}, \ref{fig:2002jul23flare} and
\ref{fig:all2d2002jul23} that the $\chit$-\surf\ (or equivalently, the
Bayesian posterior \surf\ - see Section \ref{sec:mcmc}) with respect
to this model is quite different from that seen in the 19 January 2005
flare (Figures \ref{fig:2005jan19thermal}, \ref{fig:2005jan19flare}
and \ref{fig:all2d2005jan19}).  The mode values in the \bmcmc\
marginal distributions are noticeably shifted with respect to the \MC\
distributions.  This is because the \bmcmc\ marginal distributions in
Figures \ref{fig:2002jul23thermal} and \ref{fig:2002jul23flare} are
formed by integrating over a structured seven-dimensional space
(Figure \ref{fig:all2d2002jul23}).  The mode of the one-dimensional
marginal distributions need not be at the $\varv^{MAP}$ or $\varvhat$
value.  Note however from Table \ref{tab:2002jul23} that the
$\varv^{MAP}$ value is close to the $\varvhat$ value, which is to be
expected given the priors used in setting up the Bayesian posterior
(see Appendix \ref{sec:imp}) and the close correspondence between the
$\chi^{2}$-\surf\ (Equation \ref{eqn:rchi}) and the Bayesian posterior
(Equation \ref{eqn:bsurf}).


Figures \ref{fig:2002jul23thermal} (thermal model parameters) and
\ref{fig:2002jul23flare} (non-thermal model parameters) show that the
uncertainty estimates for specific parameters can depend on the
uncertainty estimation method used.  The methods used are influencing
the uncertainty estimates for some parameters (Table
\ref{tab:2002jul23}).  These uncertainty estimates behave quite
differently from those expected from a \NormGauss\ distribution, with
the ratios of the 95\% to 68\% uncertainty estimates very different
from 1.96.  The reason for this is apparent when considering the
two-dimensional Bayesian posterior marginal distributions as shown in
Figure \ref{fig:all2d2002jul23}.  Many of the distributions are
structured, asymmetric, and show extended tails compared to those
derived from the \surf\ of the \janfive\ analysis.  The \LEC\ in
particular shows significant deviation from a simple
\NormGauss\ distribution, as does the break energy $E_{b}$ and the
slope of the spectrum above the break energy, parameterized by
$\delta_{2}$.  Many pairs of parameters have high magnitude
correlation coefficients indicating strong interdependence of one
value on another.  Further, note that the correlation of $E_{c}$ with
all other parameters is relatively weak.  This indicates the relative
independence of the \LEC\ from other features in the model, given the
data.


Figure \ref{fig:2002jul23flare}(e) and \ref{fig:all2d2002jul23} show
that below around 25 keV, all values of $E_{c}$ are approximately
equally likely, but also that $E_{c}< 25$ keV does not constrain
likely values of the emission measure $EM$, the thermal temperature
$kT$, the normalization $A$ and the lower power-law index
$\delta_{1}$.  This leads to a wide range of possible electron-\flRux\
spectra at lower energies, the effect of which leads to wide 68\% and
95\% credible intervals of Figure \ref{fig:electron:2002jul23}(a).
The uncertainty estimates for the electron \flRux\ in Figure
\ref{fig:electron:2005jan19}(a) also show a widening at lower
energies, but it is much less pronounced compared to that in Figure
\ref{fig:electron:2002jul23}(a).  The reason for this is that at lower
values of $E_{c}$, the other parameter values in the model are
constrained, and so there is a restricted range of electron \flRux\
spectra that is generated.

Figure \ref{fig:electron:2002jul23} shows the (scaled) electron
\flRux\ energy spectrum as a function of energy, along with
probability density functions for total integrated electron number
\flRux\ and electron power derived from the \bmcmc\ results.  The wide
68\% and 95\% credible intervals of Figures
\ref{fig:electron:2002jul23}(a, b) show that the electron spectrum
becomes poorly constrained at low energies.  Figures
\ref{fig:electron:2002jul23}(c) and (d) are the electron number and
power \pdfs, respectively (found by integrating the flare spectrum
electron \flRux\ spectrum from $E_{c}$ to $E_{h}$). Both are
asymmetric and show more pronounced tails when compared to the
corresponding plots for the \janfive\ data (Figures
\ref{fig:2005jan19flare}(a) and \ref{fig:electron:2005jan19}(c)).
This is due to the asymmetric \LEC\ \pdf\ which leads to a tail
extending to high values in the \pdf\ of the electron number \flRux.
Uncertainty estimates for the total number of flare-accelerated
electrons and their energy are given in Figures
\ref{fig:electron:2002jul23}(c, d).  The \pdf\ for the energy can be
integrated to determine lower limits to the energy contained in the
flare-accelerated electrons whilst simultaneously supplying a
probability estimate.  The cumulative probability distribution
function for the energy shows that there is a 95\% probability that
the energy in the flare-accelerated electrons is greater than
$10^{28.0}$ erg sec$^{-1}$, and a 68\% probability that it is greater
than $10^{28.2}$.

\begin{deluxetable}{clcccc}
\tabletypesize{\scriptsize}
\tablecolumns{5}
\tablewidth{0pt}
\tablecaption{Parameter values and uncertainty estimates derived for
    the four uncertainty estimation methods applied to the \jultwo\
    flare spectrum, as described in Section \ref{sec:obs}.  The final
    column ``Ratio'' is defined as the ratio of the $\pm95$\%
    uncertainties to the $\pm68$\% uncertainties; for an exact \NormGauss\
    distribution the entry in this column would be $1.96, 1.96$. Two
    ratios are quoted in order to reveal the presence of any relative
    asymmetry in the upper and lower uncertainty estimates, if
    present.  See Section \ref{sec:parest} for a detailed description
    of how the uncertainty estimates are found for each method.  The entry
  `\notdet' indicates that the value was not determinable
  by the method.\label{tab:2002jul23}}
\tablehead{
\colhead{Parameter} & 
\colhead{Method}    & 
\colhead{Value\tablenotemark{a}}     & 
\multicolumn{2}{c}{Uncertainties} & \colhead{Ratio} \\
\colhead{}          & 
\colhead{}          & 
\colhead{}          & 
\colhead{68\%}      & 
\colhead{95\%}      &
\colhead{}     }
\startdata
$EM$ $(\unitsEM)$           & \covmat\tablenotemark{b}  & 2.16  &  $\pm$0.08    &       \notcalc\         &   \notcalc\   \\
               & \chimap\tablenotemark{c}  &       & $\pm$0.04   &  $\pm$0.08    & 2.05, 1.99  \\
               & \MC\tablenotemark{d}      &       & -0.05, 0.03   &  -0.09, 0.07    & 1.82, 2.28             \\
               & \bmcmc\tablenotemark{e}    & 2.17  & $\pm$0.04   &  $\pm$0.08    & 1.89, 1.96\\
               & & & & & \\

$kT$ (\keV)           & \covmat\  & 3.18  &  $\pm$0.03    &      \notcalc\         & \notcalc\ \\
               & \chimap\  &       & $\pm$0.01    & $\pm$0.02    & 1.97, 2.13  \\
               & \MC\      &       & $\pm$0.01  &  -0.02, 0.03     & 2.20, 1.87  \\
               & \bmcmc\   & 3.18  & $\pm$0.01    & $\pm$0.03   &  1.93, 1.92 \\
               & & & & & \\

$A$  & \covmat\ &  0.028 & $\pm$0.004     & \notcalc\             &    \notcalc\      \\
(electron \flRux\ at 50 keV,                     & \chimap\ &        &  -0.003, 0.002 & -0.006, 0.005 & 2.15, 1.94\\
  $10^{35}$ electrons $(\mbox{sec keV})^{-1}$)                    & \MC\     &        &  -0.002, 0.003 & -0.005, 0.005 & 2.21, 1.74\\
                     & \bmcmc\  &  0.028 &  -0.003, 0.002  & -0.006, 0.004 & 2.09, 1.76\\
               & & & & & \\

$\delta_{1}$   & \covmat\       & 3.40  & $\pm$0.16       &    \notcalc\         &   \notcalc\          \\
               & \chimap\      &        & -0.14, 0.10    &  -0.36, 0.17 & 2.61, 1.78 \\
               & \MC\          &        & -0.14, 0.12    &  -0.34, 0.19 & 2.52, 1.61          \\
               & \bmcmc\       & 3.41   & -0.13, 0.08    &  -0.33, 0.13 & 2.55, 1.55 \\
               & & & & & \\

$E_{b}$ (\keV)        & \covmat\      &256 &  $\pm$135       &     \notcalc\     &   \notcalc\   \\
               & \chimap\     &     &  -77, 147    & -123, 686 & 1.59, 6.67    \\
               & \MC\         &     &  -77, 253    & -121, 1319 & 1.58, 5.22     \\
               & \bmcmc\      &269  &  -147, 5615     & -217, 1239 & 1.47, 2.01        \\
               & & & & & \\

$\delta_{2}$   & \covmat\               & 3.92  &  $\pm$0.11   &   \notcalc\        &  \notcalc\   \\
              & \chimap\               &        & -0.08, 0.13  & -0.13, 0.78 & 1.67, 5.67   \\
              & \MC\                   &        & -0.07,0.23   & -0.12, 3.27 & 1.74, 14.2     \\
              & \bmcmc\                & 3.93   & -0.11, 0.58 & -0.18, 1.92 & 1.57, 3.33     \\
               & & & & & \\

$E_{c}$ (\keV)        & \covmat\               & 32.0  &  $\pm$24.091   &      \notcalc\       & \notcalc\  \\
              & \chimap\               &        & -5.78, 5.05    & \notdet, 12.1   &  \notdet, 2.4 \\
              & \MC\                   &        & -6.86, 7.37    & -20.7, 15.9 & 3.02, 2.16  \\
              & \bmcmc\                & 31.2   &  -16.1, 11.7   & -23.1, 19.1 & 1.44, 1.63  \\
\enddata
\tablenotetext{a}{The `\covmat', `\MC' and `\chimap' methods all start
  from the same parameter value $\varhat$ where $\chit$
  is minimized. For the \bmcmc\ approach, the ``maximum a posteriori''
  $\varv^{MAP}$ value is quoted.}
\tablenotetext{b}{See Section \ref{sec:chimap} and Equation \ref{eqn:uncert:chimap} for the
  definition of the parameter uncertainty for the \covmat\ method.}
\tablenotetext{c}{See Section \ref{sec:lscov} and Equation \ref{eqn:cmat} for the
  definition of the parameter uncertainty for the \chimap\ method.}
\tablenotetext{d}{See Section \ref{sec:mc} and Equation \ref{eqn:uncert:mc} for the
  definition of the parameter uncertainty for the \MC\ method.}
\tablenotetext{e}{See Section \ref{sec:mcmc} and Equation \ref{eqn:uncert:mcmc} for the
  definition of the parameter uncertainty for the \bmcmc\ method.}
\end{deluxetable}

As was noted in Section \ref{23July2002}, a different spectral
normalization was used in the analysis of the \jultwo\ flare compared
to the \janfive\ flare.  The package OSPEX implements the spectral
normalization of the \janfive\ model spectrum using the integrated
normalization factor, $F_{0}=AE_c^{1-\delta_1}/(\delta_1 - 1)$.  This
implementation of the flare spectral model therefore introduces a
parameter dependence into the $\chit$-\surf\ between the normalization
$A$, the \LEC\ and the spectral index $\delta_{1}$.  However, since
the \LEC\ for the \janfive\ flare is relatively well defined, the
integrated \flRux\ $F_{0}$ is relatively well defined, and the MCMC
algorithm can explore the $\chit$-\surf\ as a function of $F_{0}$ and
$E_{c}$ with no difficulty.  However, the \LEC\ is not well defined
for the \jultwo\ flare, and so the range of values to $F_{0}$ is
large.  Therefore when using the implementation of Equation
\ref{eqn:powerlaw} used in the analysis of the 19 January 2005 flare,
the parameter space that must be covered by the MCMC algorithm is
large due to the inherent dependence of $F_{0}$ on $E_{c}$.  This was
found to be prohibitive to an efficient MCMC search, and so an
alternate implementation of Equation \ref{eqn:powerlaw} was created
for OSPEX (re-parameterization of the fitting function is a
recommended tactic in creating better search spaces for MCMC
\citep{citeulike:105949}).  In this implementation, the normalization
factor used to describe the spectrum is $A$, the value of the spectrum
at the pivot value $E_{p}$.  Moving to a different hypersurface for
the same problem greatly improved the efficiency of the MCMC
algorithm.

\section{Discussion}\label{sec:discussion}

\subsection{Comparison of uncertainty analyses}\label{sec:comparison}

The uncertainty analyses performed on both data-sets shows that the
shape of the $\chi^{2}$-\surf\ has a significant effect on the values
of the uncertainties found.  All the uncertainty estimates found for
the spectral parameters describing the \janfive\ flare data are
similar, regardless of the method.  The uncertainty estimates found
for the spectral parameters describing the \jultwo\ flare data depend
on the method chosen.

Since the data have a large number of counts at almost all energies,
the \surfs\ described by Equation \ref{eqn:rchi} and Equation
\ref{eqn:bsurf} are almost identical.  The two-dimensional marginal
distributions for the \jultwo\ flare data (Figure
\ref{fig:all2d2002jul23}) shows structures which are not simple
two-dimensional \NormGauss\ distributions, and, since the two
\surfs\ described by Equation \ref{eqn:bsurf} and Equation
\ref{eqn:rchi} are almost identical, the $\chit$-\surf\ must have
structures which are not simple two-dimensional \NormGauss\ distributions.
This means that one or more of the assumptions that lead to the
assertion that the probability distribution for $\delta\vars_{obs}$ is
a multivariate \NormGauss\ distribution around $\varhat$ does not hold for
this model applied to these flare data (Section \ref{sec:lscov}). The
non-\NormGauss\ distribution shapes of Figure \ref{fig:all2d2002jul23}
suggest that the assumption that the spectral model is linear (or at
least locally so within the range of the desired uncertainty
calculation) is not satisfied (\citeauthor{1992nrfa.book.....P}
\citeyear{1992nrfa.book.....P}, p. 690).  Hence, the \covmat\ and
\chimap\ methods cannot be expected to give reliable and consistent
estimates in this case.

The shape of the $\chit$-\surf\ also influences the results of the
\MC\ method.  This can be seen in the results for the \LEC\ in the
\jultwo\ data-set (Figure \ref{fig:2002jul23flare}e).  It is expected
that below a given energy $E_{plateau}$, all values of the \LEC\ are
equally likely.  This is because in this energy range the number of
counts due to thermal emission greatly exceed the number due to the
\FIES, and so changing one value of $E_{c}$ over another makes no
difference to the fit to the data - the value of $\chi^{2}$, or
equivalently, the Bayesian posterior probability, are unaffected.
Therefore, all values below $E_{plateau}$ are equally likely\footnote{
  $E_{plateau}$ can also be interpreted as the energy below which no
  further information is available that can be used to better
  constrain a lower limit to the \LEC.}.  The \MC\ method results do
not show this; the results are clustered around the best-fit value and
do not show the extension to lower energies as expected.  Hence the
uncertainty estimate arising from the \MC\ method does not conform to
our prior expectation of what it should report.

In contrast, the Bayesian posterior \surf\ for the \janfive\ shows
simple \NormGauss\-like one-dimensional distributions (and so the
assumptions behind the \covmat\ and \chimap\ methods are approximately
true) and give similar answers.  The \MC\ method (Section
\ref{sec:mc}) relies on finding local minima to simulated data which
is statistically similar to the original data.  This method works well
in the \janfive\ analysis as the shape of the \surf\ (Figure
\ref{fig:all2d2005jan19}) is dominated by a nearly \NormGauss\ single
minimum, a feature the method repeatedly finds in all the similar
$\chit$-\surf.  The \chimap\ method does agree with the \bmcmc\ result
in that the \chimap\ method does indicate that below a certain value
($E_{plateau}$), all values of the \LEC\ are equally likely.  However,
the method cannot give a lower limit to the 95\% uncertainty estimate
since at no point does $\delta \chi^{2}=4$ for $E<E_{c}^{min}$
(Section \ref{sec:chimap}).

The \bmcmc\ method samples the parameter space via the posterior
probability and the \MCMC\ algorithm (Section \ref{sec:mcmc}).  The
Bayesian interpretation of the posterior probability means that the
parameter samples are found in proportion to how well they describe
the data (values of $\vars$ that have lower probability are less
likely explanations of the data).  The method does not make any
assumptions about the nature of the \surf, as the other three methods
do.  Hence it agrees with the results from the methods of Section
\ref{sec:nlls} when applied to simple \surfs\ where the assumptions
made by those methods are valid, but generates different results when
those assumptions do not hold.  Therefore, the \bmcmc\ method can, in
principle, be used without having to invoke any special knowledge of
the shape of the \surf\ and without making some simplifying
assumptions.


\subsection{Probability density functions of the parameters of the
  \jultwo\ electron spectrum model}\label{sec:jul23discuss} 

\begin{figure}
  \includegraphics[width=0.502\textwidth,clip=]{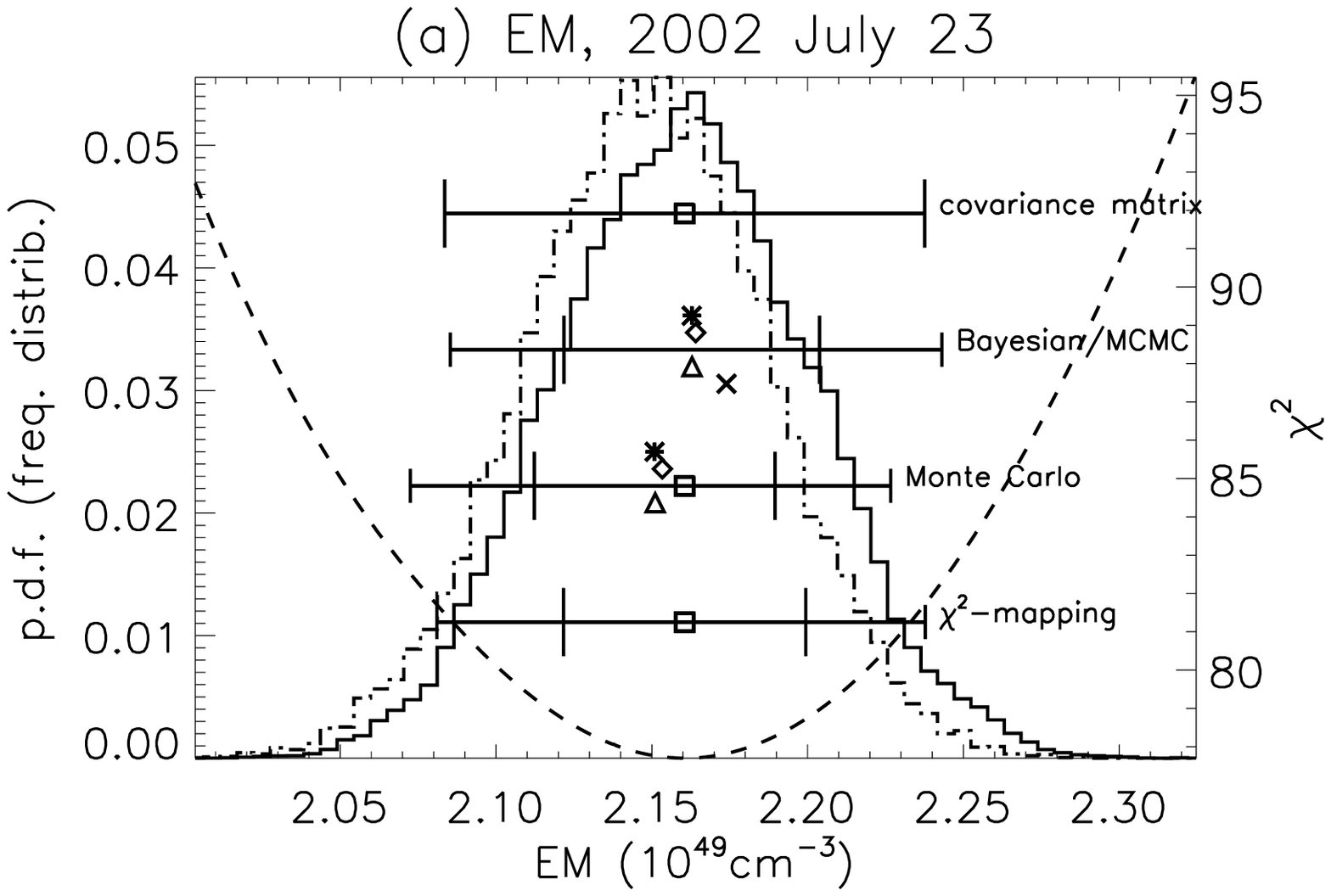}
  \includegraphics[width=0.502\textwidth,clip=]{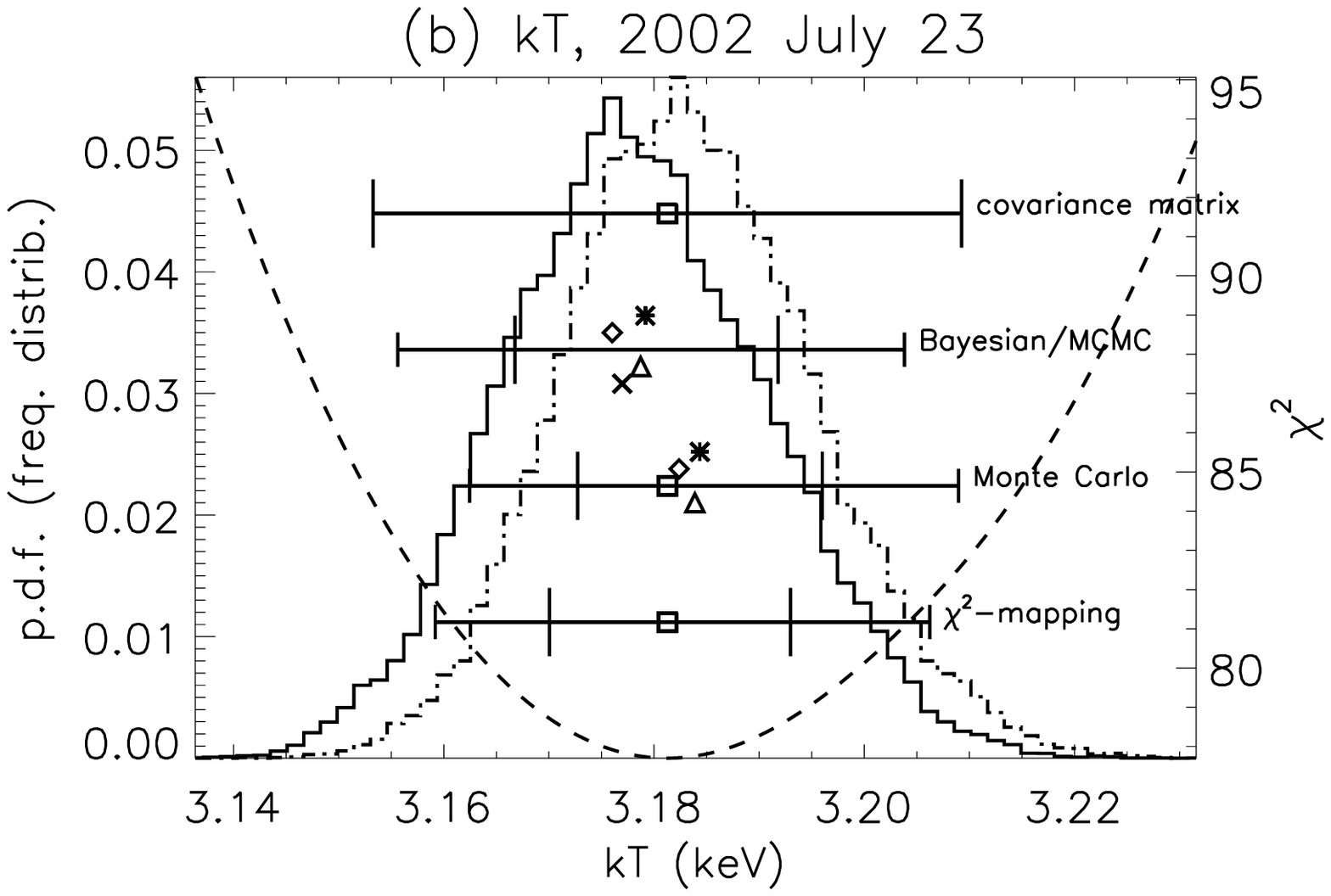}

  \caption{Results from each of the four uncertainty analysis methods
    (Section \ref{sec:parest}) for (a) $EM$ and (b) $kT$ from the
    model fit to the \jultwo\ flare data. These
    plots follow the same convention as Figure
    \protect\ref{fig:2005jan19thermal}.  See Section \ref{sec:res} for
    more detail on these results.}\label{fig:2002jul23thermal}
\end{figure}

\begin{figure}
  \includegraphics[width=0.502\textwidth,clip=]{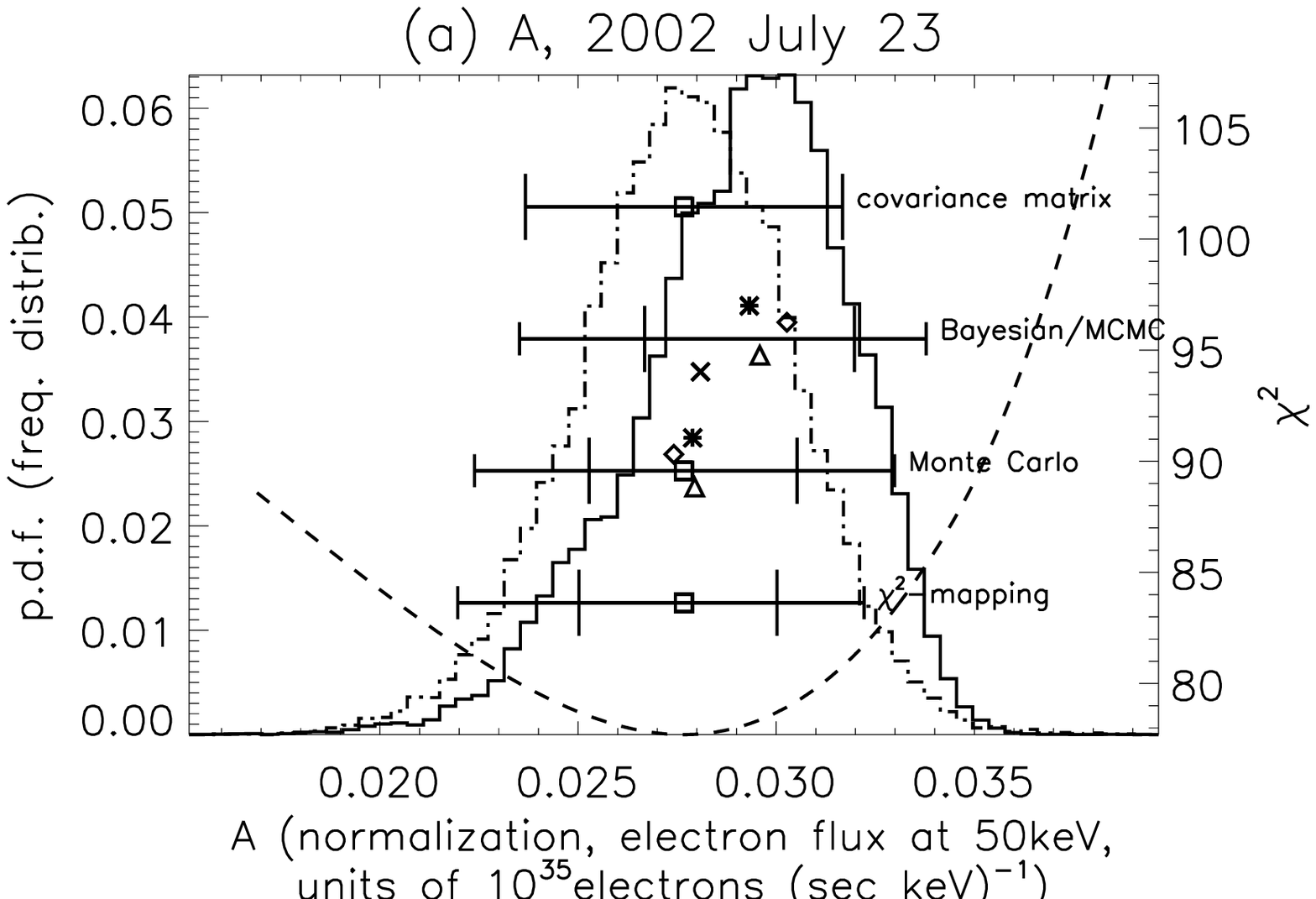}
  \includegraphics[width=0.502\textwidth,clip=]{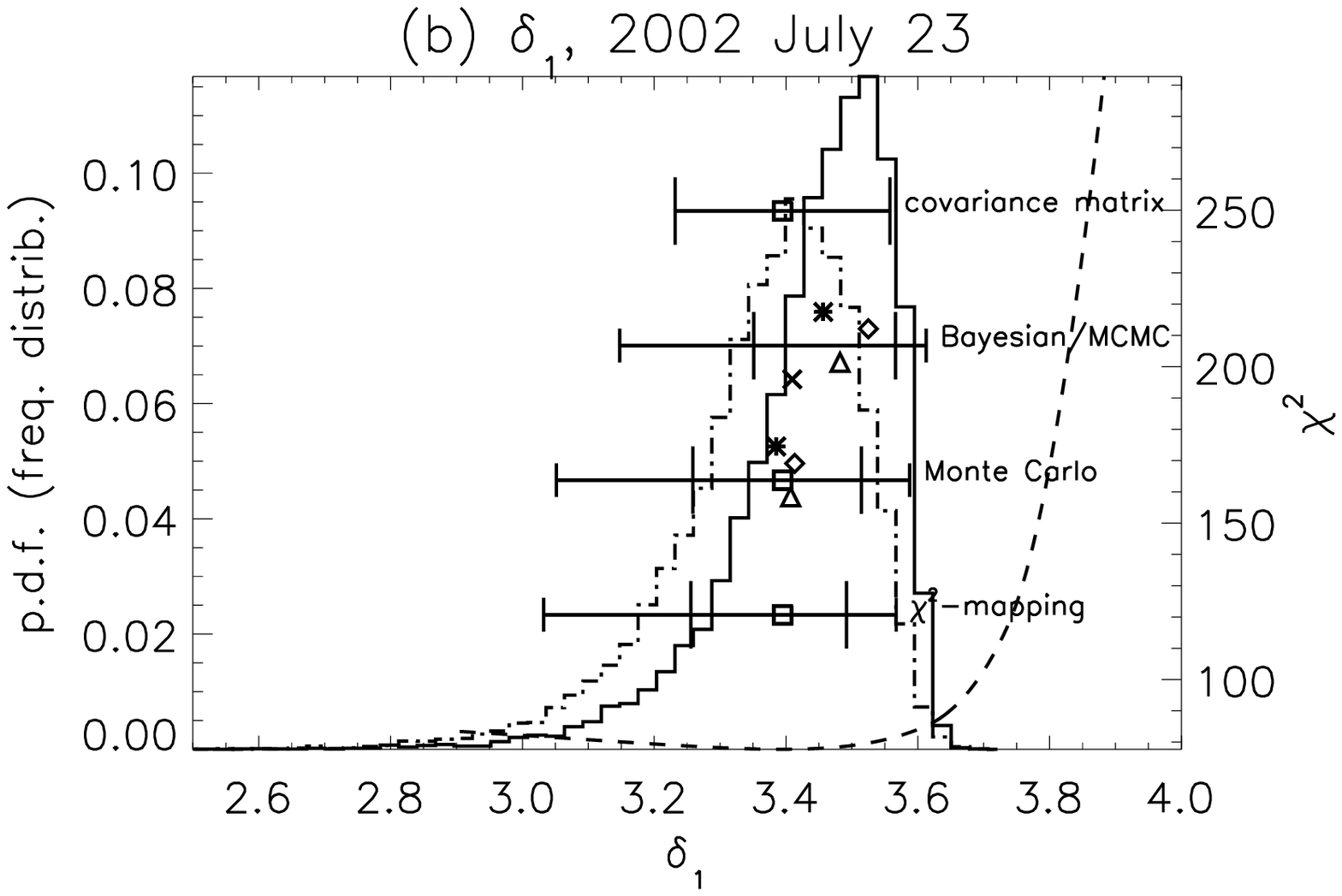}
  \includegraphics[width=0.502\textwidth,clip=]{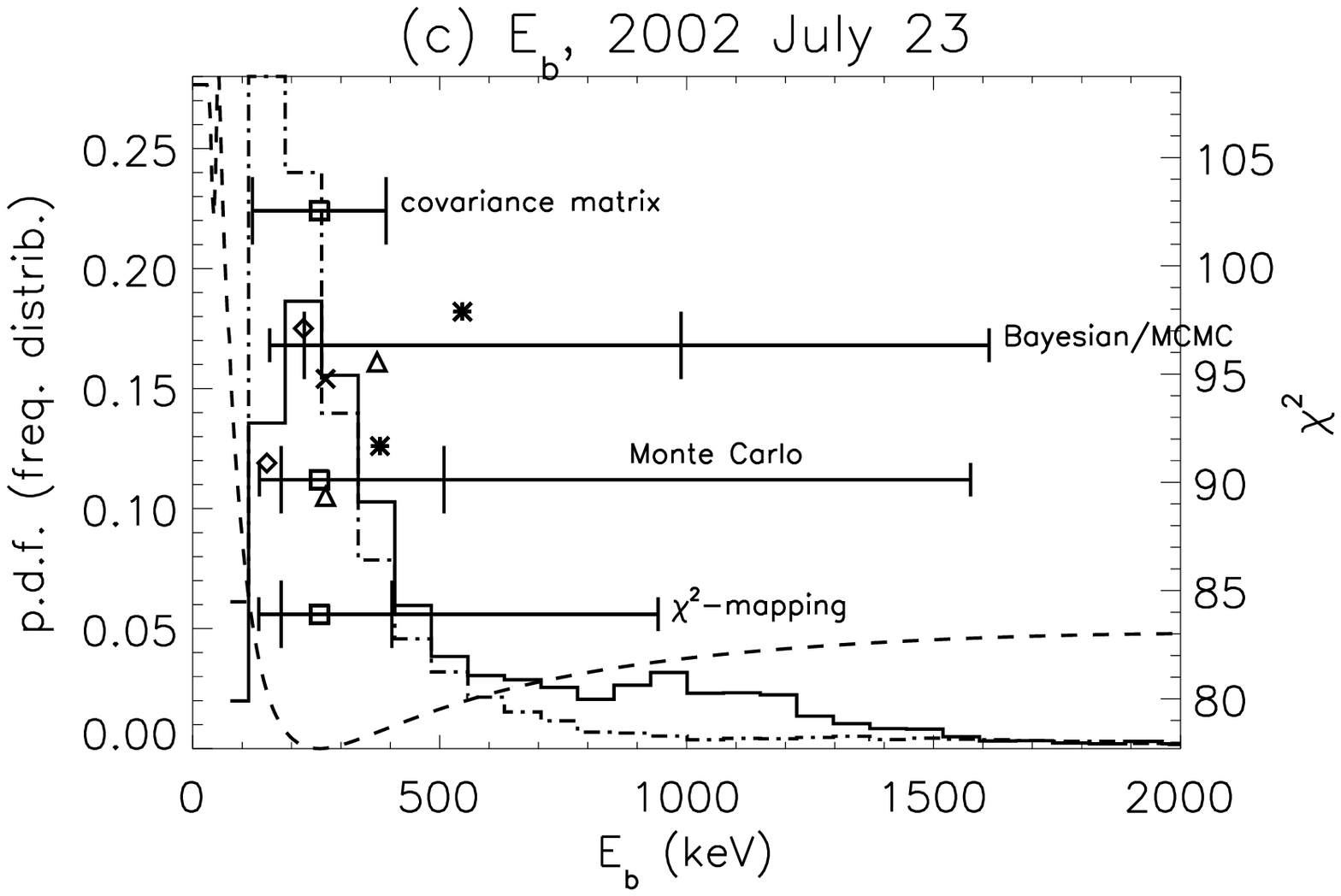}
  \includegraphics[width=0.502\textwidth,clip=]{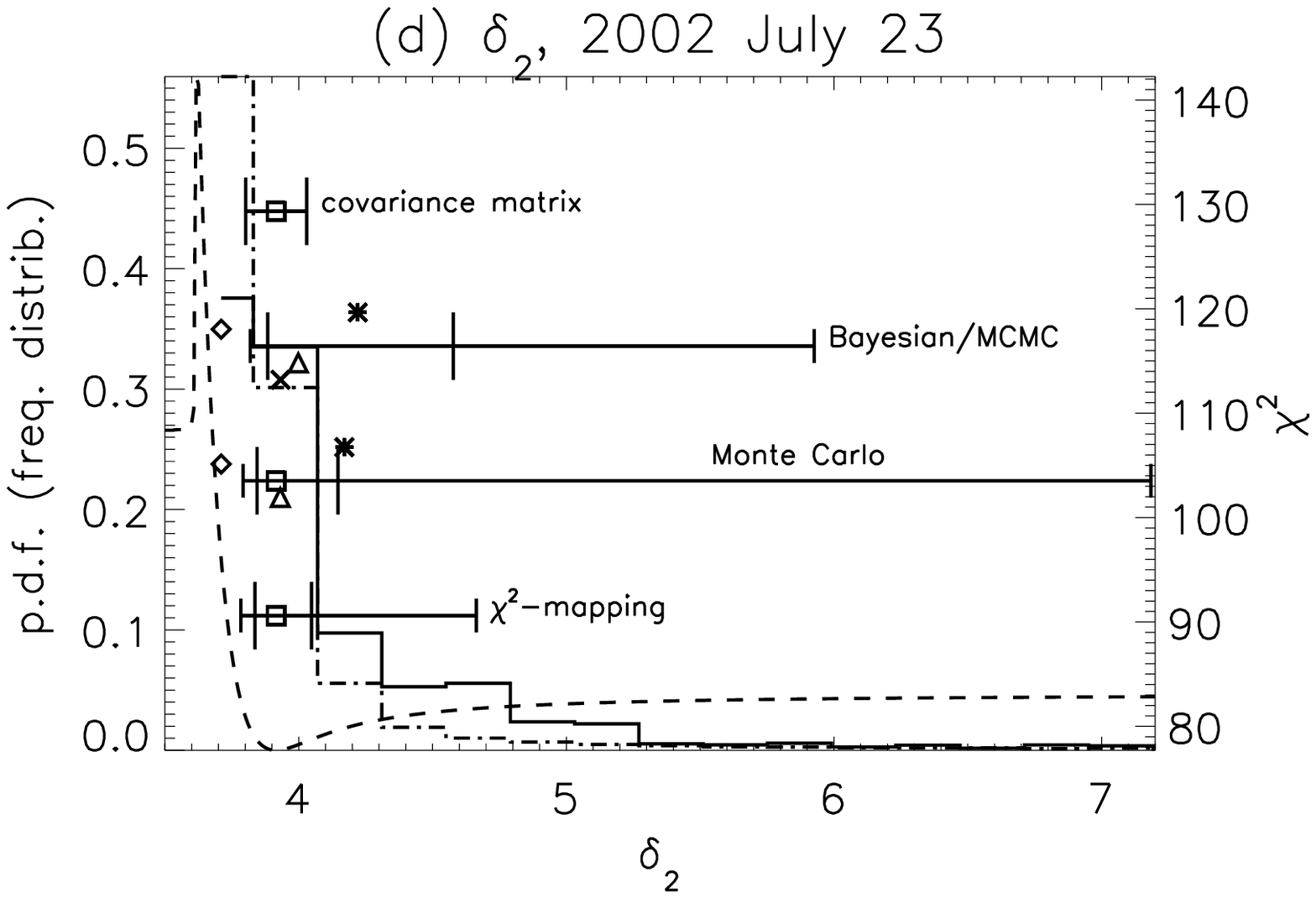}
  \centerline{
    \includegraphics[width=0.502\textwidth,clip=]{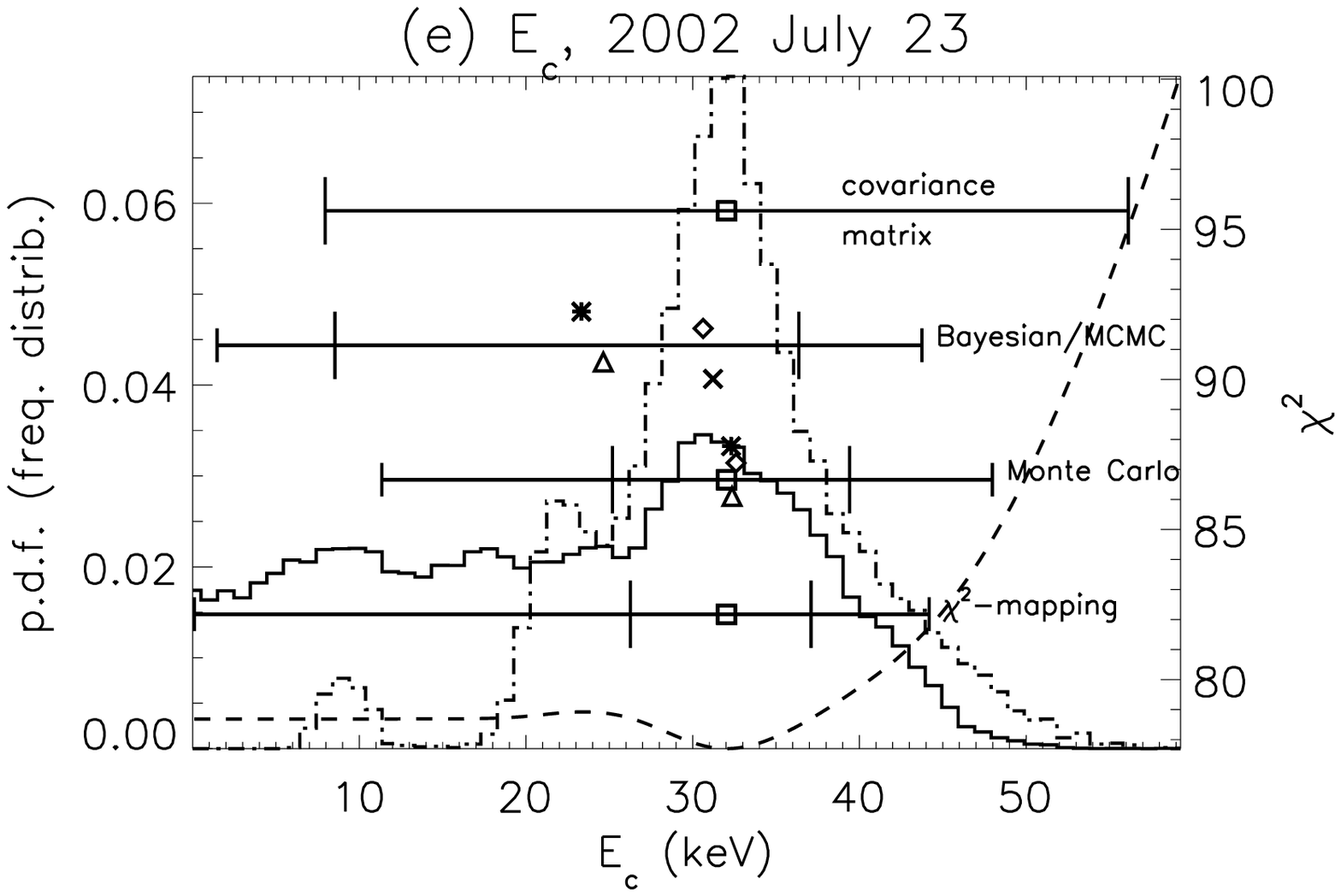}
    }
  \caption{Results from each of the four uncertainty analysis methods
    (Section \ref{sec:parest}) for (a) $A$, (b) $\delta_{1}$, (c)
    $E_{b}$, (d) $\delta_{2}$ and (e) $E_{c}$, from the model fit to the
    \jultwo\ flare data.  These plots follow the same convention as
    Figure \protect\ref{fig:2005jan19thermal}. See Section
    \ref{sec:res} for more detail on these
    results.}\label{fig:2002jul23flare}
\end{figure}

\begin{figure}
  \centerline{
    \includegraphics[width=0.900\textwidth,clip=]
{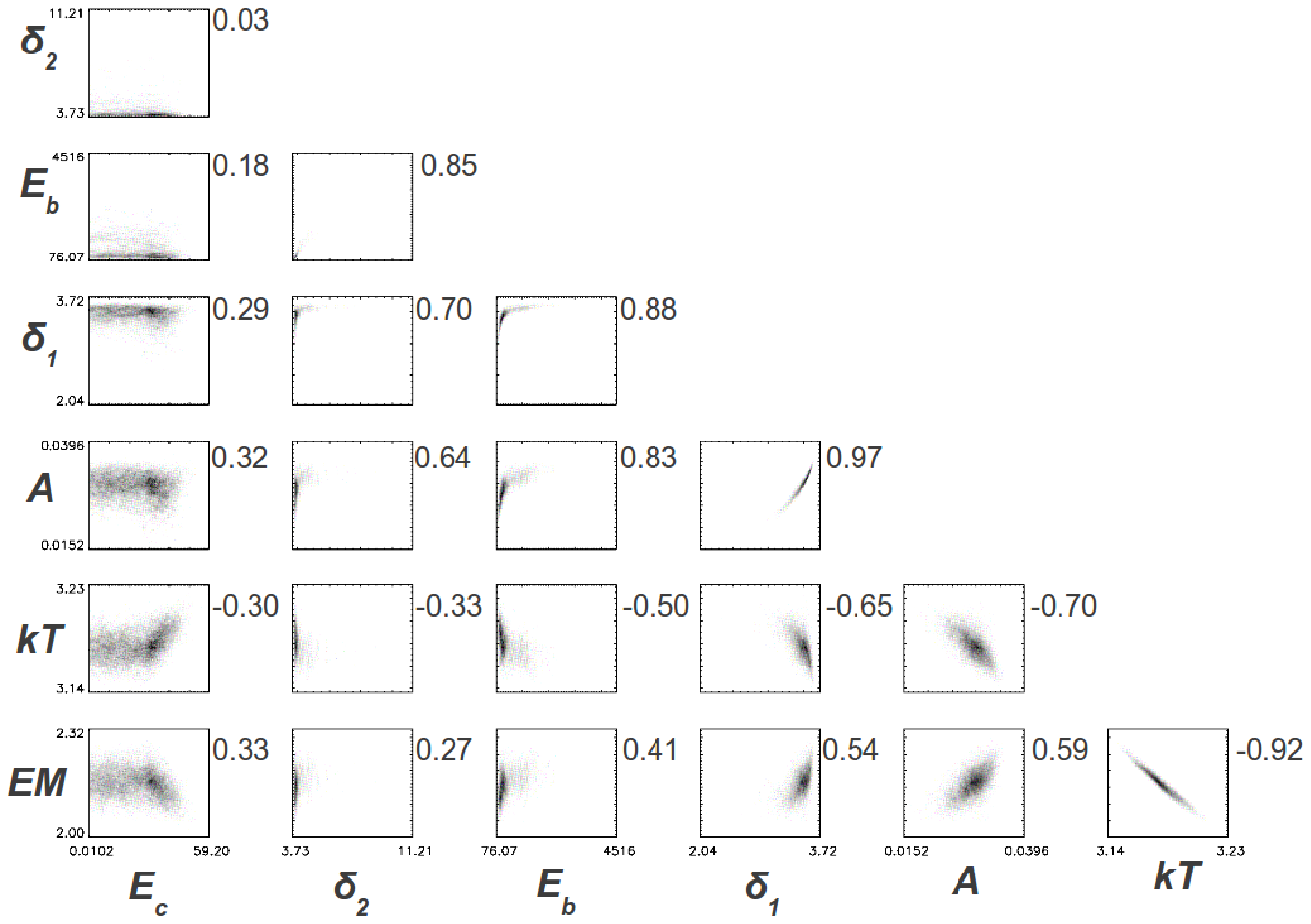}
    }
    \caption{Two dimensional marginal \pdfs\ for the parameters of the
      model used to fit the spectrum of the \jultwo\ flare.  In
      contrast to similar distributions plotted in
      Fig. \ref{fig:all2d2005jan19} for the \janfive\ flare, some
      distributions are highly asymmetric within the parameter ranges
      found.  The number on the upper right of each plot is the
      Spearman rank correlation coefficient for the abscissa versus
      the ordinate.  There are many more moderately and strongly
      (anti-) correlated pairs of parameters for this flare model
      compared to the \janfive\ flare model.  For some pairs of
      parameters (for example $\delta_{1}$ versus $A$ and $\delta_{2}$
      versus $E_{b}$), the proportion of the space taken up by the
      high probability volume is relatively small, and for others (for
      example, $E_{c}$ versus $A$), it is relatively large.  For the
      model applied to this flare spectrum, many of the resulting
      \pdfs\ do not show \NormGauss\ distribution shapes.  This
      indicates that the \surf\ for the model fit to these flare data
      has a more complicated structure than the \surf\ of the model
      fit to the \janfive\ flare.  \label{fig:all2d2002jul23}}
\end{figure}

\begin{figure}
\centerline{
  \plottwo{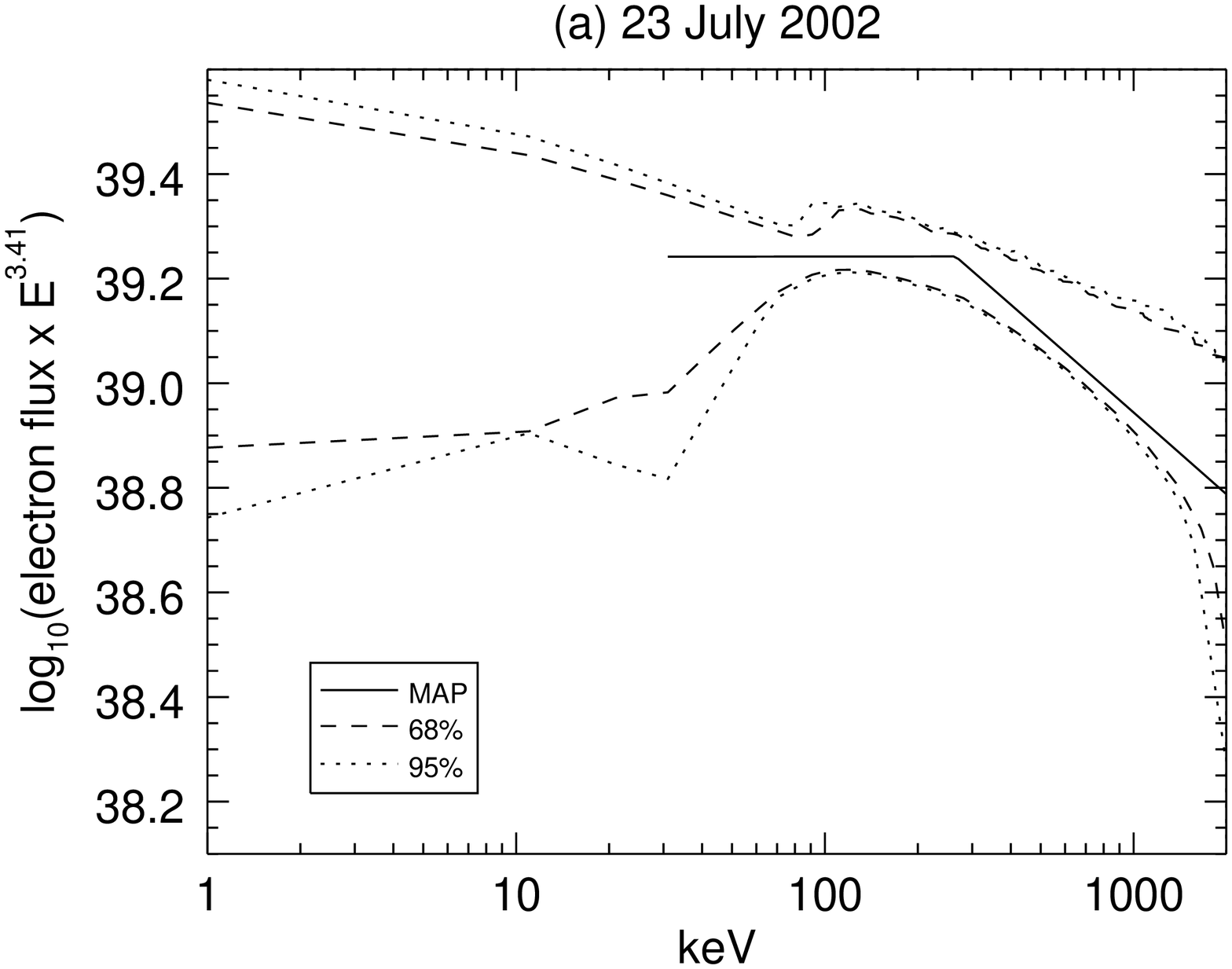}{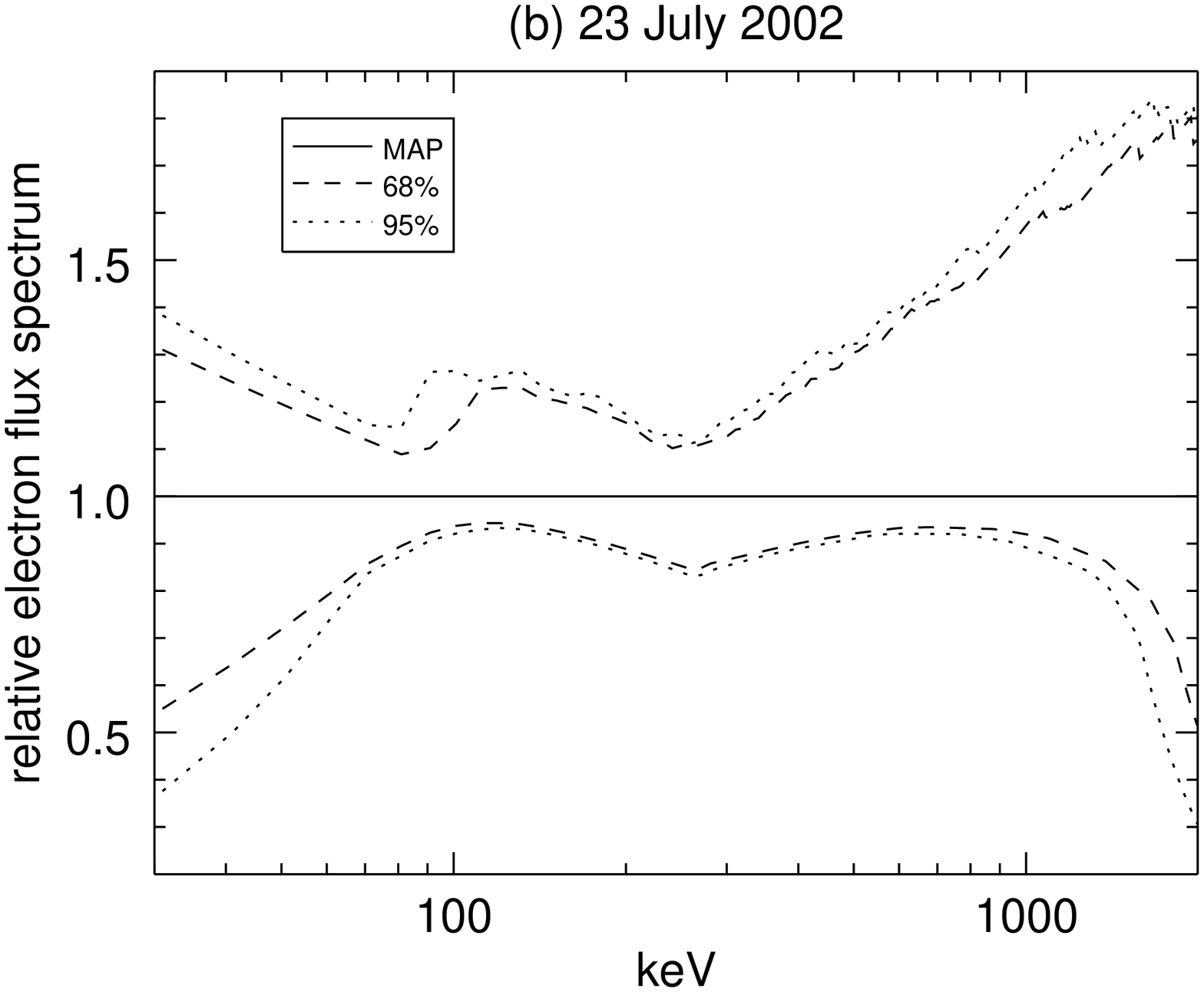}
}
\centerline{
  \plottwo{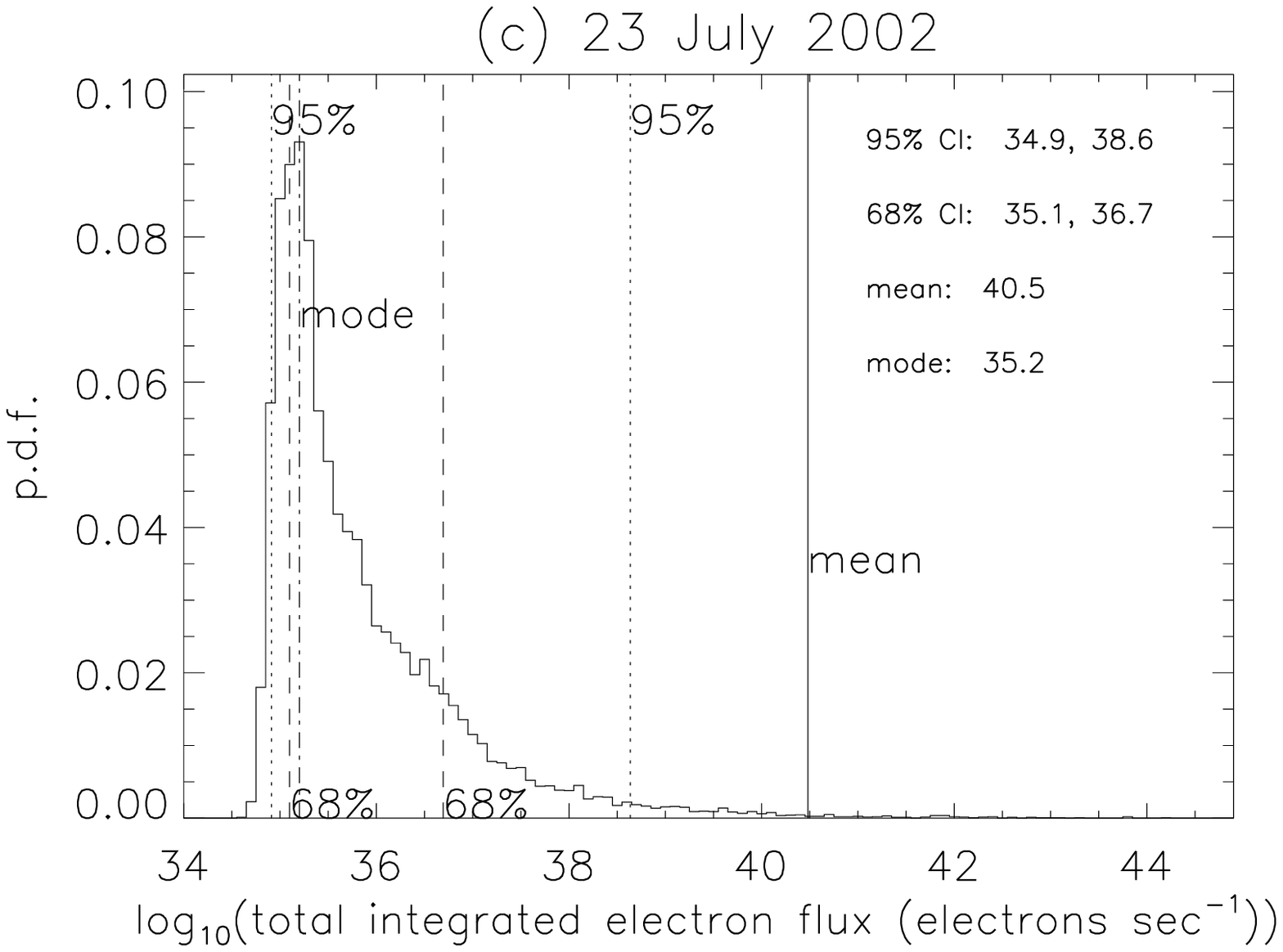}{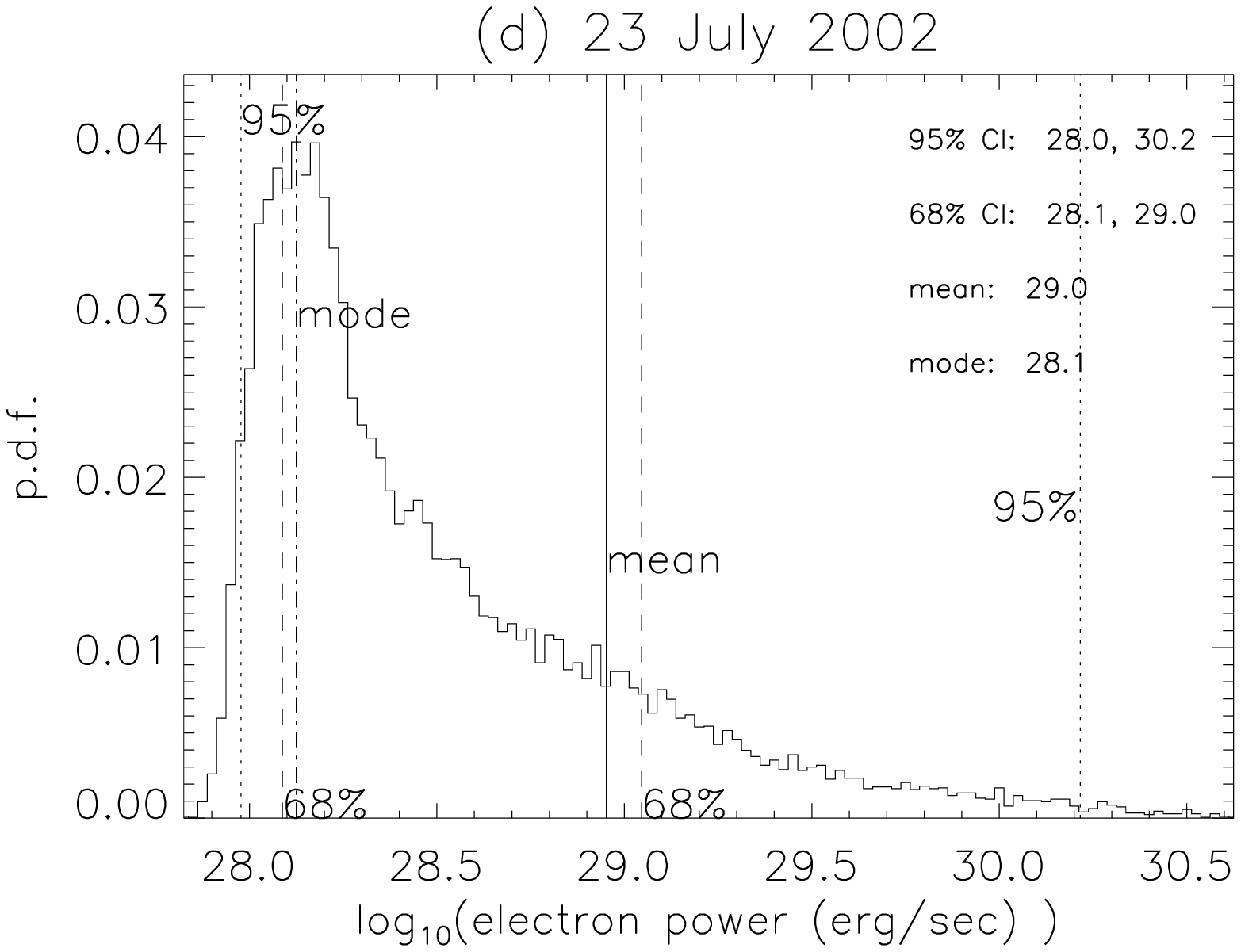}
}
  \caption{Electron spectrum results for the flare-injected electrons
    arising from the \bmcmc\ method for the \jultwo\ flare.  (a)
    Electron spectrum (\flRux\ (in units of
    $\mbox{erg}$ $\mbox{keV}^{-1}$ $\mbox{s}^{-1}$) multiplied by
    $E^{3.38}$) with 68\% and 95\% credible interval spectra indicated
    by the dashed and dotted lines, respectively.  The electron
    \flRux\ spectrum corresponding to $\varv^{MAP}$ is indicated by
    the solid line. (b) 68\% and 95\% credible intervals (dashed and
    dotted lines, respectively) relative to the $\varv^{MAP}$ electron
    \flRux\ spectrum.  (c) Flare injected electron number
    \flRux\ \pdf, with 68\% and 95\% credible intervals indicated. (d)
    Flare injected electron power \pdf, with 68\% and 95\% credible
    intervals indicated.  In plots (c) and (d) the distribution
    mean/mode is indicated by the solid/dot-dashed vertical line.}
  \label{fig:electron:2002jul23}
\end{figure}

Figures \ref{fig:2002jul23thermal} and \ref{fig:2002jul23flare} show
the marginal \pdfs\ of the parameter values arising from a
\bmcmc\ treatment of the data analysis problem.  It is notable that
the distributions for $E_{b}, \delta_{2}$ and $E_{c}$ are distinctly
different from more symmetrical and \NormGauss\ distribution-like
distributions of the other parameters in the fit.  The break energy
$E_{b}$ and the power law index above the break $\delta_{2}$ are
highly correlated (Figure \ref{fig:all2d2002jul23}) over a wide range
of values.  As $E_{b}$ increases, the value of $\delta_{2}$ increases.
The mild curvature of the spectrum implied by these \pdfs\ is
consistent with a wide range of near power-law electron \flRux\ spectrum
models, leading to an ill-defined value for $E_{b}$ and softer
power-law indices at higher values of $E_{b}$.  A count spectrum that
appears to come from emission that is mildly curved with respect to
the radiation from the thick-target interaction of a flare-injected
electron \flRux\ spectrum with a power law distribution could arise from
an inaccurate \Xray\ albedo correction \citep{2006A&A...446.1157K} or
from a non-uniform ionization within the target plasma
\citep{2009ApJ...705.1584S, kontar2002}.

The \LEC\ also has an interesting \pdf\ (also reproduced by the
\chimap\ analysis, Figure \ref{fig:2002jul23flare}e).  There is a peak
in the \bmcmc\ \LEC\ \pdf\ at 31 keV, and a tail at lower energies
where the thermal emission of the plasma dominates over the emission
due to the flare-injected electron \flRux.  We wish to estimate how much
more likely the \LEC\ is close to the peak, compared to other parts of
the \pdf. An estimate can be generated using the following procedure.
If the \pdf\ of the \LEC\ were a \NormGauss\ distribution
$N(E_{cutoff},\sigma)$ (where $N(a,b)$ is a \NormGauss\ distribution
centered at $a$ with standard deviation $b$), then the total
probability that $E_{c}$ lies in the range $E_{cutoff}-\sigma,
E_{cutoff}+\sigma$ is about 68\%. The maximum probability that $E_{c}$
lies in a $2\sigma$ wide range of values that does not overlap with
the range $E_{cutoff}-\sigma, E_{cutoff}+\sigma$ is about 16\%.
Therefore the value of $E_{c}$ is about 4 times more likely to be in
the range $E_{cutoff}-\sigma, E_{cutoff}+\sigma$ than in a $2\sigma$
wide range of values that does not overlap with the range
$E_{cutoff}-\sigma, E_{cutoff}+\sigma$.  Fitting the peak of the
\pdf\ of Figure \ref{fig:2002jul23flare}(e) with a \NormGauss\ distribution
yields a width $\sigma$ of about 5 keV.  Applying the estimation
procedure above on the \pdf\ of Figure \ref{fig:2002jul23flare}(e)
with $\sigma =5$ keV, it is found that $E_{c}$ is about 1.3 times more
likely to be in the range 25-35 keV than in any other continuous
window of values 10 keV wide.  This is weak evidence for a peak in the
range 25-35 keV.

Therefore, the \pdf\ is interpreted as providing evidence for the
existence of an observable \LEC\ just above the region where the
thermal emission dominates.  If the \LEC\ was at higher energies, then
the \pdf\ for $E_{c}$ would resemble more closely the \pdf\ seen in
Figure \ref{fig:2005jan19flare}(c) for the January 19 flare and
therefore lower possible values to $E_{c}$ would lead to lower
posterior probabilities (worse fits).  If the \LEC\ was present at
energies where the thermal emission dominates, then no peak in the
\pdf\ for $E_{c}$ would be seen.  Lower values would account for more
of the flare-injected spectrum, and so lower values would be more
probable.  The probability $p(E_{c})$ would eventually plateau at some
energy $E_{plateau}$ since the emission due to the flare-injected
electron \flRux\ would be far less than the emission due to the thermal
plasma below $E_{plateau}$, making all values of $E_{c}$ equally
likely, as there is nothing to distinguish one value from another.
However the observed $p(E_{c})$ is a combination of both; a peak in
the \pdf\ with an approximately constant probability density at lower
energies.


\subsection{Flare electron number and energy \pdfs}\label{sec:energyelectron}

The \bmcmc\ method allows for the construction of \pdfs\ for each
flare (Figures \ref{fig:2005jan19flare}(a) and
\ref{fig:electron:2005jan19}(c), \ref{fig:electron:2002jul23}(c, d) of
the number of flare-accelerated electrons and the energy they carry,
fully expressing the correlated dependence of one variable on another
(Figures \ref{fig:all2d2005jan19}, \ref{fig:all2d2002jul23}).  Since
the result is another \pdf, credible intervals for the number of
electrons and their energy can also be calculated.  In contrast,
taking the set of 68\% upper model parameter uncertainty estimates (or
the other model parameter uncertainty estimates) from the the other
methods cannot be used to calculate the corresponding 68\% upper
uncertainty estimate for the number of electrons and their energy.
This is because there is no guarantee that that point on the
$\chit$-\surf\ has a significant non-zero probability (or
equivalently, lies in a hightly probable region of the model parameter
\surf).  In relatively simple \surfs\ this may be true, but in highly
correlated \surfs\ such as in the analysis of the \jultwo\ flare
presented here, it may not be.  As far as we are aware, this is the
first time that flare electron number and energy \pdfs\ have been
estimated from data.

A significant difference between the two flares studied is the
uncertainty with which the model parameters are known.  This leads to
significant differences in how well the gross properties of the flare
are known.  The \LEC\ is not well constrained for the \jultwo\ flare,
leading to 68\% and 95\% credible intervals in the flare electron
number and energy \pdfs\ that span orders of magnitude.  Notably, the
\jultwo\ \pdfs\ are highly asymmetric and so lower values of flare
electron number and energies are much less likely than higher values.
It is interesting to note that there is a peak in the energy \pdf\ for
the \jultwo\ flare, even although there is a non-zero probability for
$E_{c}$ down to the lower limit given by the prior for the \LEC.  This
is due to the peak in the marginal \pdf\ of $E_{c}$, which therefore
defines a more probable total flare energy than those arising from the
lower probability range $E_{c}<E_{plateau}$.

The estimate of the actual number of electrons and the energy they
carry is also dependent on systematic errors related to the
calibration of each of RHESSI detectors with each other.  As was noted
above, the systematic errors in the individual PHA bins are small
compared to the systematic error in the overall sensitivity of each
detector \citep{2009ApJ...699..968M, 2011ApJ...731..106S}.  This means
that the {\it shape} of the flare-accelerated electron spectrum
suffers from a smaller error compared to the integral under the curve
of the flare-accelerated spectrum.  We therefore expect that the broad
qualities of the shapes of the flare electron number and energy
distributions will remain unchanged for each of the two flared
studied; the \janfive\ results will remain approximately symmetric,
and the \jultwo\ results will remain quite asymmetric.  We estimate
that allowing for a 10\% - 30\% error in knowledge of the sensitivity
of each detector would smooth out the distribution peak, and add
another 0.1-0.2 in the logarithm (approximately) of the widths of the
\pdfs.  This estimated uncertainty is substantially more than the 95\%
estimated uncertainty in the case of the \janfive\ flare, but is
substantially less than the 95\% estimated uncertainty for the
\jultwo\ flare.  This suggests that the uncertainty in the true value
of the \LEC\ is a more important limiting factor in understanding the
electron and energy content in RHESSI-observed flares than the
detector calibration uncertainty.

\subsection{Expanding the analysis}\label{sec:expand}

It is common in RHESSI data analysis to remove a background component
from the observed count data to yield an estimate of the counts due
solely to the flare.  This background-subtracted data is then used in
further analysis.  Strictly, models for the background and the flare
should be fit simultaneously since the observed counts are due to the
background and the flare simultaneously.  Therefore, the first
improvement we will make is to fit both the flare response and
background simultaneously.  This will be done by including a simple
parameterization of the pre- and post-flare hard \Xray\ \flRux\ observed
by RHESSI into the flare model.  The parameters of the background
model will also require their own priors.  The inclusion of a
background model in the fit is expected to have an effect an higher
energies, where the signal-to-noise ratio of the flare-accelerated
electrons are smaller, such as in smaller flares.

The analyses presented here made use of data from one single detector.
Our second improvement to the existing analysis will be to including
data from more than one detector, which will increase the signal to
noise ratio.  In order to use data from more than one detector,
information about the relative calibration of each detector will have
to be included.  This will be incorporated into priors for each
detector that express the degree of uncertainty in their calibration.
Since each detector is observing the same flare, the flare model will
be the same across detectors.  The posterior will be a product of the
priors for the flare model plus background, a likelihood function for
each detector, and a prior function expressing the degree of
uncertainty in their calibration.  The resulting posterior will
express the increased knowledge that comes with a larger number of
counts, but also the uncertainty in their relative calibration.

We note also that Bayesian data analysis provides a framework that can
be used to compare the explanatory power of different models of the
data whilst taking into account the number and type of variables in
each model \citep{2005blda.book.....G}.  We will use Bayesian model
comparison techniques to determine if RHESSI data can distinguish
between different effects that may contribute to the observed spectra.
In particular, we will re-analyze the \jultwo\ data presented here
using a model that incorporates the non-uniform ionization of the
thick-target plasma \citep{2009ApJ...705.1584S, 2003ApJ...595L.123K}.
Such a model produces a curvature in flare-accelerated electron
spectrum which may explain the high correlation between the break
energy $E_{b}$ and the value of $\delta_{2}$ (Section
\ref{sec:res:2002jul23}).

\section{Conclusions}\label{sec:conc}
This paper describes in some detail four methods that can be used to
estimate the uncertainties in parameters of flare models fit to RHESSI
hard \Xray\ flare data.  Three of the four methods -- \covmat, \MC, and
\chimap\ -- measure \ssizes\ in the $\chit$-\surf\ (or related
\surfs) and call them uncertainty estimates.  We have shown that care
must be taken in relying upon these uncertainty measurements, as we
have seen that they need not agree with our expectation of what an
uncertainty estimate should report, or with each other.  The fourth
method, Bayesian data analysis, can answer the question ``what is the
uncertainty in this parameter?'' by calculating a \pdf\ for that
parameter through the marginalization procedure of Section
\ref{sec:marginal} without making any further assumptions about the
number of counts in each bin (see Section \ref{sec:mcmc}).  The fourth
method broadly agrees with the other three in the case of the
\janfive\ flare.  Each method generates different uncertainty
estimates for the \jultwo\ flare.

The source of the different uncertainty estimates is the shape of the
$\chit$-\surf\ parameterized by the flare model.  \surfs\ that broadly
conform to the assumptions underlying the \covmat, \MC, and
\chimap\ methods yield consistent uncertainty estimates that agree
with each other and those from the \bmcmc\ approach.  Conversely,
\surfs\ that break those assumptions yield method-dependent results.
The \bmcmc\ approach makes no assumptions on the nature of the \surf.
Further, the position of the \LEC\ in relation to the region where
thermal \Xray\ emission dominates is crucial in determining the shape
of the \surf.  Most flares are thought to have a \LEC\ close to or at
the region of thermal emission dominance.  The \bmcmc\ method
presented here handles both flare analyses without regard to the
location of the \LEC, and makes no assumption about the
$\chit$-\surf\ or Bayesian posterior probability \surf.  The
\bmcmc\ method was the only method to generate an uncertainty estimate
of the \LEC\ that reflects our intuition of how it is constrained by
the data, for both flares studied.  Since the \chimap\ approach does
partially map the space around $\varhat$, it is perhaps the best of
the three non-Bayesian based methods that can give an indication that
the $\chi^{2}$-\surf\ contains features that are not similar to \NormGauss\
distribution shapes.  If the $\chi^{2}$-\surf\ does contain features
not anticipated by the \covmat, \MC, and \chimap\ methods, then we
suggest a \bmcmc\ approach is warranted if reliable uncertainty
estimates are desired.

The \jultwo\ flare shows evidence for the existence of a \LEC\ in the
range 25--35 keV, just above the region where the thermal emission
dominates.  The \pdf\ of the \LEC\ shows significant non-zero
probability below 25 keV, and zero probability above 50 keV. This peak
is important, as it leads to highly asymmetric \pdfs\ for the total
number of flare electrons accelerated by the flare, and the energy
they carry, in which the upper limit to these quantities are poorly
constrained.  In each of these quantities, the 95\% upper credible
limit is orders of magnitude larger than the MAP value, whilst the
95\% lower limit is within one order of magnitude of the MAP value.
In comparison, the MAP values for the same quantities of the
\janfive\ flare lie are approximately centered within a tenth of a
decade.  This points to the importance of the \LEC\ \pdf\ in
determining the quality of our knowledge of the gross properties of
the flare.

Further work will involve improving the modeling of RHESSI
observations by including data from other RHESSI detectors,
incorporating the simultaneous fitting of the background emission at
the same time as the flare model, and testing different models of
flare emission for the same flare.

\acknowledgments

This work was supported by a NASA ROSES award made under the
opportunity NNH09ZDA001N-SHP entitled ``Investigation of the low energy
cutoff in solar flares'', and by the HESPE (High Energy Solar Physics
Data in Europe) collaboration.  We are grateful to D. van Dyk and
C. A. Young for their helpful suggestions.  CHIANTI is an Atomic
Database Package for Spectroscopic Diagnostics of Astrophysical
Plasmas.  It is a collaborative project involving the Naval Research
Laboratory (USA), the University of Florence (Italy), the University
of Cambridge and the Rutherford Appleton Laboratory (UK).



{\it Facilities:} \facility{RHESSI}.



\appendix

\section{Parallel tempering Markov chain Monte Carlo algorithm}\label{sec:pt}

A significant problem in MCMC is ensuring that the posterior is
explored sufficiently.  The first MCMC algorithms used in this study
did not generate the expected marginal probability distribution of the
low-energy cutoff $E_{c}$ for the flare of \jultwo.  The distribution
arising from these MCMC algorithms showed a single peak with
$p(E_{c})=0$ below some value.  The expected distribution contains a
plateau region of approximately constant non-zero probability density
for values $E_{c} < E_{plateau}$ for some value of $E_{plateau}$
determined from the data (see also Section \ref{sec:comparison}).  The
difference between the expected distribution and those derived from
the MCMC algorithm may be due to either insufficient exploration of
the posterior by the MCMC algorithm, or to some previously unexpected
feature in the flare spectrum.  To test these explanations, a new MCMC
algorithm was implemented to more fully explore the parameter space of
the posterior distribution.

The {\it parallel tempering algorithm} allows one to explore the
parameter space by optionally making easier moves in related spaces
\citep{2005blda.book.....G}. Parallel tempering is based on {\it
  simulated tempering}.  This scheme mimics the physical process of
annealing, whereby a metal is heated and cooled in order to obtain a
more crystalline and therefore lower energy structure.  By analogy,
simulated tempering uses a set of discrete values of a temperature
parameter $T$ to label and describe flatter versions of the original
posterior distributions.  The value $T=1$ is reserved for the the
original posterior distribution.  Higher values of $T$ correspond to
flatter distributions.  In simulated tempering, the distribution is
`warmed up' by increasing $T$.  In these flatter versions, it is
easier for the sampler to jump out of local minima and explore the
full posterior to find the global minimum.  Inferences are drawn from
the $T=1$ sampler.

As above, let $p(H|\sobs,\information)$ be the target posterior distribution we want
to sample; by Bayes' theorem
\begin{equation}
p(H|\sobs,\information ) \propto p(H|\information) \times p(\sobs|H,\information )
\end{equation}
where we have dropped the normalization factor $1/p(\sobs|\information)$.  
Other posterior distributions at different annealing temperatures
$\beta\equiv 1/T$ are constructed as
\begin{eqnarray}
\pi(H|\sobs,\information ,\beta)   & = & p(H|\information) p(\sobs|H,\information )^{\beta} \\
                       & = & p(H|\information)\exp\left( \beta \log\left[ p(\sobs|H,\information ) \right] \right)
\end{eqnarray}
where $0<\beta \le 1$.  The parameter $\beta$ varies from 0 to 1;
$\beta=1$ corresponds to the original, target distribution, with lower
values corresponding to flatter (higher temperature) versions of the
target distribution.

In parallel tempering, multiple MCMC chains are run in parallel at
$n_{T}$ temperatures $\{1, \beta_{0}, \beta_{1}, ..., \beta_{n_{T}}\}$
for $n_{T}>1$.  At intervals, proposals are made to swap the parameter
states at adjacent but randomly selected temperatures.  For example,
at iteration $t$, suppose that the sampler at $\beta_{i}$ has a
parameter $H_{t,i}$, and $\beta_{i+1}$ has a parameter state
$H_{t,i+1}$.  These are the candidate parameter states for swapping.
The swap is accepted with probability
\begin{equation}
r = \min\left\{ \frac
{\pi\left( H_{t,i+1}| \sobs,\beta_{i},\information \right) \pi\left( H_{t,i}| \sobs,\beta_{i+1},\information \right) }
{\pi\left( H_{t,i}| \sobs,\beta_{i},\information \right) \pi\left( H_{t,i+1}| \sobs,\beta_{i+1},\information \right) }
\right\}.
\end{equation}
The swap is accepted if $U_{1} \approx \mbox{Uniform}[0,1] \le r$,
that is, if a number $U_{1}$ drawn from a uniform random distribution
between zero and 1, is less than or equal to $r$.  If the swap is
accepted, then the parameter states are swapped: the chain indexed $i$
now has parameter state $H_{t,i+1}$, and the chain indexed $i+1$ now
has parameter state $H_{t,i}$.  This swapping process propagates
information across the parallel simulations.  At higher temperatures,
the algorithm can explore very different locations in the posterior
parameter space.  At lower temperatures, the algorithm can improve
local knowledge of the space around minima.  Swapping allows highly
probable parameter states to propagate down to lower temperatures
where they can be explored locally.  The swap itself need not be
proposed at every iteration.  \citet{2005blda.book.....G} implements
an example parallel tempering algorithm by allowing a swap on average
once every $n_{s}$ iterations: the swap is only performed if the value
of $U_{2}$, drawn from a uniform distribution between zero and 1, is
less than or equal to $1/n_{s}$.

Each of the MCMC chains uses the Metropolis-Hastings algorithm
\citep{2005blda.book.....G} to explore each $\pi(H|\sobs,\information
,\beta)$. \NormGauss\ distributions were used as the proposal distributions
for Metropolis-Hastings algorithm.  Widths for each proposal
distribution were found after making several shorter exploratory runs
of the $\beta=1$ chain with an adaptive algorithm that varied the
proposal distribution with to generate an acceptance ratio in the
range $0.16 \rightarrow 0.30$ \citep{citeulike:105949}.  For each
variable $\varv$ in each spectral model, a uniform prior is assumed,
that is, $p(\varv)= 1/(\varv_{1} - \varv_{0})$ for $\varv_{0} \le
\varv \le \varv_{1}$ and $p(\varv)=0$ otherwise.  The lower
($\varv_{0}$) and upper ($\varv_{1}$) values are constants.  The
limits($\varv_{0}$) and upper ($\varv_{1}$) and the proposal
distribution step-size are given in Tables \ref{tab:priors:props}.

As is described in the main text, the parallel tempering MCMC
algorithm produces marginal distributions of $E_{c}$ for the \jultwo\
flare consistent with expectations.  The parallel tempering MCMC
algorithm described here was used in the analysis of both the
\janfive\ and \jultwo\ flares.


\section{Implementation of the parallel tempering Markov chain Monte Carlo algorithm}\label{sec:imp}
The results described in the paper arise from implementing the
parallel tempering algorithm described in Section \ref{sec:pt}. Five
temperatures in the algorithm are used: $\beta = {1, 0.75, 0.5, 0.25,
  0.01}$. Each simulation takes 50,000 samples (five times as many
samples as the \MC\ approach of Section \ref{sec:mc}).  The simulation
is run ten times with a different starting point chosen uniformly
randomly in the volume $\vars_{0}-5\mathbf{s}, \vars_{0}+5\mathbf{s}$,
where $\mathbf{s}$ is the size of the proposal distribution step size.
The proposal distribution step size is the square root of the diagonal
elements of the \covmat\ of a least squares fit calculated at
$\vars_{0}$.  The last half of the samples are considered post
burn-in, and are retained.  Convergence between and within the 10
simulation runs is assessed using the $R$ measurement from
\citet{citeulike:105949}.  In all cases, the $R$-measurement was below
approximately 1.1, which may be taken as indicating convergence
\citep{citeulike:105949,2001ApJ...548..224V}.

\begin{deluxetable}{cccc}
\tabletypesize{\scriptsize}
\tablecaption{Details of the prior variable ranges and the proposal
  distribution step-size used in the
  \bmcmc\ analysis of the \janfive\ and \jultwo\ flare data.   Priors
  for each variable are uniform within the stated ranges.  Each
  proposal distribution is \NormGauss, with width as indicated.  See Section \ref{sec:obs}
  for more detail on the choice of model, and Appendix \ref{sec:pt}
  for more detail on the implementation of the \bmcmc\ analysis.
  the proposal distributions are all \NormGauss.\label{tab:priors:props}}
\tablewidth{0pt}
\tablehead{
\colhead{Flare} & 
\colhead{Parameter} & 
\colhead{Prior range} & 
\colhead{Proposal distribution width}
}
\startdata
\textbf{\janfive} & $EM$& $0.77\rightarrow 6.94$& 0.01\\
 & $kT$                                 & $0.68\rightarrow 6.08$& 0.01\\
 & $F_0$ & $0.01\rightarrow 1000$& 0.0004\\
 & $\delta_{1}$ & $1.1\rightarrow 20$& 0.002\\
 & $E_{c}$      & $6.8\rightarrow 290$& 0.17\\
 & $G_{1}$                                       & $3334\rightarrow 33347$& 821\\
 & $G_{2}$                                       & $1279\rightarrow 12790$& 283\\
&&&\\
\textbf{\jultwo} & $EM$ & $0.9\rightarrow 8.14$ & 0.004\\
 & $kT$ & $0.5\rightarrow 8.0$ & 0.001\\
 & $A$ & $0.002\rightarrow 0.3$ & 0.0003\\
 & $\delta_{1}$ & $1.1\rightarrow 50$ & 0.014\\
 & $E_{b}$ & $50 \rightarrow 32000$ & 7.5\\
 & $\delta_{2}$& $1.1 \rightarrow 50$ & 0.007 \\
 & $E_{c}$& $0.01 \rightarrow 50$ & 0.57 \\
\enddata
\end{deluxetable}

\section{Normality of the marginal distributions}\label{app:normaltests}



\begin{figure}
  \includegraphics[width=0.276\textwidth,clip=]{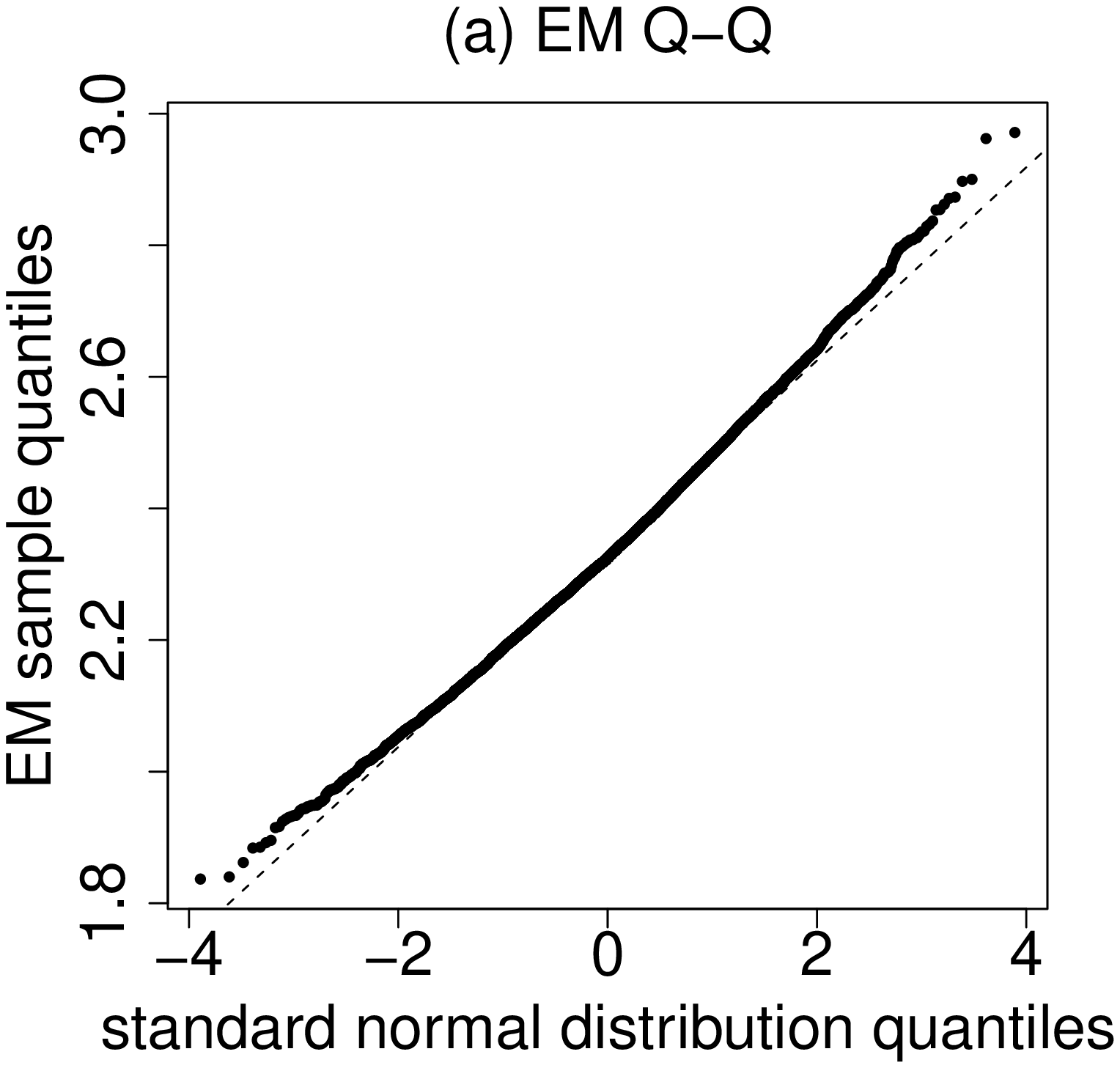}
  \includegraphics[width=0.276\textwidth,clip=]{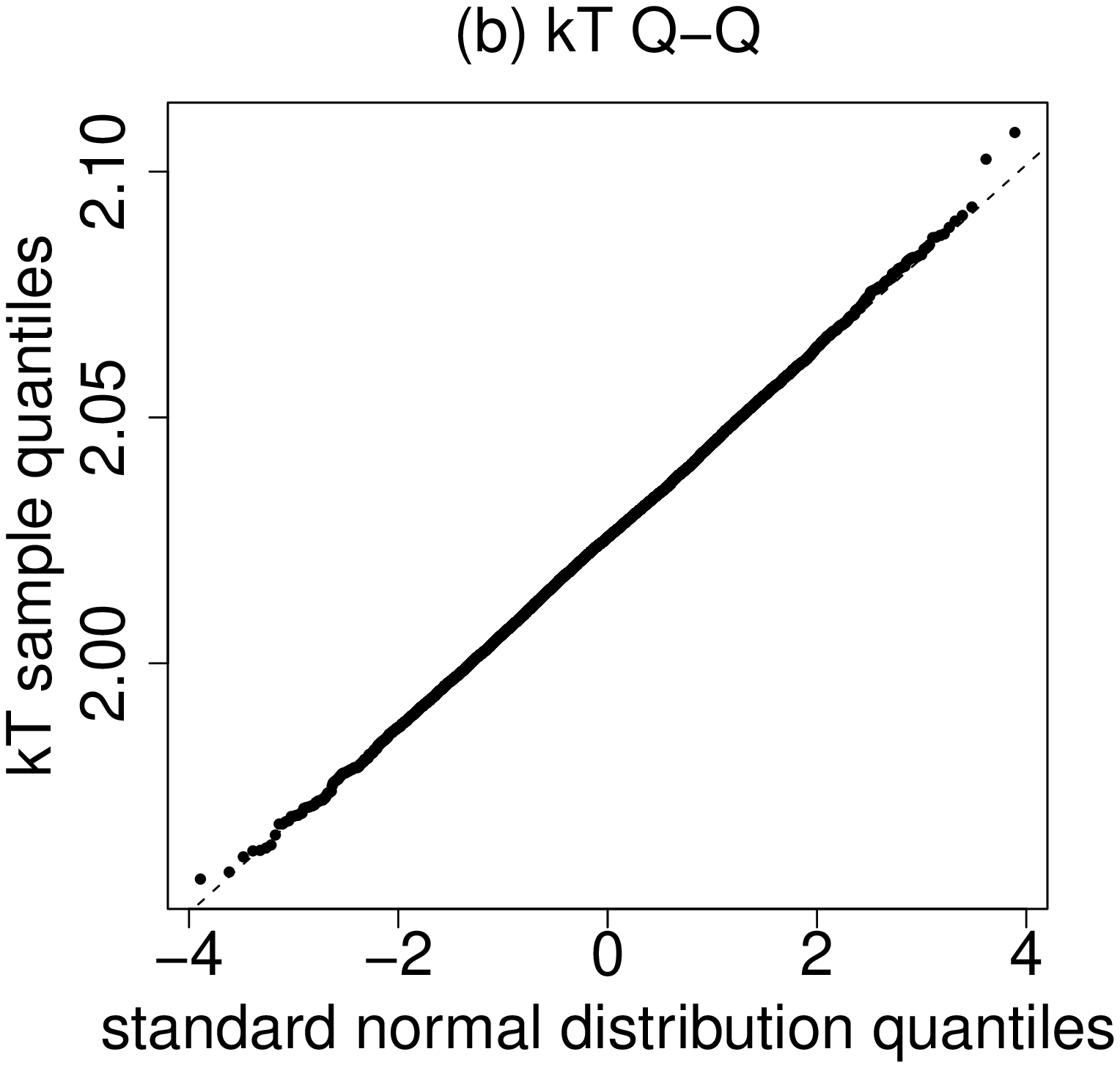}
  \includegraphics[width=0.276\textwidth,clip=]{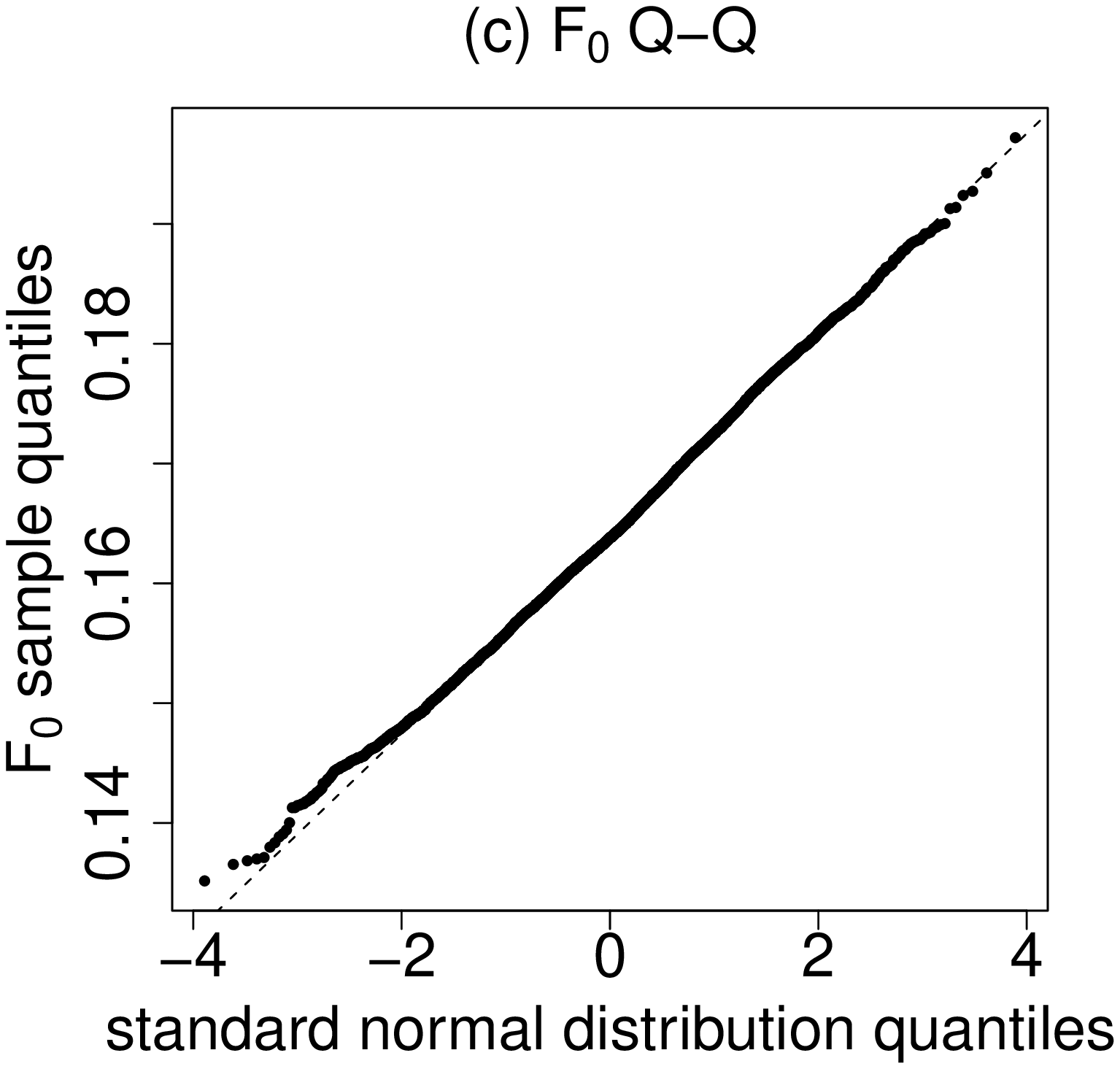}
  \includegraphics[width=0.276\textwidth,clip=]{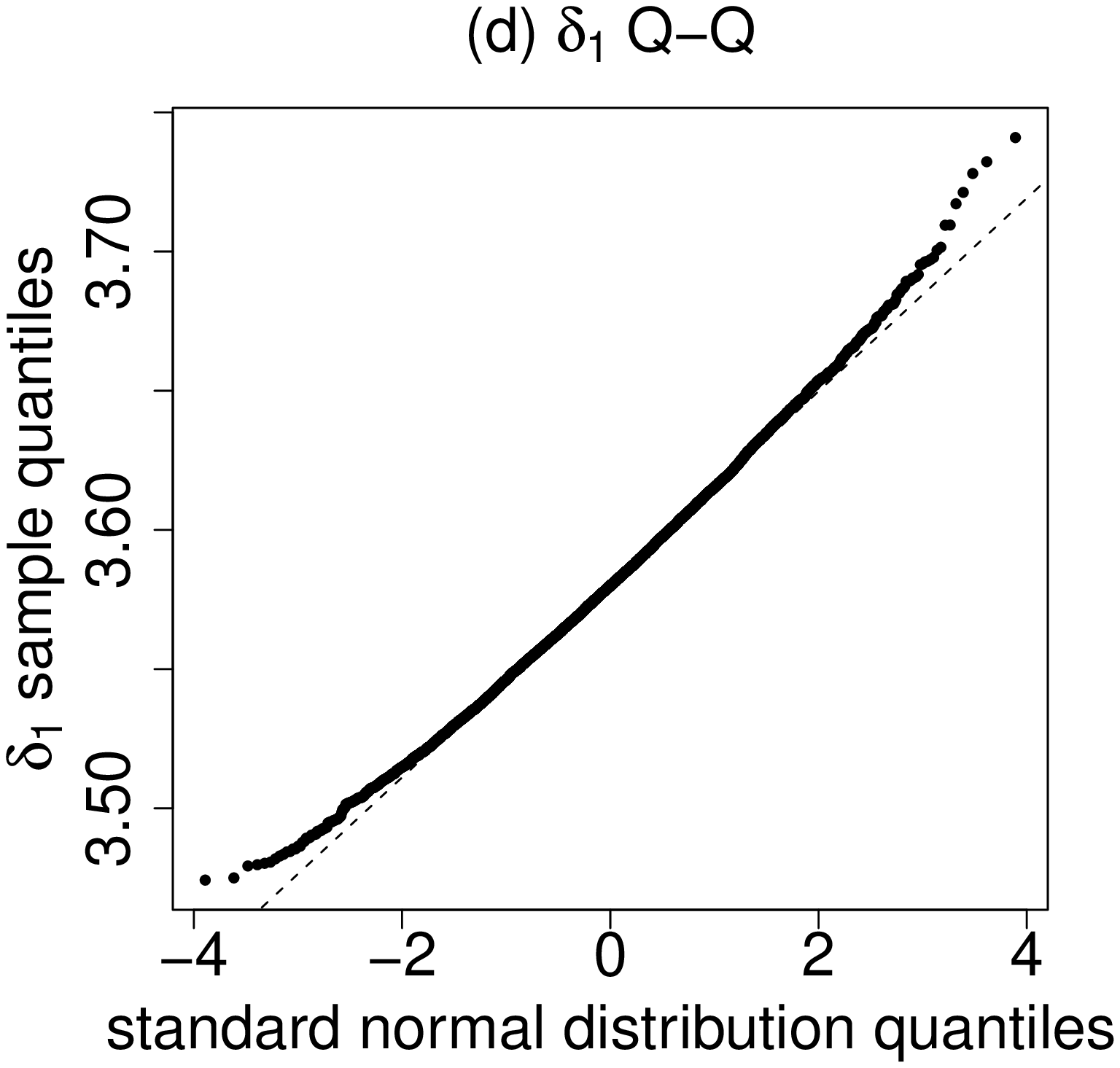}
  \includegraphics[width=0.276\textwidth,clip=]{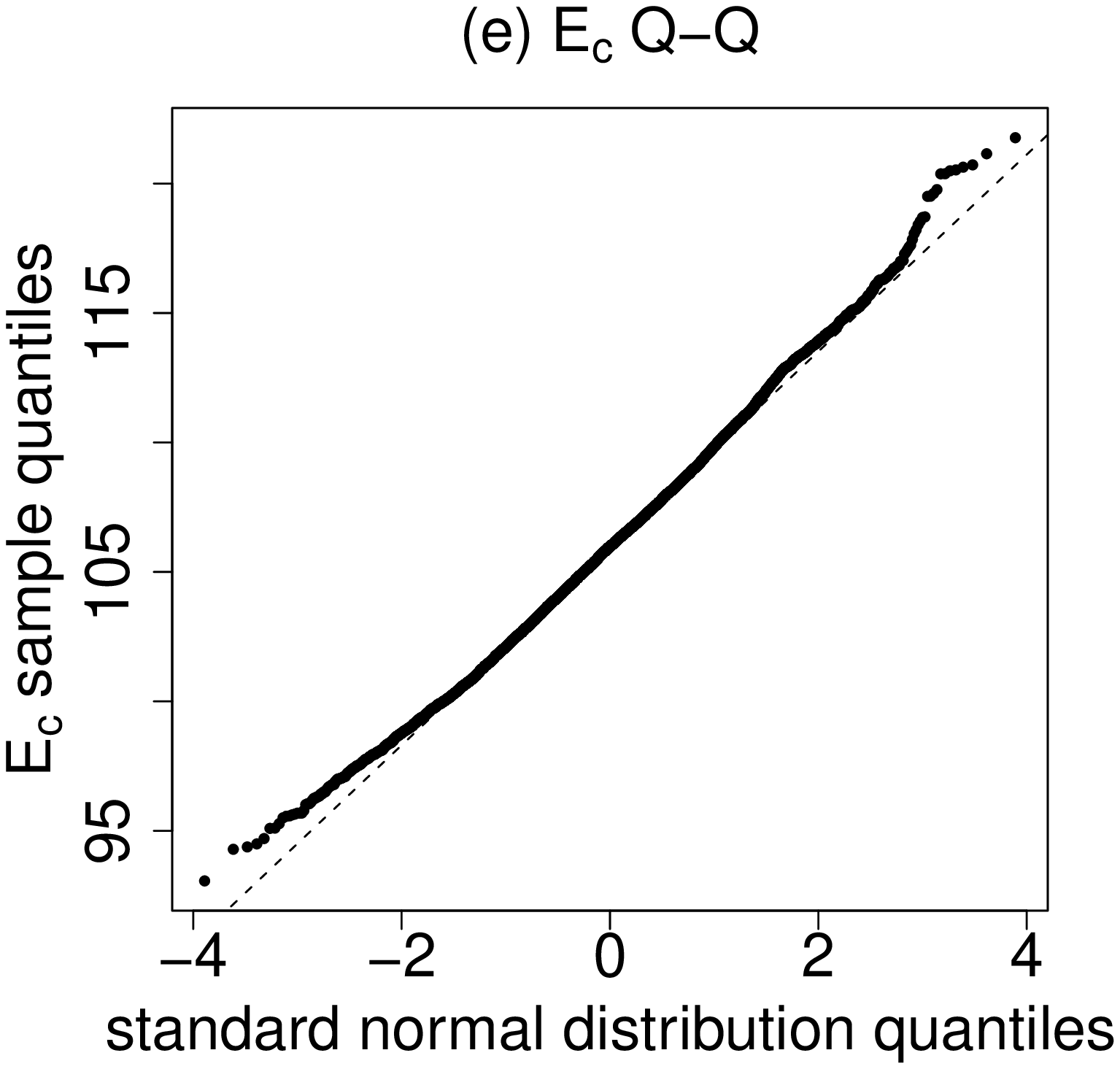}
  \includegraphics[width=0.276\textwidth,clip=]{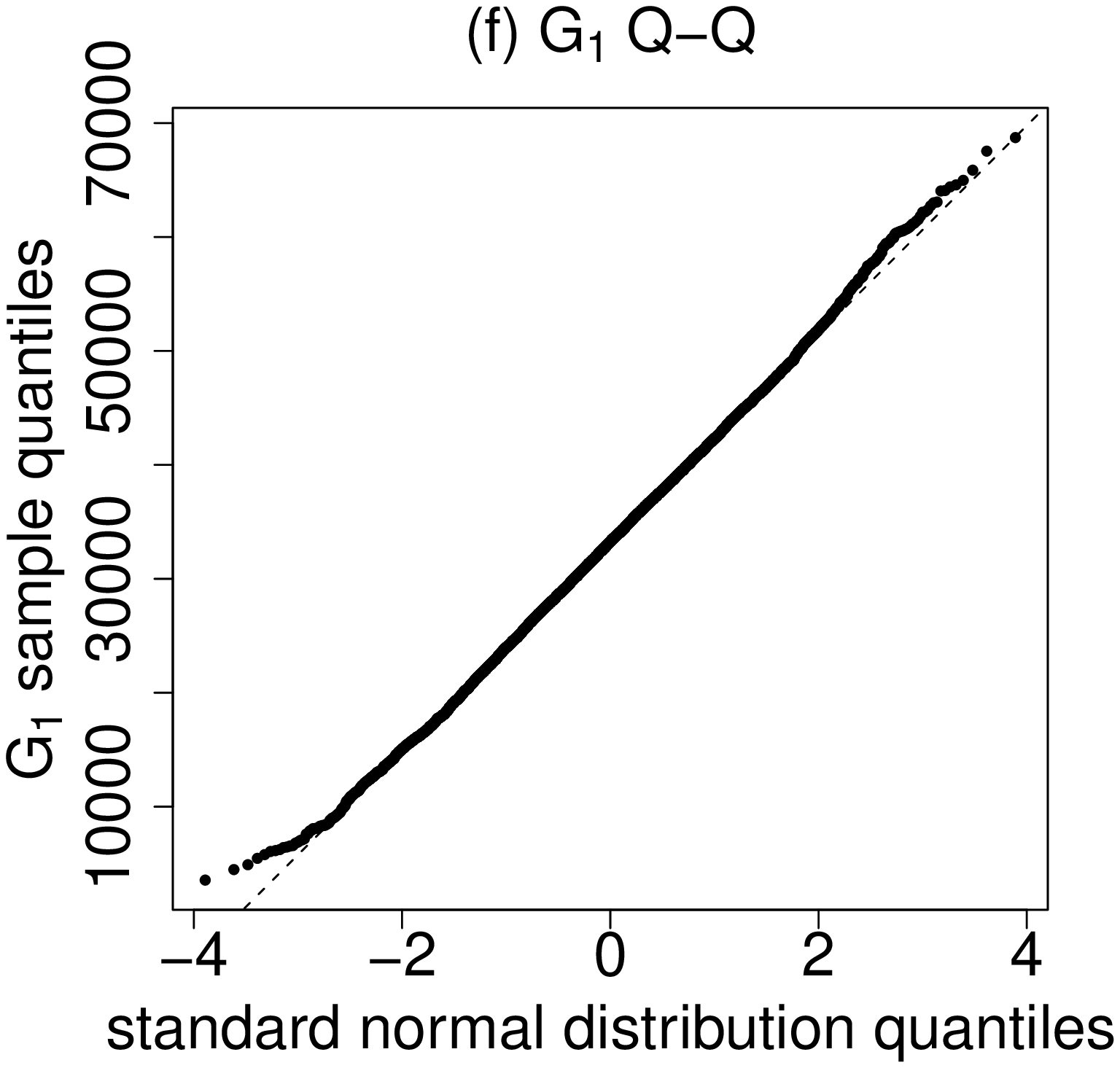}
  \centerline{
  \includegraphics[width=0.276\textwidth,clip=]{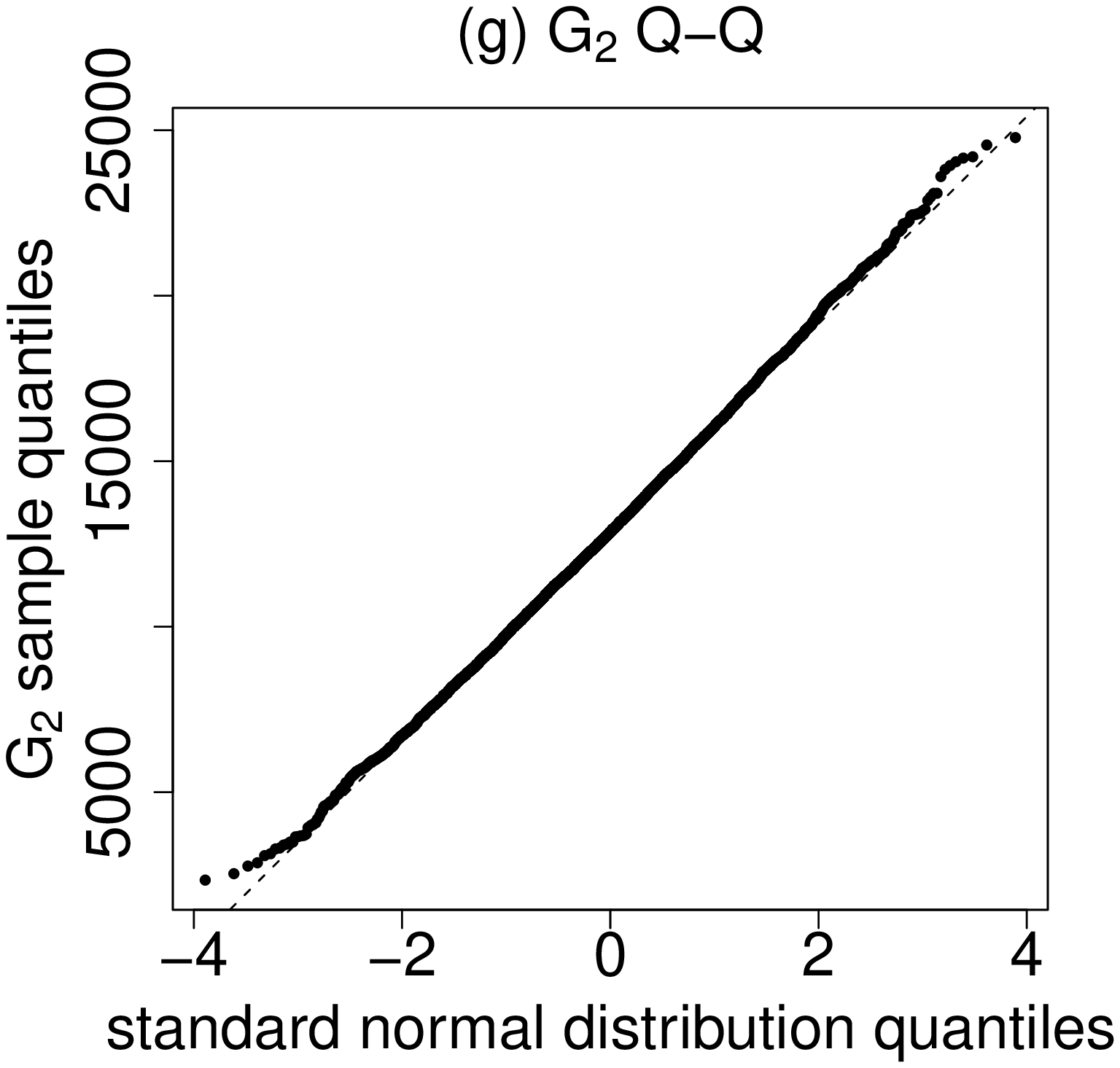}
  }
  \caption{Q-Q plots for the \bmcmc\ samples of the \janfive\ model
    spectrum parameter values.  All parameters are approximately
    \NormGaussly\ distributed in the range $-2,+2$ quantiles about the
    estimated mean.  The tails of the distributions show deviations
    away from a true \NormGauss\ distribution.  Curvature of the sample
    distribution at negative quantiles indicates that the tail is
    thinner than that expected from the sample \NormGauss\ distribution
    $N\left(\hat{\varv_{i}},\hat{\sigma_{\varv_{i}}^{2}}\right)$.
    Similarly, curvature of the sample distribution at large positive
    quantitles indicates that the tail is fatter than that expected
    from the sample \NormGauss\ distribution
    $N\left(\hat{\varv_{i}},\hat{\sigma_{\varv_{i}}^{2}}\right)$. }\label{fig:2005jan19qq}
\end{figure}

\begin{figure}
  \includegraphics[width=0.276\textwidth,clip=]{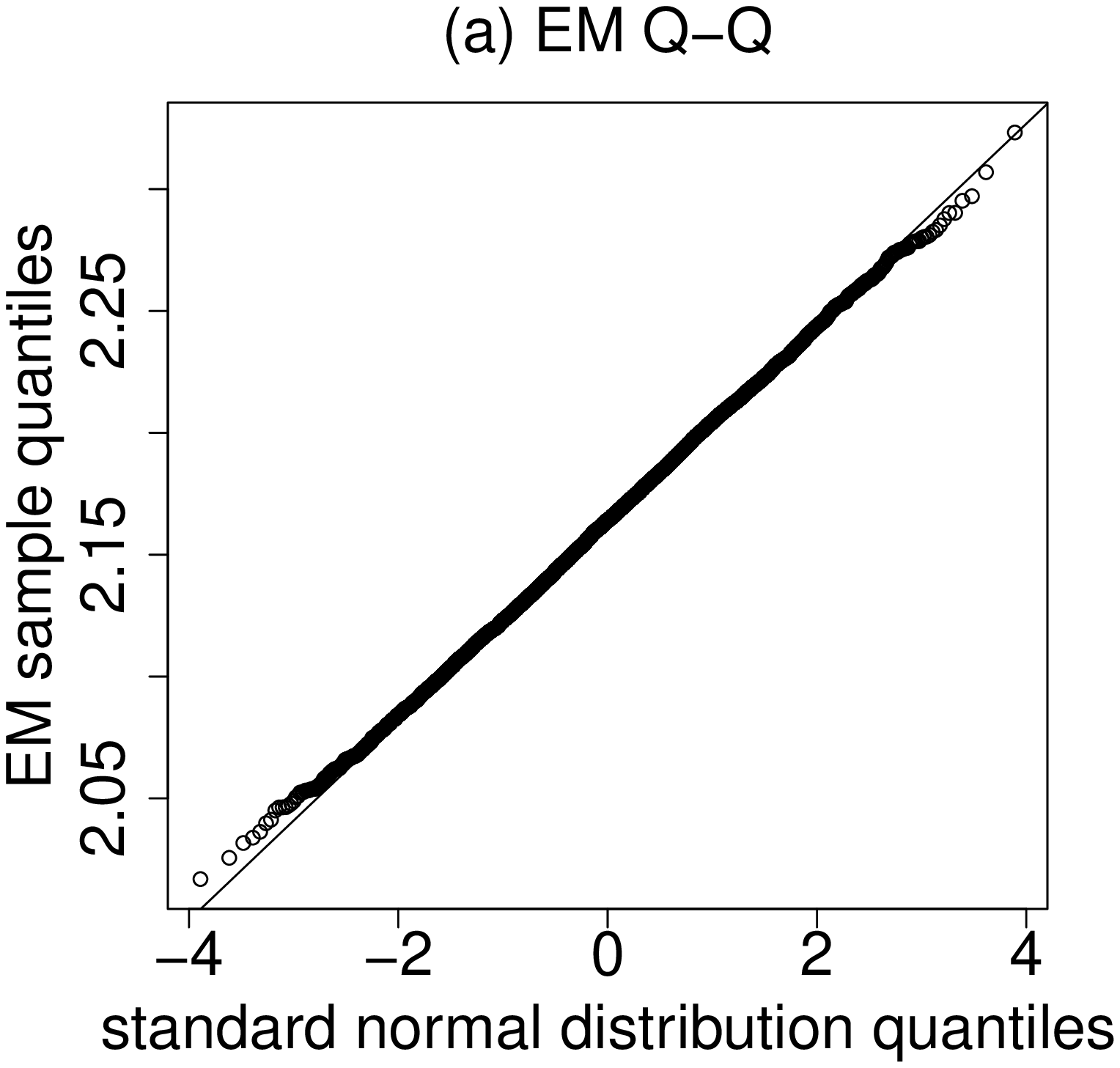}
  \includegraphics[width=0.276\textwidth,clip=]{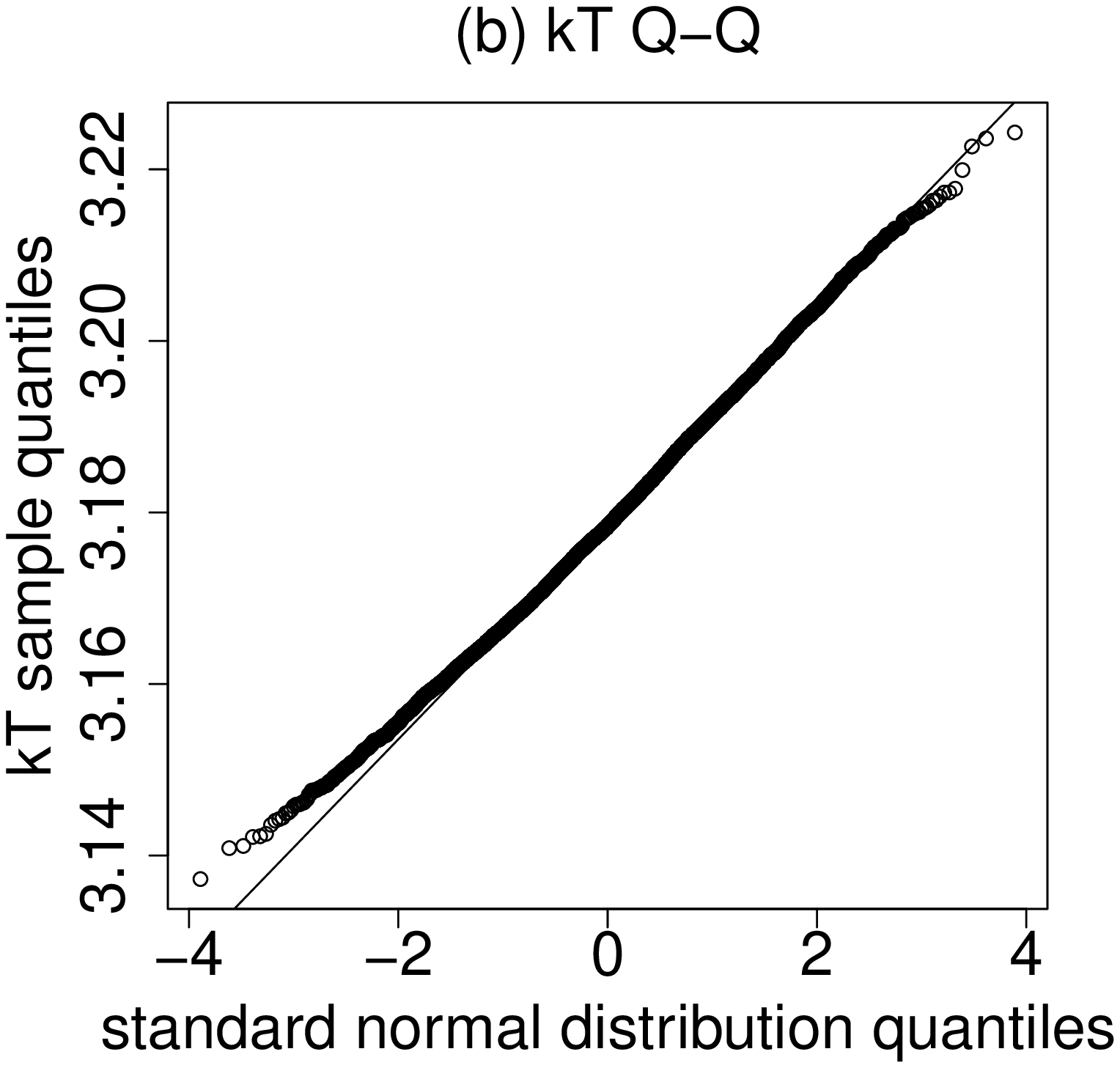}
  \includegraphics[width=0.276\textwidth,clip=]{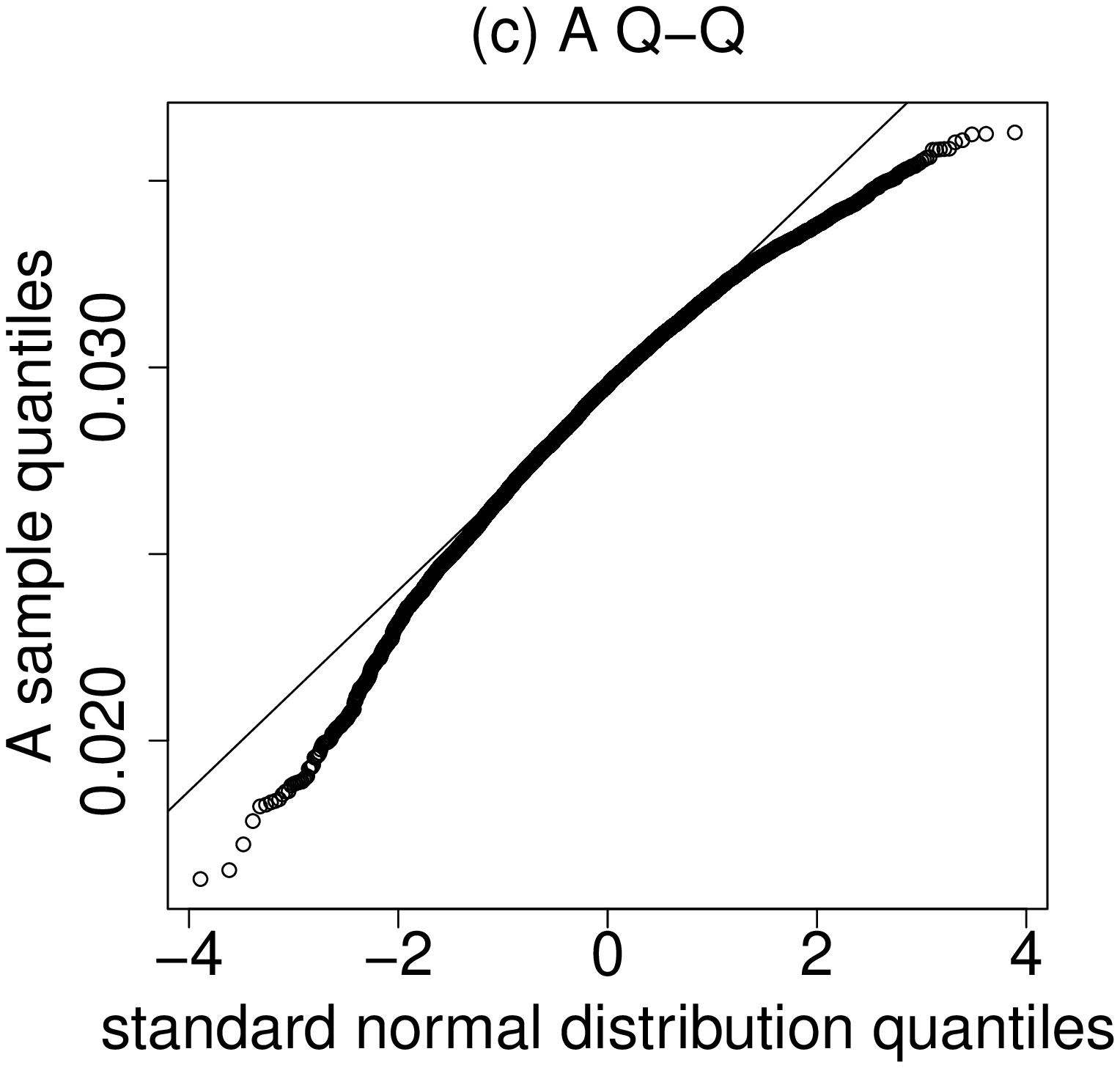}
  \includegraphics[width=0.276\textwidth,clip=]{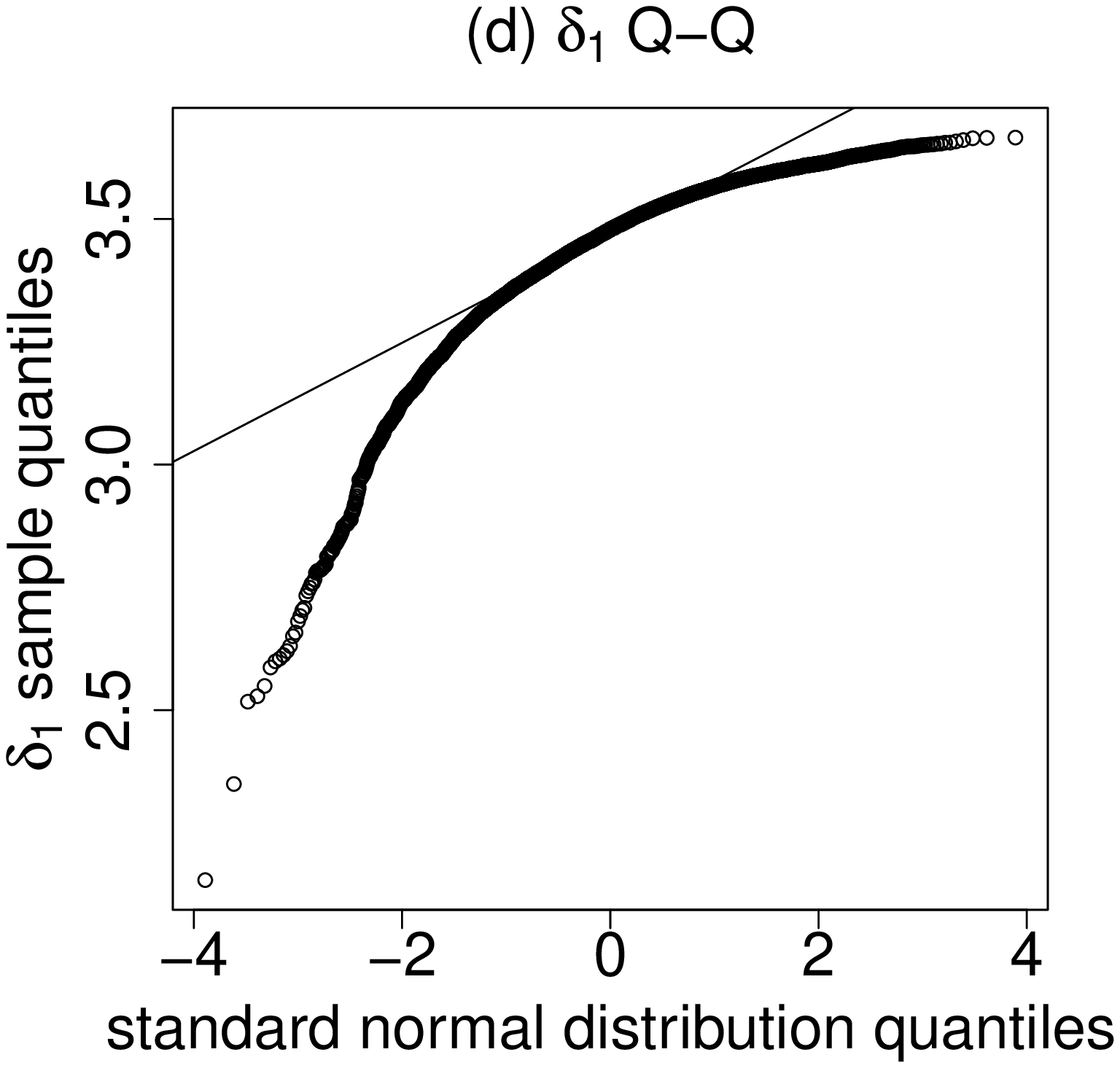}
  \includegraphics[width=0.276\textwidth,clip=]{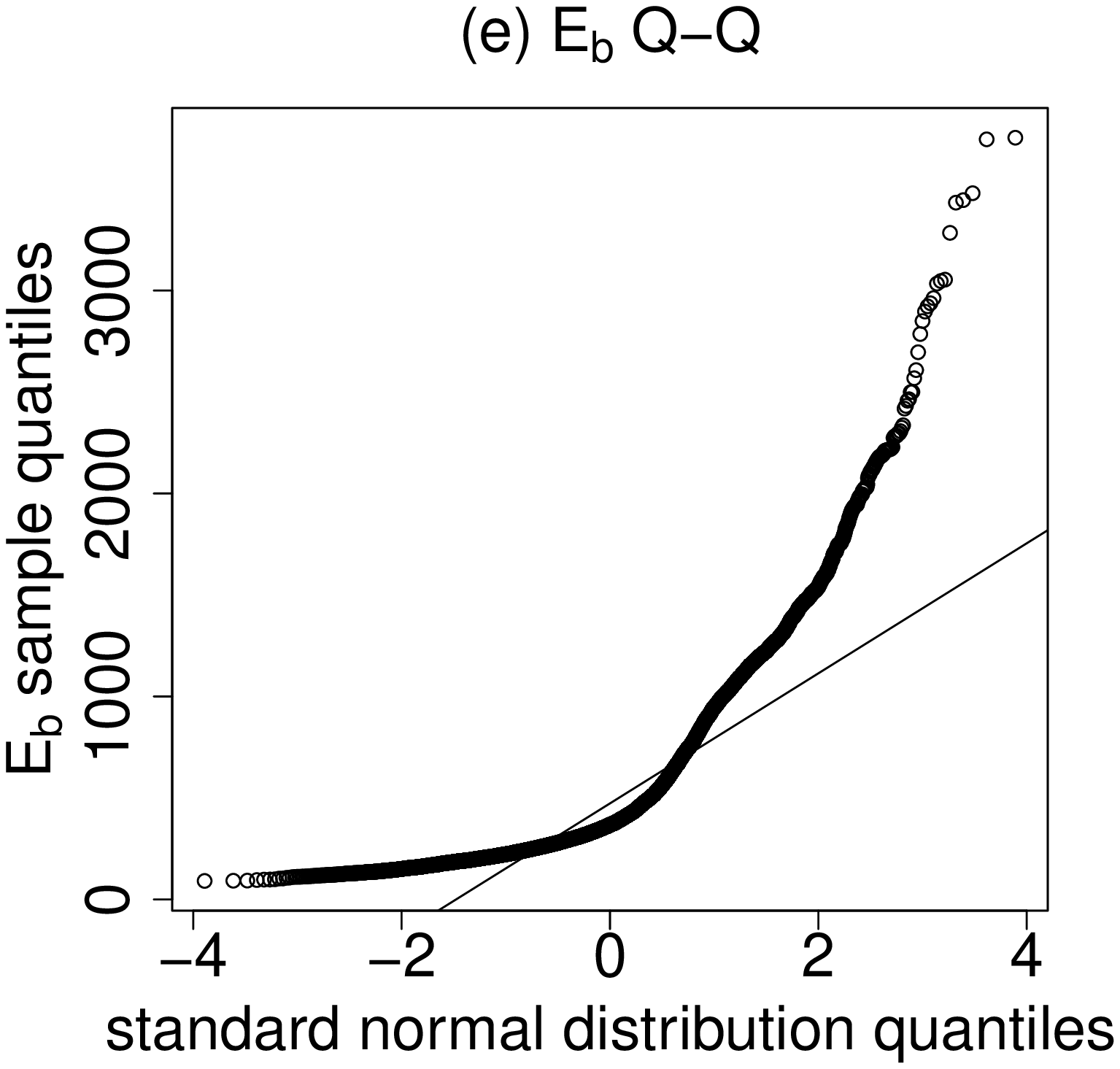}
  \includegraphics[width=0.276\textwidth,clip=]{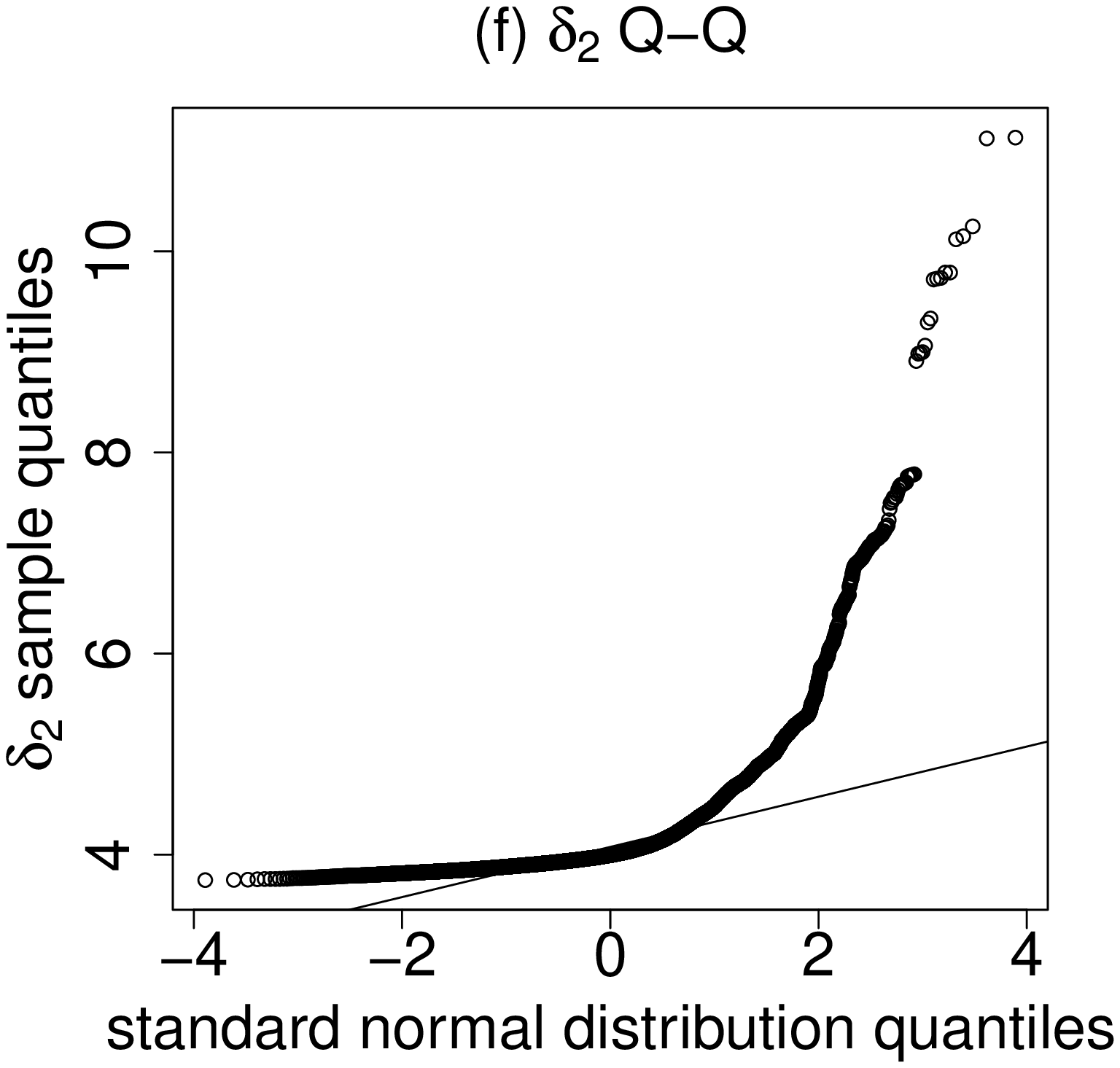}
  \centerline{
  \includegraphics[width=0.276\textwidth,clip=]{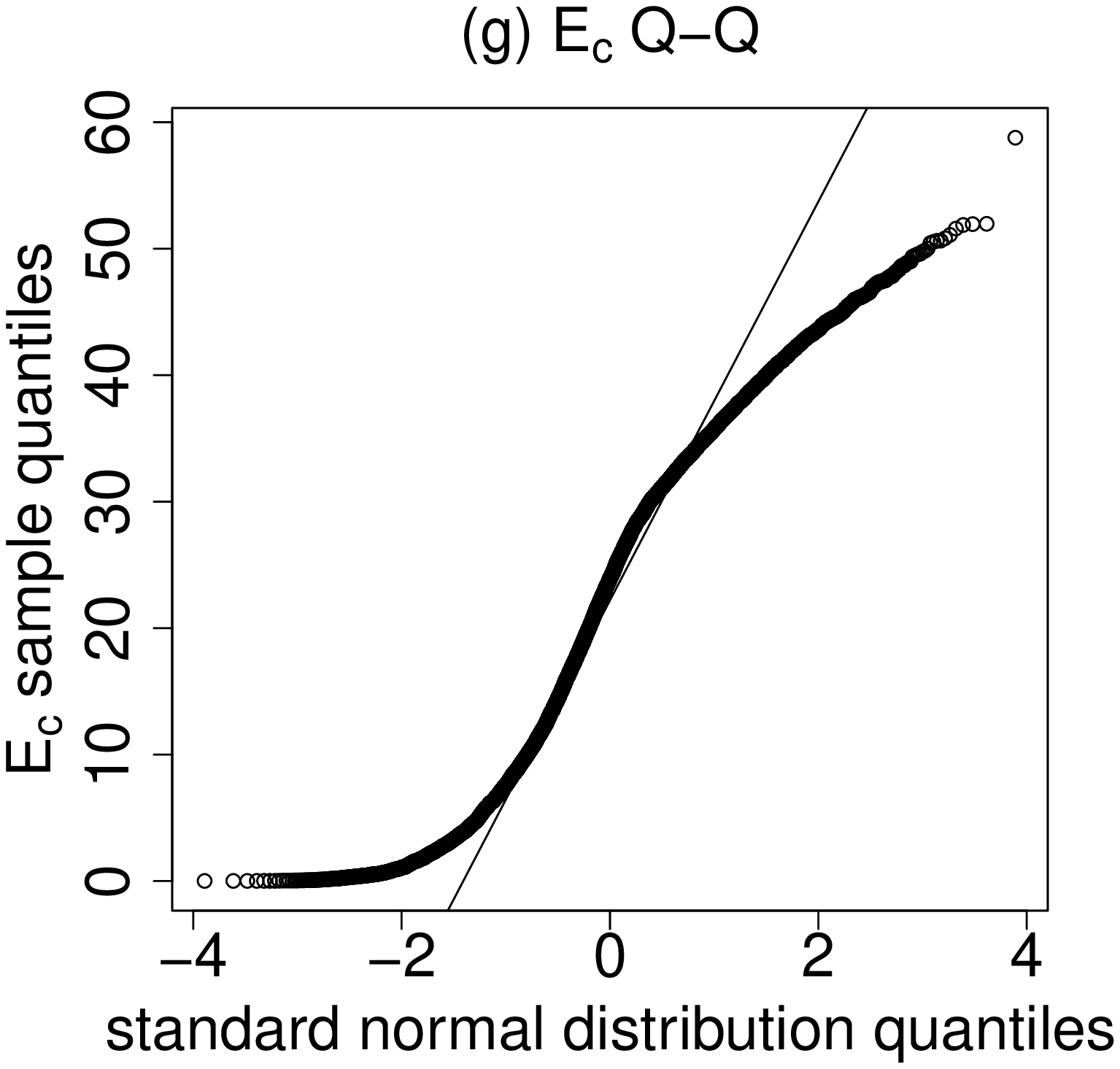}
  }
  \caption{Q-Q plots for the \bmcmc\ samples of the \jultwo\ model
    spectrum parameter values.  The two thermal parameters of the
    model $EM$ and $kT$ appear to be approximately \NormGaussly\
    distributed; the remaining non-thermal parameters ($A$,
    $\delta_{1}$, $E_{b}$, $\delta_{2}$ and $E_{c}$)  are clearly not
    \NormGaussly\ distributed.}\label{fig:2002jul23qq}
\end{figure}

The normality of the univariate marginal distributions was assessed
using Q-Q (quantile-quantile) plots (Figures \ref{fig:2005jan19qq} and
\ref{fig:2002jul23qq}). A Q-Q plot is a graphical method of comparing
two different distributions, and is constructed as follows.  The
cumulative distribution function of a random variable $X$ is defined
as
\begin{equation}
F_{X}(x) = P(X\le x)
\end{equation}
that is, the probability that the random variable $X$ takes on a value
less than or equal to $x$.  The function $F_{X}(x)$ is monotonically
increasing in the range zero to one.  The inverse of $F_{X}$ is called
the quantile function, $Q$, and is defined as
\begin{equation}
Q_{X}(r) = x \mbox{ if } F_{X}(x) = r.
\end{equation}
If $F_{X}$ is a one-to-one function, the inverse $Q$ is uniquely
determined.  If the function $F_{X}$ is not one-to-one the inverse $Q$
can be defined as the weighted average of all relevant points.  The
definition of the quantile function applies to random variables or
sample distributions.  The Q-Q plots shown in Figures
\ref{fig:2005jan19qq} and \ref{fig:2002jul23qq} show a set of
open circles and a straight line.  A circle is plotted at the point
where the abscissa and the ordinate are the values of quantile
functions for the standard \NormGauss\ distribution $N(0,1)$ and the
marginal distribution, for a given value of probability $r$.  A
straight line is drawn through the points defined by the quantile
functions for the standard \NormGauss\ distribution and a \NormGauss\ distribution
$N\left(\hat{\varv_{i}},\hat{\sigma_{\varv_{i}}^{2}}\right)$, where
\[
\hat{\varv_{i}} = \frac{1}{N_{S}}\sum_{j=1}^{N_{S}}[\varv_{i}]_{j}
\] and
\[
  \hat{\sigma_{\varv_{i}}^{2}}=
    \frac{1}{N_{S}-1}\sum_{j=1}^{N_{S}}\left\{[\varv_{i}]_{j}-\hat{\varv_{i}}\right\}^{2}
\]
are estimated from the $N_{S}$ samples of the parameter $\varv_{i}$,
$\varlim{i}$.  The straight line enables an assessment of how closely
the marginal distribution follows a \NormGauss\ distribution, and where any
deviations occur.  The quantile function for the standard \NormGauss\
distribution function is called the probit function and is defined as
\begin{equation}
\mbox{probit}(r) = \sqrt{2}\mbox{ erf}^{-1}(2r-1), r\in(0,1)
\end{equation}
where $\mbox{erf}^{-1}(x)$ is the inverse error function.  The probit
function gives the value of a $N(0,1)$ random variable associated with
specified cumulative probability $r$, for example:
\begin{equation}
\mbox{probit}(0.025) \simeq -1.96 \simeq -\mbox{probit}(0.975).
\end{equation}
Therefore, and conveniently, the abscissa in the Q-Q plots can be
understood as multiples of the standard deviation away from the mean.
The Q-Q plots were implemented using the `R' statistical computing
environment, available from the R Project for Statistical Computing
\citep{rproject}.


\bibliography{brd}

\end{document}